\documentclass[opre,nonblindrev]{informs3_modified1} % current default for manuscript submission

\DoubleSpacedXI % Made default 4/4/2014 at request
\usepackage{endnotes}
\let\footnote=\endnote
\let\enotesize=\normalsize
\def\notesname{Endnotes}%
\def\makeenmark{$^{\theenmark}$}
\def\enoteformat{\rightskip0pt\leftskip0pt\parindent=1.75em
  \leavevmode\llap{\theenmark.\enskip}}

\usepackage{amsmath,amsfonts,amssymb,graphicx,bm}%,arydshln}
\usepackage[font=small,labelfont=bf]{caption}
\usepackage{algorithm}
\usepackage{algorithmic}
\usepackage{graphicx}
\usepackage{caption}
\usepackage{subcaption}
\usepackage{array}
\usepackage{makecell}
\newcolumntype{P}[1]{>{\centering\arraybackslash}p{#1}}
\newcommand{\R}{\mathbb{R}}
\usepackage{natbib}
 \bibpunct[, ]{(}{)}{,}{a}{}{,}%
 \def\bibfont{\small}%
\TheoremsNumberedThrough     % Preferred (Theorem 1, Lemma 1, Theorem 2)
\ECRepeatTheorems

\EquationsNumberedThrough    % Default: (1), (2), ...
\begin{document}
%%%%%%%%%%%%%%%%

%\setcounter{page}{1}
%\DOI{10.1287/xxxx.0000.0000}%

\RUNAUTHOR{Bai et al.}

\RUNTITLE{Calibrating Over-Parametrized Simulation Models}

% \TITLE{Bundling Information Goods of Decreasing Value}
\TITLE{Calibrating Over-Parametrized Simulation Models: A Framework via Eligibility Set}

\ARTICLEAUTHORS{%
\AUTHOR{Yuanlu Bai}
\AFF{Department of Industrial Engineering and Operations Research, Columbia University, New York, NY 10027, \EMAIL{yb2436@columbia.edu}} %, \URL{}}
\AUTHOR{Tucker Balch}
\AFF{JP Morgan AI Research, New York, NY 10179, \EMAIL{tucker.balch@jpmchase.com}}
\AUTHOR{Haoxian Chen}
\AFF{Department of Industrial Engineering and Operations Research, Columbia University, New York, NY 10027, \EMAIL{hc3136@columbia.edu}}
\AUTHOR{Danial Dervovic}
\AFF{JP Morgan AI Research, London, UK, \EMAIL{danial.dervovic@jpmchase.com
}}
\AUTHOR{Henry Lam}
\AFF{Department of Industrial Engineering and Operations Research, Columbia University, New York, NY 10027, \EMAIL{henry.lam@columbia.edu}}
\AUTHOR{Svitlana Vyetrenko}
\AFF{JP Morgan AI Research, Palo Alto, CA 94304, \EMAIL{svitlana.s.vyetrenko@jpmchase.com}}

} % end of the block

% the model parameters in the underlying dynamics
\ABSTRACT{%
Stochastic simulation aims to compute output performance for complex models that lack analytical tractability. To ensure accurate prediction, the model needs to be calibrated and validated against real data. Conventional methods approach these tasks by assessing the model-data match via simple hypothesis tests or distance minimization in an ad hoc fashion, but they can encounter challenges arising from non-identifiability and high dimensionality. In this paper, we investigate a framework to develop calibration schemes that satisfy rigorous frequentist statistical guarantees, via a basic notion that we call eligibility set designed to bypass non-identifiability via a set-based estimation. We investigate a feature extraction-then-aggregation approach to construct these sets that target at multivariate outputs. We demonstrate our methodology on several numerical examples, including an application to calibration of a limit order book market simulator (ABIDES).

%  by using machine learning tools to extract important summaries from the simulated and real outputs followed by an aggregation to endow statistical guarantees
% for financial market model with these data hypothesis testing parameters in models often contain input parameters that need to be properly calibrated to ensure accurate prediction. When output

% Simulation models are ubiquitous for prediction and decision-making across scientific and management disciplines and in model-based learning. As the need to capture increasingly granular system dynamics surges, simulation models also become increasingly complex, often exhibiting over-parametrization and high-dimensionality on inputs and outputs that challenge the reliability of their calibration and downstream tasks. We study a general recipe to address these issues by integrating so-called distributionally robust optimization, a recent branch of robust optimization methods to handle uncertainty in stochastic problems, and high-dimensional hypothesis testing. We present the theoretical guarantees for our framework and illustrate how our approach applies on synthetic and real-data examples in calibrating model parameters and providing confidence bounds on important model quantities.
% Enter your abstract
}%

%\KEYWORDS{deterministic inventory theory; infinite linear programming duality;
%  existence of optimal policies; semi-Markov decision process; cyclic schedule}

%\KEYWORDS{butter, margarine, silliness} \HISTORY{This paper was
%first submitted on April 12, 1922 and has been with the authors for
%83 years for 65 revisions.}

\maketitle
%%%%%%%%%%%%%%%%%%%%%%%%%%%%%%%%%%%%%%%%%%%%%%%%%%%%%%%%%%%%%%%%%%%%%%

% Samples of sectioning (and labeling) in OPRE
% NOTE: (1) \section and \subsection do NOT end with a period
%       (2) \subsubsection and lower need end punctuation
%       (3) capitalization is as shown (title style).
%
%\section{Introduction.}\label{intro} %%1.
%\subsection{Duality and the Classical EOQ Problem.}\label{class-EOQ} %% 1.1.
%\subsection{Outline.}\label{outline1} %% 1.2.
%\subsubsection{Cyclic Schedules for the General Deterministic SMDP.}
%  \label{cyclic-schedules} %% 1.2.1
%\section{Problem Description.}\label{problemdescription} %% 2.

% Text of your paper here

\section{Introduction}

Stochastic simulation aims to compute output performances for complex systems that lack analytical tractability. Other than prediction, this evaluation capability facilitates downstream decision-making tasks including optimization, feasibility test and sensitivity analysis, by changing the design parameter value and monitoring the output from the simulation. Simulation modeling as such has been widely used across multiple disciplines in operations research and management \citep{kelton2000simulation,banks2000dm}, finance \citep{glasserman2013monte}, intelligent systems \citep{go2004accurate,vzlajpah2008simulation,zhao2016accelerated} and model-based learning \citep{ruiz2018learning}.

% Often times, This approach is especially useful when data are only indirectly observed in an aggregate level, and the decision maker requires one to peek into scenarios with essentially no past data, which deems standard machine learning approaches inapplicable.

% They operate by emulating system dynamics believed to adequately resemble the unobserved reality. Once the model is built, a user can discern the outcomes under various scenarios that are unobserved in the past by running the outputs with properly changed input configurations. This

The reliability of simulation models, nonetheless, depends crucially on how well these models capture reality, which in turn calls for the correctness of calibration. The latter refers to the task of choosing the model parameter values, and the associated uncertainty quantification on the inference errors. Often times, data represented by the output of the model are only observed at an aggregate level, without full information on the detailed underlying dynamics of the considered system. This is both the strength and challenge of simulation modeling: By building model for the underlying dynamics, one could answer questions that require looking into unobserved scenarios  by suitably tuning the design configuration, which is beyond what standard machine learning or statistics can provide. On the other hand, without direct data, there could be no consistency in estimating these model parameters, and the reliability of the ultimate model prediction could be in doubt.

% This literature takes the perspective of combining the tasks of validation and calibration iteratively
In stochastic or discrete-event simulation, calibration has been studied mostly assuming direct data on the parameter of interest. In this situation, the focus is on assessing the impact of the statistical error in fitting the parameter that is propagated to the simulation output, known as the input uncertainty or input error (see, e.g., the surveys \citealt{barton2002panel,henderson2003input,chick2006bayesian,barton2012tutorial,song2014advanced,lam2016advanced,nelson2013foundations} Chapter 7, and \citealt{corlu2020stochastic}). This problem has drawn wide attention among the simulation community in recent years, studying methods ranging from the bootstrap \citep{barton1993uniform,cheng1997sensitivity,song2015quickly,lq2018}, finite-difference and delta method \citep{cheng1998two,lin2015single,song2019input}, robust optimization \citep{glasserman2014robust,hu2012robust,ghosh2019robust,lam2016robust,lam2018sensitivity} and relatedly empirical likelihood \citep{lam2016empirical}, sectioning \citep{glynn2017} and Bayesian methods \citep{chick2001input,zouaoui2003accounting,barton2013quantifying,zhu2020risk}. 

% no direct data on the parameters are observable but instead
Nonetheless, when the simulation model validity is checked against output data instead of assuming direct input data, the problem is considerably much more difficult \citep{nelson2016some}. This problem arises routinely in the simulation model construction process, yet a rigorous study appears quite open. Conventionally, this problem is treated under the umbrella of model calibration and validation \citep{sargent2005verification,kleijnen1995verification} which suggests an iterative model refinement process. Once a simulation model is built, it is tested against real data, via two-sample statistical hypothesis tests that compare simulated outputs and real outputs \citep{balci1982some} or Turing tests \citep{schruben1980establishing} (`` validation"). If it is determined from these tests that the model is inadequate, the model would be refined by either expanding the system logic or tuning the model parameters (``calibration"). These two steps are reiterated until the model is satisfactory. Though intuitive, this approach is ad hoc, potentially time-consuming and, moreover, there is no guarantee of a satisfactory model at the end \citep{goeva2019optimization}. The ad-hoc-ness arises because just by locating model parameter values that match the simulated versus real outputs in terms of simple hypothesis tests, there is no guarantee that 1) there exists a unique set of parameter values that gives the match and 2) the simulation model is good enough for output dimensions different from the one being tested.

Issue 1) above regards how to locate model parameter values, and contains two sub-issues on the existence and uniqueness respectively. The existence of matching parameter values means either the model is correctly specified, or it is parametrized in a rich enough manner so that its degree of freedom is more than necessary to include the ``truth", i.e., the model is \emph{over-parametrized}. In the latter case, uniqueness may not hold as there could be more than one set, or even potentially many sets, of parameter values that give the exact match. This issue is known as \emph{non-identifiability} \citep{tarantola2005inverse}. Though this resembles overfitting in machine learning, its underlying challenge and handling methodology are fundamentally different: Whereas one can diagnose overfitting by measuring generalization errors on test sets, in simulation modeling the main goal is to change the design parameters to values that are unobserved before, which by definition have no available test data to start with. 

Issue 2), on the other hand, concerns the adequacy in assessing the match between simulated and real data. Challenges arise from the stochasticity and high-dimensionality of outputs in complex simulation models. In stochastic simulation, the outputs are random by definition and are often in the form of serial trajectory. In the literature this stochasticity and high dimensionality is usually avoided by simply using the means of certain outputs as a summary and assessing the match with real data via classical low-dimensional two-sample hypothesis tests \citep{sargent2005verification,kleijnen1995verification}. However, even if these tests deem an accurate match, it does not necessarily mean the model parameter takes the correct value, because we may have ignored the many possible other output dimensions which, related to the first issue above, exacerbates the non-identifiability problem. 

% Note that, comparing with standard regression or classificationthe outputs from simulation could be multivariate time series or other high-dimensional objects. Under these immense dimensionality issues, one needs to brace for a realistic target accuracy in calibration. That is, even though we may not recover the type of guarantees offered by (arguably easier) regression tasks, we still aim to achieve a level of statistical confidence that allows us to reliably conduct downstream decision-making tasks.

%  where the output is low-dimensional (but potentially high for input features). Indeed, 
% When a simulation model contains a large number of parameters, with the hope of providing enough expressiveness, it can be \emph{over-parametrized} so that many different parameter values could generate outputs that behave similar to the real output data. This leads to so-called non-identifiability . 

Our goal in this paper is to provide a systematic framework to address the two issues above, focusing on \emph{over-parametrized regime with high-dimensional stochastic outputs}. In other words, we presume that the considered simulation model is sufficiently expressive in resembling the true hidden dynamics, but it can contain many parameters. The real data and the model outputs can be multivariate or in the form of time series. Under these presumptions, we first properly define the target as to find what we call an \emph{eligibility set}, namely the set of all possible parameter values such that the real and simulated outputs match, in the idealized situation when there are infinite amount of data on both ends. Note that depending on the over-parametrization and the matching criterion, non-identifiability can arise, and in this case it is theoretically impossible to locate the true parameter value. The eligibility set is defined as the theoretically best (i.e., smallest) set of parameter values that one can possibly aim to obtain under the given degree of non-identifiability. This set contains the true value, and the smaller it is, the more accurate we know regarding the truth. In this way, instead of enforcing a consistent estimation as in classical statistics, we relax the estimation target to a set in order to bypass the non-identifiability issue.

%  We use unsupervised learning approaches to pick up the most important these summaries  effectiveness of these summaries 
The eligibility set provides a theoretical limit on what we can achieve. When finite data are available, estimation error kicks in and we could only attain a correct eligibility set up to a certain statistical confidence. Moreover, the data also affects how we produce the matching criterion. Ideally, we want our estimation to be correct (i.e., the set we create indeed contains the truth with high confidence) and non-conservative (i.e., a small set). To this end, we study a framework using feature extraction-then-aggregation as follows. In the extraction phase, we identify the important ``summaries" for the outputs, where a summary is regarded as effective if it possesses the ability to distinguish an incorrect model output from the truth. In this paper, we simply use unsupervised learning tools, such as penultimate layer extraction from neural network models, to locate the important output characteristics as summaries. While there are considerably other potential better approaches, we will see how our suggestion applies well to examples to a reasonable extent. 

In the next phase, we compare the match between real and simulated data by testing the similarities of these features. Since the number of features could be large in order to well represent the high-dimensional outputs, the individual hypothesis tests on these features need to be properly aggregated to control familywise error. Here, good choices of the statistics and the aggregation method are based on the tractability in calibrating the acceptance-rejection threshold or analyzing its $p$-value (``correct"), and a small resulting Type II error probability on mistakenly accepting a wrong value (``non-conservative"). Based on these considerations, we propose an approach using an aggregation of the Kolmogorov-Smirnov (KS) statistics, which leads to the following analytic tool: Given a parameter value, we can say, with a statistical confidence, whether this value is inside the eligibility set. If it is not, then the value is not the truth with the prescribed confidence. If it is, then the value is in the smallest possible set that includes the truth (though no conclusion is drawn on whether this value is the truth). In short, the more values are ``rejected" or equivalently the smaller is our constructed eligibility set, the more accurate we can locate the truth.

% Therefore, the calibration of such an agent-based model is non-trivial. 
We test our methodology on the calibration of the Agent-Based Interactive Discrete-Event Simulation Environment (ABIDES)~\citep{byrd2020abides,vyetrenko2019get}, a detailed model to emulate market order dynamics. Faced with an increasing demand of a simulation tool to analyze market participant’s behaviors, ABIDES is designed to emulate the interaction of the individual agents with an exchange agent (e.g., mimicking NASDAQ) and output the emergent time series of the bid-ask prices and order volumes. ABIDES enables the simulation of tens of thousands of trading agents, many securities, and can involve various behavioral types of trading agents. The trading agents and exchange agents communicate through a kernel to request the information of the limit order book and perform trading order placement. The calibration of the ABIDES model here means finding a configuration of agents such that the generated price series is similar to observed data -- arising from either an ABIDES simulation with unknown parameters or historical price series. To test the power of our approach, we measure the similarity in terms of stylized facts of the limit order book that can be used as realism metrics \citep{vyetrenko2019get}. The specific stylized facts that we examine include heavy tails and aggregation normality, absence of autocorrelations of the log return distribution, and volatility clustering. In other words, we show that our calibration framework, which operates in ``black-box" regime without knowing the explicit realism metrics in advance, is able to select models that match data with respect to them. 
% We demonstrate that not only can our calibration framework rediscover the true and some close configurations but also the calibrated configurations behave similarly to the truth in terms of the realism metrics.

% . To verify the performance of our calibration framework, we test some of the
%  addition to the calibration against the time series, matching

%  resulting probability of rejecting so that we can decide for each given parameter value its ``likelihood" of being the truth
% Our choice is based on the readiness in calibrating the acceptance-rejection threshold of the aggregate statistic, or in other words the ability to analyze and compute its $p$-value. With these developments,
% whether this aggregation allows statistically correct identification of the threshold e feature tests could be combined in a statistically correct manner. 
% To explain, our goal is to first extracting the important our goal is to , The by setting up a framework, at least to some degree
\subsection{Comparisons with Existing Approaches}\label{sec:comparisons}
We compare our work with several existing approaches and discuss our relative strengths and limitations. First, as mentioned previously, our approach aims to provide a systematic framework to carry out validation-calibration for stochastic simulation that rigorizes the ad hoc suggestions in the past. To this end, we also point out a related line of work on discrete-event or queueing inference that leverages analytic solutions or heavy-traffic approximations of the model to recover input parameters \citep{larson1990queue,whitt1981approximating,mandelbaum1998estimating,bingham1999non,hall2004nonparametric,moulines2007}. Different from these works, our approach does not rely on specific structures of the discrete-event model and assume only a simulatable black-box input-output relation. The closest to our approach is \citet{goeva2019optimization} that uses distributionally robust optimization (DRO) \citep{goh2010distributionally,delage2010distributionally,wiesemann2014distributionally,ben2013robust} to compute bounds for performance measures. Like us, they consider a match between real and simulated data using the KS statistics. However, they do not handle high-dimensional outputs and assume a single (distributional) input and as such, do not explicitly use the concept of eligibility set nor the feature extraction framework that we investigate here. We also mention that a preliminary conference version of the current work has appeared in \citet{bai2020calibrating}, but without the elaborate investigation on the eligibility set guarantees, the feature-based approach to handle high-dimensional outputs, and the detailed numerics.

Second, in computer experiments, the calibration problem admits a long-standing history and is also known as simulation-based or likelihood-free inference, as the likelihood function of data is only accessible via the (black- or grey-box) simulation model \citep{cranmer2020frontier}. This is regarded as an inverse problem in scientific fields as one attempts to recover physical parameters from surrogate models built on physical laws \citep{wunsch1996ocean,shirangi2014history,mak2018efficient}. The predominant approach uses a Bayesian perspective, by placing priors on the unknown input model parameters and computing posterior for inference \citep{kennedy2001bayesian,currin1991bayesian,craig2001,bayarri2007,higdon2008}. Recent work such as \citet{kisamori2020simulator} further improves the extrapolation capability by using covariate shift techniques. The Bayesian approach is advantageously automatic in addressing non-identifiability, as the posterior distribution is universally well-defined. However, as the likelihood function is not directly available, the posterior can often be updated only via approximate Bayesian computation (ABC) \citep{marin2012approximate,robert2016approximate}, by running acceptance-rejection on simulated samples that is computationally costly and requires ad hoc tuning \citep{marjoram2003markov,robert2011lack}. In contrast to this literature, our approach aims for a frequentist guarantee, hinging on our notion of eligibility set to bypass non-identifiability. In terms of computation, we utilize feature extraction to produce a matching instead of looking at the full likelihood, and we do not use the rejection steps in ABC that could require intense computation and tuning.

We regard our work as the first systematic attempt to address model calibration under non-identifiability and high-dimensionality in stochastic simulation. We expect much follow-up work to improve our framework, and currently there are two admitted limitations. First is our presumption on over-parametrized models which means that the ``truth" is recovered with the correct parameter values. Often times, however, the model, even when richly parametrized, could still deviate from the true dynamics and incur a systematic model error -- a discrepancy with the truth even when the parameters are best calibrated.  
This latter error is handled commonly via a Bayesian perspective  by incorporating an associated prior (e.g., \citealt{kennedy2001bayesian,conti2010bayesian}, and also \citealt{plumlee2016learning,lam2017improving} which consider stochastic outputs), but is also addressed from a ``distance minimization" view among the frequentists \citep{tuo2015efficient,tuo2016theoretical,wong2014frequentist}. One can also address this by ensuring sufficient expressiveness of the model \citep{cranmer2020frontier}, which is the perspective we take in this paper.

% even though we attempt to address high-dimensional output, 
Our second perceived limitation is that, when the input dimensionality is high, it could be difficult to identify the full eligibility set (with confidence). Our resulting analytic tool allows one to check whether a given parameter value is ``eligible" or not, but in high-dimensional situations it is difficult to check for all points, even when discretization is used. In this situation, a potential remedial approach is to define the goal as to compute confidence bounds on target performance measure, and use robust optimization to compute these bounds by treating the eligibility set as the constraint, playing the role as the so-called uncertainty set or ambiguity set (see Section \ref{sec:remedy}). Such optimization is a high-dimensional simulation optimization problem for whom its computation approaches will be relegated to future work.

\section{Calibration Framework}
\label{sec:framework}
% \subsection{Problem Setting and Notations}
We consider a family of output probability distributions $\{P^{\theta}:\theta\in\Theta\}$ on $\mathbb R^m$ where $\theta$ is an unknown $d$-dimensional parameter on the parameter space $\Theta\subset\mathbb R^d$. The unknown true parameter is denoted as $\theta_0\in\Theta$. We suppose real-world output data $X_1,\dots,X_N$ drawn from $P^{\theta_0}$ is available. On the other hand, given any $\theta\in\Theta$, we can generate a simulated random sample $Y_1^{\theta},\dots,Y_n^{\theta}$ from $P^{\theta}$ via a black-box machine. Note that, besides simulatability, we do not make any further assumptions on the relationship between $\theta$ and $P^{\theta}$, which can have a complicated structure as exhibited by most large-scale simulation models. Also note that the setup above implies implicitly that the simulation model represented by $P^\theta$ is expressive enough, i.e., over-parametrized, so that it covers the true output distribution.

% or even a black box.

% The latter condition is mild and generally satisfied in stochastic simulation. It is noteworthy that . Our framework applies to the case where the relationship is

Our goal is to infer the true value of $\theta$ from the real output data and our simulation capability. For convenience, we use $P_N^{\theta_0}$ and $P_n^{\theta}$ to denote the empirical probability distributions determined by the real output data set $X_1,\dots,X_N$ and the simulated output data set $Y_1^{\theta},\dots,Y_n^{\theta}$ respectively. That is, 
\begin{equation*}
    P_N^{\theta_0}(\cdot)=\frac{1}{N}\sum_{k=1}^N \delta_{X_k}(\cdot),\ \ P_n^{\theta}(\cdot)=\frac{1}{n}\sum_{j=1}^n \delta_{Y_j^{\theta}}(\cdot)
\end{equation*}
where $\delta$ denotes the Dirac measure. Naturally, we also use $F^{\theta_0}$ and $F^{\theta}$ to denote the distribution functions of $P^{\theta_0}$ and $P^{\theta}$, and correspondingly $F_N^{\theta_0}$ and $F_n^{\theta}$ to denote the respective empirical distribution functions.

%  with the real output data and the simulated output data.
\subsection{Remedying Non-Identifiability via Eligibility Sets}\label{sec:remedy}
To infer $\theta_0$, the standard statistical approach would be to obtain a point estimate as close to $\theta_0$ as possible. However, when the simulation model is over-parametrized as in the setup above, there could be many values other than $\theta_0$ that give rise to the same output pattern, an issue known as non-identifiability. More precisely, the observability of output-level data allows us to only obtain information on the output distribution $P^{\theta_0}$. Consider any statistical distance $d(\cdot,\cdot)$ between two probability distributions that is valid (in particular, the distance is zero if two distributions are identical). The best we can achieve is to find $\theta$ such that $d(P^{\theta},P^{\theta_0})=0$. If there exists $\theta_0'\in \Theta$ such that $\theta_0\ne\theta_0'$ but $d(P^{\theta_0'},P^{\theta_0})=0$, the output data form fundamentally prevents us from identifying which one is the true parameter. This non-identifiability stems from the unobservability of the detailed dynamics that create these data.

% In other words, Therefore, it is challenging to obtain a consistent estimator for $\theta_0$ under the over-parameterized regime. , adopting the terminology from robust optimization (\cite{ben2009robust,bertsimas2011theory}) (as explained further below)

Our idea to bypass the non-identifiability issue is to construct a region that contains the true parameter value instead of trying to get a ``best'' point estimation. We call this region the \emph{eligibility set}. To start with, imagine we have infinitely many real data so that $P^{\theta_0}$ is fully known. In this case, the set $\{\theta\in\Theta:d(P^{\theta},P^{\theta_0})=0\}$ clearly contains the true parameter value $\theta_0$, and moreover it contains the most refined information on the parameter value coming from the output data. 

In practice where we only have a finite real data set of size $N$, we construct a statistically confident eligibility set by relaxing the distance from zero to a small threshold, namely
\begin{equation}
     \mathcal E=\{\theta\in\Theta:d(P^{\theta},P_N^{\theta_0})\leq \eta\}\label{uncertainty set elementary}
 \end{equation}
where $\eta\in\R^+$ is a suitable constant. This construction is highly intuitive: If $P^{\theta}$ is sufficiently close to the empirical distribution $P_N^{\theta_0}$, then we take $\theta$ as an acceptable candidate, and vice versa. The use of \eqref{uncertainty set elementary} is highlighted in the following simple lemma:
\begin{lemma}
If $d(\cdot,\cdot)$ satisfies
\begin{equation}
    \mathbb{P}(d(P^{\theta_0},P_N^{\theta_0})\leq \eta)\geq 1-\alpha\label{threshold guarantee}
\end{equation} 
where $\mathbb P$ refers to the probability with respect to the real data $X_1,\ldots,X_N$, then  $\mathcal E=\{\theta\in\Theta:d(P^{\theta},P_N^{\theta_0})\leq \eta\}$ is a $(1-\alpha)$-level confidence region for $\theta_0$, i.e.,
\begin{equation}
\mathbb{P}(\theta_0\in\mathcal E)\geq 1-\alpha.\label{conclusion}
\end{equation}
Moreover, if \eqref{threshold guarantee} holds asymptotically as $N\to\infty$, then  \eqref{conclusion} holds asymptotically as well.
\label{simple uncertainty lemma}
\end{lemma}
\proof{Proof of Lemma \ref{simple uncertainty lemma}.}
The result follows by noting that $d(P^{\theta_0},P_N^{\theta_0})\leq \eta$ implies $\theta_0\in\mathcal E$.\hfill\Halmos
\endproof

The main implication of Lemma \ref{simple uncertainty lemma} is that, in order to obtain a statistically valid confidence set for parameter $\theta$, it suffices to construct a statistical distance $d(\cdot,\cdot)$ in the nonparametric space. Suppose we have knowledge on the sampling distribution on $d(P^{\theta_0},P_N^{\theta_0})$, then we can obtain $\eta$ as the quantile of this distribution so that  $\{Q:d(Q,P_N^{\theta_0})\leq\eta\}$ is a $(1-\alpha)$-level confidence region for $P^{\theta_0}$, consequently giving rise to a confidence region for $\theta_0$ by ``lifting" from the nonparametric to parametric space. The nonparametric viewpoint here is beneficial as it is typically extremely difficult to find the right parametric class for simulation models (hence the term ``likelihood-free" inference as discussed in Section \ref{sec:comparisons}). On the other hand, it is relatively easy to construct $d(\cdot,\cdot)$ such that we know its universal sampling distribution of $d(Q,Q_N)$, for general output distribution $Q$ and its associated empirical distribution $Q_N$.

An easy but handy observation, which we will utilize heavily in Section \ref{sec:construction}, is that $d(\cdot,\cdot)$ above can be a semi-metric, i.e., $d(P,Q)$ can be 0 not only when $P=Q$ but potentially when $P\neq Q$. Because of this, we can extract a certain dimension or transformation of the output in constructing this $d$. For instance, let $\Pi P$ be a the distribution of a transformation of the output generated from $P$. Then we can consider $d(\Pi P,\Pi Q)$ for any statistical distance $d$ defined on the suitable probability distribution space. If we adopt such a semi-metric, then the non-identifiability should correspondingly be defined with respect to this semi-metric between $P$ and $Q$.

% Clearly, if 
% That is, we use a non-parametric distance $d$ to construct a non-parametric confidence region, which boils down to a parametric one in the parametric setting. This approach puts objective These uncertain parameters taken over an uncertainty set.  eligibility set actually can be viewed as an uncertainty set as well.  

Before moving to more details on $d(\cdot,\cdot)$ and $\eta$, we discuss the relation of \eqref{uncertainty set elementary} to the robust optimization (RO) literature  \citep{ben2009robust,bertsimas2011theory}. The latter advocates decision-making under the worst-case scenario as a principled approach to handle optimization under uncertainty, where the worst case is taken over the uncertain parameters in the model. In computing the worst case, RO places the target performance measure in the objective, and constrain the uncertain parameters to lie in the so-called uncertainty set or ambiguity set, which is a set believed to contain the true parameter with high confidence. In DRO, the uncertain parameters are the underlying probability distributions in a stochastic problem \citep{goh2010distributionally,wiesemann2014distributionally,delage2010distributionally}. In this setting, a common approach to construct the uncertainty set is a neighborhood ball measured by statistical distance such as the Kolmogorov-Smirnov (KS) distance \citep{bertsimas2018robust} that we will utilize, $\phi$-divergence \citep{petersen2000minimax,ben2013robust,glasserman2014robust,lam2016robust,lam2018sensitivity,bayraksan2015data} and Wasserstein distance \citep{esfahani2018data,blanchet2016quantifying,gao2016distributionally}. It can also be constructed based on moment matching \citep{delage2010distributionally,ghaoui2003worst,hu2012robust} or imposition of shape information \citep{popescu2005semidefinite,li2016ambiguous,van2016generalized}. 
Like DRO, our set \eqref{uncertainty set elementary} is a confidence region that is constructed to cover the truth with high confidence and, although our target parameters are finite-dimensional, the set is created via scrutinizing the unknown distributions as in DRO. However, different from this literature, our uncertainty set is constructed by matching the model and the real data at the \emph{output} level as a way to bypass non-identifiability. 

% Moreover, if the dimension of $\theta$ is high, it is often difficult to fully compute $\mathcal U$, and $\psi(P^{\theta_0})$ is viewed as a summary of our inference.
In addition to the similarity between eligibility set and uncertainty set, the worst-case notion in (D)RO that gives rise to bounds on performance measure also plays a role in our framework. Although we do not pursue in this paper, this role is important for two reasons. First is that when the input is high-dimensional, it is difficult to compute the entire eligibility set $\mathcal E$, and focusing on bounds for relevant performance measure simplifies the problem.  Second, in simulation analysis our goal often is to evaluate a target performance measure or conduct a downstream task that utilizes such an evaluation. For concreteness, say this target is $\psi(P^{\theta_0})$ where $\psi$ is a function dependent on $P^{\theta_0}$. The optimization pair
\begin{equation}
    \begin{split}
        \text{maximize/minimize }&\psi(P^{\theta})\\
        \text{subject to }& \theta\in\mathcal E
    \end{split}\label{RO}
\end{equation}
which utilizes the eligibility set $\mathcal E$ as the feasible region, results in valid confidence bounds for $\psi(P^{\theta_0})$. We rigorize this as:
\begin{lemma}
If $\mathcal E$ is a $(1-\alpha)$-level confidence set for the true parameter $\theta_0$, i.e., $\mathbb{P}(\theta_0\in\mathcal E)\geq 1-\alpha$ where $\mathbb P$ refers to the probability with respect to the data, then the optimal values of \eqref{RO}, denoted $Z^*$ and $Z_*$ respectively, satisfy
$$\mathbb P(Z_*\leq\psi(P^{\theta_0})\leq Z^*)\geq1-\alpha$$
That is, $Z^*$ and $Z_*$ constitute $(1-\alpha)$-level upper and lower confidence bounds for $\psi(P^{\theta_0})$.\label{DRO simple lemma}
\end{lemma}
\proof{Proof of Lemma \ref{DRO simple lemma}.}
The result follows by noting that $\theta_0\in\mathcal E$ implies $Z_*\leq\psi(P^{\theta_0})\leq Z^*$, so that
$$\mathbb P(Z_*\leq\psi(P^{\theta_0})\leq Z^*)\geq\mathbb{P}(\theta_0\in\mathcal E)\geq 1-\alpha.$$\hfill\Halmos
\endproof

That is, by combining the nonparametric distance $d$ in Lemma \ref{simple uncertainty lemma} and the DRO formulation in Lemma \ref{DRO simple lemma}, we arrive at confidence bounds for target performance measure $\psi(P^{\theta_0})$, from which one can utilize for many tasks such as optimization and feasibility analyses.

\subsection{Constructing Eligibility Sets: An Elementary Case}\label{sec:simple}
To implement our proposed framework, we need to resolve two questions. First is the choice of distance measure $d$ and the associated constant $\eta$ in order to achieve the guarantee depicted in Lemma \ref{simple uncertainty lemma}. Second, at least in elementary cases where we want to approximate the entire $\mathcal E$, we need to be computationally able to determine whether $d(P^{\theta},P_N^{\theta_0})\leq\eta$ or not for any $\theta\in\Theta$.

For the first question, a good choice of $d$ should satisfy that (i) the resulting uncertainty set is non-conservative, i.e., it is as small as possible in terms of ``volume"; (ii) the associated threshold $\eta$ that satisfies \eqref{threshold guarantee} is obtainable; (iii) $d$ is efficient to compute. For one-dimensional output, we consider the Kolmogorov-Smirnov (KS) distance
\begin{equation*}
    d(P_1,P_2)=\sup_{x\in\R}|F_1(x)-F_2(x)|
\end{equation*}
where $P_1,P_2$ are probability distributions and $F_1,F_2$ are respectively their cumulative distribution functions. Since $d(P^{\theta_0},P_N^{\theta_0})$ is exactly the KS statistic for the goodness-of-fit for $F^{\theta_0}$, its sampling distribution is well-known and the threshold can be readily set as $\eta=q_{1-\alpha}/\sqrt{N}$, where $q_{1-\alpha}$ is the $(1-\alpha)$-quantile of $\sup_{t\in [0,1]}|BB(t)|$ and $BB(\cdot)$ denotes a standard Brownian bridge. This choice of $(d,\eta)$ satisfies \eqref{threshold guarantee} asymptotically as $N$ increases, and thus Theorem \ref{simple uncertainty lemma} applies.
% version of the confidence property.

% the standard corresponding choice of 
% For the second question, 

In practice, the simulation size is also finite as constrained by computational capacity. This requires us to use $P_n^{\theta}$ and $F_n^{\theta}$ to approximate $P^{\theta}$ and $F^{\theta}$, and hence $d(P_n^{\theta},P_N^{\theta_0})$ to approximate $d(P^{\theta},P_N^{\theta_0})$. To summarize, the eligibility set for $\theta$ is constructed as 
\begin{equation}
    \mathcal E=\left\{\theta\in\Theta:\sup_{x\in\R}|F_n^{\theta}(x)-F_N^{\theta_0}(x)|\leq\frac{q_{1-\alpha}}{\sqrt{N}}\right\}
    \label{eqn:thetahat}
\end{equation}
where $q_{1-\alpha}$ is the $(1-\alpha)$-quantile of $\sup_{t\in[0,1]}|BB(t)|$. Given any $\theta\in\Theta$, we can generate a simulated sample $Y_1^{\theta},\dots,Y_n^{\theta}\sim P^{\theta}$ and check whether it is in $\mathcal E$ defined by \eqref{eqn:thetahat}. If there are infinitely many values in the parameter space, then we can choose a finite number of values, say $\theta_1,\dots,\theta_m\in\Theta$ as ``representatives'', which can be either deterministic grid points or randomly sampled. 
% In the high-dimensional case, they can be chosen using stochastic gradient-based search with randomly sampled initial points.

Putting everything together, we propose Algorithm \ref{alg:eligibilityset} to compute the eligibility set for the unknown parameter $\theta$. In the following, we present some theoretical results that guide the amount of simulation runs needed to produce the guarantees offered by Lemma \ref{simple uncertainty lemma}, and the conservativeness level of the resulting eligibility set. Theorem \ref{thm:confidence} first gives a sufficient condition on the simulation size.

\begin{algorithm}
\caption{Constructing eligibility set of $\theta$.}
\begin{algorithmic}[1]
\REQUIRE The real output data $X_1,\cdots,X_N$. The number of candidate $\theta$'s $m$. The simulation replication size $n$. The confidence level $1-\alpha$.
\ENSURE An approximate eligibility set $\hat{\mathcal E}_m$.
\STATE Generate $\theta_1,\cdots,\theta_m\in \Theta$;
\FOR{$i=1,\cdots,m$}
\STATE Generate an i.i.d. random sample of simulated data  $Y_1^{\theta_i},\cdots,Y_n^{\theta_i}\sim P^{\theta_i}$;\\
\STATE Compute 
$c_i=\sup_{x\in\R}\left|\frac{1}{n}\sum_{j=1}^n I\left(Y_j^{\theta_i}\leq x\right)-\frac{1}{N}\sum_{l=1}^N I(X_l\leq x)\right|$;
\ENDFOR 
% \STATE Construct $\hat{\mathcal E}_m=\{\theta_i:c_i\leq q_{1-\alpha}/\sqrt{N}\}$;
\RETURN The eligibility set $\hat{\mathcal E}_m=\{\theta_i:c_i\leq q_{1-\alpha}/\sqrt{N}\}$.
\end{algorithmic}
\label{alg:eligibilityset}
\end{algorithm}

% To make this approximation valid, we require that the simulation size $n$ is much larger than the data size $N$, for which we will provide a theoretical guarantee. Since we use $F_n^{\theta}$ to approximate $F^{\theta}$, additional randomness is introduced and thus we need to show that this substitution is still valid. 

% \subsection{Theoretical Guarantees}  in the above simple setup. These results present
\begin{theorem}
Suppose that $X_1,\cdots,X_N$ is an i.i.d. true sample from $P^{\theta_0}$ and $Y_1^{\theta_0},\cdots,Y_n^{\theta_0}$ is an i.i.d. simulated sample from $P^{\theta_0}$. $F_N^{\theta_0}$ and $F_n^{\theta_0}$ are respectively the empirical distribution functions of the two random samples. If $n=\omega(N)$ as $N\rightarrow\infty$, then 
\begin{equation*}
    \liminf_{n,N\rightarrow\infty}\mathbb{P}\left(\sup_{x\in\R}|F_n^{\theta_0}(x)-F_N^{\theta_0}(x)|\leq\frac{q_{1-\alpha}}{\sqrt{N}}\right)\geq 1-\alpha.
\end{equation*}
\label{thm:confidence}
\end{theorem}
From the view of hypothesis testing, Theorem \ref{thm:confidence} analyzes the probability of Type I error in validating the simulation model outputs. Here, Type I error refers to the eligibility set $\hat{\mathcal E}_m$ excluding the true parameter $\theta_0$ given that $\theta_0$ is chosen as a representative. The theorem states that the Type I error probability is asymptotically bounded by $\alpha$ as long as the simulation size $n$ is of larger order than the real data size $N$. In this sense, we may regard the discretized eligibility set $\hat{\mathcal E}_m$ computed with Algorithm \ref{alg:eligibilityset} as an approximate confidence region for $\theta_0$. 

On the other hand, we also analyze Type II error, i.e., some representative that does not have the same output distribution as the truth is accepted. Theorem \ref{thm:error} provides an upper bound for the probability that $F^{\theta}\neq F^{\theta_0}$ yet $\theta$ is eligible.
\begin{theorem}
Suppose that $X_1,\cdots,X_N$ is an i.i.d. true sample from $P^{\theta_0}$ and $Y_1^{\theta},\cdots,Y_n^{\theta}$ is an i.i.d. simulated sample from $P^{\theta}$. $F_N^{\theta_0}$ and $F_n^{\theta}$ are respectively the empirical distribution functions of the two random samples. $F^{\theta_0}$ and $F^{\theta}$ denote the cumulative distribution functions of $P^{\theta_0}$ and $P^{\theta}$. Suppose that $\sup_{x\in\R}|F^{\theta}(x)-F^{\theta_0}(x)|>0$. For any $\varepsilon_1,\varepsilon_2>0$ such that $\varepsilon_1+\varepsilon_2<\sup_{x\in\R}|F^{\theta}(x)-F^{\theta_0}(x)|$, if 
\begin{equation*}
  N>\left(\frac{q_{1-\alpha}}{\sup_{x\in\R}|F^{\theta}(x)-F^{\theta_0}(x)|-\varepsilon_1-\varepsilon_2}\right)^2,  
\end{equation*}
 then 
\begin{equation*}
    \mathbb{P}\left(\sup_{x\in\R}|F_n^{\theta}(x)-F_N^{\theta_0}(x)|\leq q_{1-\alpha}/\sqrt{N}\right)\leq 2\left(e^{-2n\varepsilon_1^2}+e^{-2N\varepsilon_2^2}\right).
\end{equation*}
\label{thm:error}
\end{theorem}
Theorem \ref{thm:error} states that if a representative $\theta$ gives a different output distribution from $\theta_0$, then for sufficiently large $N$, we have a finite-sample upper bound for the probability that $\theta$ is still included in $\hat{\mathcal E}_m$. Note that the smaller is $\sup_{x\in\R}|F^{\theta}(x)-F^{\theta_0}(x)|$, i.e. the closer is the simulation output distribution to the truth, the larger are the required simulation size $n$ and real data size $N$ in order to guarantee that $\theta$ is distinguished from $\theta_0$ with high probability, which coincides with intuition. With Theorem \ref{thm:error}, we may easily develop a Type II error guarantee for Algorithm \ref{alg:eligibilityset}, as shown in Theorem \ref{thm:guarantee}.
\begin{theorem}
We follow Algorithm \ref{alg:eligibilityset} to obtain $\hat{\mathcal E}_m$. For any $\varepsilon,\varepsilon_1,\varepsilon_2>0$ such that $\varepsilon_1+\varepsilon_2<\varepsilon$, if 
\begin{equation*}
    N>\left(\frac{q_{1-\alpha}}{\varepsilon-\varepsilon_1-\varepsilon_2}\right)^2,
\end{equation*}
then
\begin{equation*}
    \mathbb{P}\left(\exists i=1,\cdots,m \text{ s.t. } \sup_{x\in\R}|F^{\theta_i}(x)-F^{\theta_0}(x)|>\varepsilon,\theta_i\in \hat{\mathcal E}_m\right)\leq 2m\left(e^{-2n\varepsilon_1^2}+e^{-2N\varepsilon_2^2}\right).
\end{equation*}
\label{thm:guarantee}
\end{theorem}
Here, Type II error probability is characterized by the probability that there exists some representative $\theta_i$ such that $\sup_{x\in\R}|F^{\theta_i}(x)-F^{\theta_0}(x)|>\varepsilon$ (i.e. the output distribution given by $\theta_i$ is not too close to the truth) yet it is accepted in the eligibility set $\hat{\mathcal E}_m$. Theorem \ref{thm:guarantee} gives a finite-sample upper bound for this probability. With chosen $\varepsilon,\varepsilon_1,\varepsilon_2$ and sufficiently large real data size $N$, we could suitably choose the simulation size $n$ and the number of representatives $m$ such that with high probability, the eligibility set $\hat{\mathcal E}_m$ only includes representatives that are close enough to the truth in terms of output distribution. In this way, Theorem \ref{thm:guarantee} provides guidance on how to choose $n$ and $m$ according to the real data size $N$. For example, we have the following corollaries. 
\begin{corollary}
We follow Algorithm \ref{alg:eligibilityset} to obtain $\hat{\mathcal E}_m$. If $\log m=o(N)$ and $n=\Omega(N)$ as $N\rightarrow \infty$, then for any $\varepsilon>0$,
\begin{equation*}
    \lim_{m,n,N\rightarrow\infty}\mathbb{P}\left(\exists i=1,\cdots,m \text{ s.t. } \sup_{x\in\R}|F^{\theta_i}(x)-F^{\theta_0}(x)|>\varepsilon,\ \theta_i\in \hat{\mathcal E}_m\right)=0.
\end{equation*}
\label{cor:guarantee1}
\end{corollary}
\begin{corollary}
We follow Algorithm \ref{alg:eligibilityset} to obtain $\hat{\mathcal E}_m$. If $m=o(N)$ and $n=\Omega(N)$ as $N\rightarrow \infty$, then
\begin{equation*}
    \lim_{m,n,N\rightarrow\infty}\mathbb{P}\left(\exists i=1,\cdots,m \text{ s.t. } \sup_{x\in\R}|F^{\theta_i}(x)-F^{\theta_0}(x)|>\sqrt{\frac{\log m}{m}},\ \theta_i\in \hat{\mathcal E}_m\right)=0.
\end{equation*}
\label{cor:guarantee2}
\end{corollary}

Corollary \ref{cor:guarantee1} shows that if $\log(m)=o(N)$ and $n=\Omega(N)$ as $N\to\infty$, then for any $\varepsilon>0$, asymptotically almost surely any eligible representative $\theta_i$ in $\hat{\mathcal E}_m$ satisfies that $\sup_{x\in\R}|F^{\theta_i}(x)-F^{\theta_0}(x)|\leq \varepsilon$. If further we have $m=o(N)$, then Corollary \ref{cor:guarantee2} suggests that asymptotically almost surely any $\theta_i$ in $\hat{\mathcal E}_m$ satisfies $\sup_{x\in\R}|F^{\theta_i}(x)-F^{\theta_0}(x)|\leq \sqrt{\log m/m}$, where $\sqrt{\log m/m}$ shrinks in $m$. Intuitively, as long as the simulation size is sufficient and the number of representatives is moderate, we could confidently conclude that any eligible representative is hardly distinct from the truth.

Finally, we introduce a two-sample variant of the eligibility set. We say $\theta$ is eligible if 
\begin{equation*}
    \sup_{x\in\R}|F_n^{\theta}(x)-F_N^{\theta_0}(x)|\leq \sqrt{\frac{n+N}{nN}}\sqrt{-\frac{1}{2}\log(\alpha/2)}.
\end{equation*}
Here, we are considering the two-sample KS statistic $\sqrt{\frac{nN}{n+N}}\sup_{x\in\R}|F_n^{\theta}(x)-F_N^{\theta_0}(x)|$ and approximating $q_{1-\alpha}$ with $\sqrt{-\frac{1}{2}\log(\alpha/2)}$. We summarize the two-sample version calibration framework in Algorithm \ref{alg:eligibilityset_variant}.
\begin{algorithm}
\caption{Constructing eligibility set of $\theta$ (two-sample variant).}
\begin{algorithmic}[1]
\REQUIRE The real output data $X_1,\cdots,X_N$. The number of candidate $\theta$'s $m$. The simulation replication size $n$. The confidence level $1-\alpha$.
\ENSURE An approximate eligibility set $\hat{\mathcal E}_m^v$.
\STATE Generate $\theta_1,\cdots,\theta_m\in \Theta$;
\FOR{$i=1,\cdots,m$}
\STATE Generate an i.i.d. random sample of simulated data  $Y_1^{\theta_i},\cdots,Y_n^{\theta_i}\sim P^{\theta_i}$;\\
\STATE Compute 
$c_i=\sup_{x\in\R}\left|\frac{1}{n}\sum_{j=1}^n I\left(Y_j^{\theta_i}\leq x\right)-\frac{1}{N}\sum_{l=1}^N I(X_l\leq x)\right|$;
\ENDFOR 
% \STATE Construct $\hat{\mathcal E}_m=\{\theta_i:c_i\leq q_{1-\alpha}/\sqrt{N}\}$;
\RETURN The eligibility set $\hat{\mathcal E}_m^v=\left\{\theta_i:c_i\leq \sqrt{\frac{n+N}{nN}}\sqrt{-\frac{1}{2}\log(\alpha/2)}\right\}$.
\end{algorithmic}
\label{alg:eligibilityset_variant}
\end{algorithm}

Similarly, we first provide a theoretical guarantee for Type I error probability. 
\begin{theorem}
    Suppose that $X_1,\cdots,X_N$ is an i.i.d. true sample from $P^{\theta_0}$ and $Y_1^{\theta_0},\cdots,Y_n^{\theta_0}$ is an i.i.d. simulated sample from $P^{\theta_0}$. $F_N^{\theta_0}$ and $F_n^{\theta_0}$ are respectively the empirical distribution functions of the two random samples. If $n=N\geq 4$, then 
    \begin{equation*}
        \mathbb{P}\left(\sup_{x\in\R}|F_n^{\theta_0}(x)-F_N^{\theta_0}(x)|\leq \sqrt{\frac{n+N}{nN}}\sqrt{-\frac12\log(\frac{\alpha}{2})}\right)\geq 1-1.085\alpha.
    \end{equation*}
    If $n=N\geq 458$, then 
    \begin{equation*}
        \mathbb{P}\left(\sup_{x\in\R}|F_n^{\theta_0}(x)-F_N^{\theta_0}(x)|\leq \sqrt{\frac{n+N}{nN}}\sqrt{-\frac12\log(\frac{\alpha}{2})}\right)\geq 1-\alpha.
    \end{equation*}
\label{thm:confidence_variant}
\end{theorem}

In contrast to Theorem \ref{thm:confidence}, Theorem \ref{thm:confidence_variant} provides finite-sample bounds for the Type I error probability instead of asymptotic guarantees. Moreover, while Theorem \ref{thm:confidence} requires that $n$ grows in a higher order than $N$, here we let $n=N$. From the theorem, as long as $n=N\geq 4$, Type I error probability is close to achieving the nominal level $\alpha$, and as long as $n=N\geq 458$, Type I error probability is fully controlled by $\alpha$. 

Regarding Type II error probability, we also have the following theorems.
\begin{theorem}
Suppose that $X_1,\cdots,X_N$ is an i.i.d. true sample from $P^{\theta_0}$ and $Y_1^{\theta},\cdots,Y_n^{\theta}$ is an i.i.d. simulated sample from $P^{\theta}$. $F_N^{\theta_0}$ and $F_n^{\theta}$ are respectively the empirical distribution functions of the two random samples. $F^{\theta_0}$ and $F^{\theta}$ denote the cumulative distribution functions of $P^{\theta_0}$ and $P^{\theta}$. Suppose that $\sup_{x\in\R}|F^{\theta}(x)-F^{\theta_0}(x)|>0$. For any $\varepsilon_1,\varepsilon_2>0$ such that $\varepsilon_1+\varepsilon_2<\sup_{x\in\R}|F^{\theta}(x)-F^{\theta_0}(x)|$, if 
\begin{equation*}
  \frac{nN}{n+N}>\frac{-\frac12\log(\alpha/2)}{(\sup_{x\in\R}|F^{\theta}(x)-F^{\theta_0}(x)|-\varepsilon_1-\varepsilon_2)^2},  
\end{equation*}
 then 
\begin{equation*}
    \mathbb{P}\left(\sup_{x\in\R}|F_n^{\theta}(x)-F_N^{\theta_0}(x)|\leq \sqrt{\frac{n+N}{nN}}\sqrt{-\frac{1}{2}\log(\alpha/2)}\right)\leq 2\left(e^{-2n\varepsilon_1^2}+e^{-2N\varepsilon_2^2}\right).
\end{equation*}
\label{thm:error_variant}
\end{theorem}

\begin{theorem}
We follow Algorithm \ref{alg:eligibilityset_variant} to obtain $\hat{\mathcal E}_m^v$. For any $\varepsilon,\varepsilon_1,\varepsilon_2>0$ such that $\varepsilon_1+\varepsilon_2<\varepsilon$, if 
\begin{equation*}
    \frac{nN}{n+N}>\frac{-\frac12\log(\alpha/2)}{(\varepsilon-\varepsilon_1-\varepsilon_2)^2},
\end{equation*}
then
\begin{equation*}
    \mathbb{P}\left(\exists i=1,\cdots,m \text{ s.t. } \sup_{x\in\R}|F^{\theta_i}(x)-F^{\theta_0}(x)|>\varepsilon,\theta_i\in \hat{\mathcal E}_m^v\right)\leq 2m\left(e^{-2n\varepsilon_1^2}+e^{-2N\varepsilon_2^2}\right).
\end{equation*}
\label{thm:guarantee_variant}
\end{theorem}

The implications of Theorem \ref{thm:error_variant} and \ref{thm:guarantee_variant} are similar to Theorem \ref{thm:error} and \ref{thm:guarantee}, except that now $n$ and $N$ need to together satisfy some conditions in order that the finite-sample bounds stated in the theorems are valid. We could also derive corollaries on how to choose $m$ and $n$ according to $N$ to control Type II error probability. Indeed, Corollary \ref{cor:guarantee1} and \ref{cor:guarantee2} still hold if we replace $\hat{\mathcal E}_m$ with $\hat{\mathcal E}_m^v$. That is, we have the following corollaries:
\begin{corollary}
We follow Algorithm \ref{alg:eligibilityset_variant} to obtain $\hat{\mathcal E}_m^v$. If $\log m=o(N)$ and $n=\Omega(N)$ as $N\rightarrow \infty$, then for any $\varepsilon>0$,
\begin{equation*}
    \lim_{m,n,N\rightarrow\infty}\mathbb{P}\left(\exists i=1,\cdots,m \text{ s.t. } \sup_{x\in\R}|F^{\theta_i}(x)-F^{\theta_0}(x)|>\varepsilon,\ \theta_i\in \hat{\mathcal E}_m^v\right)=0.
\end{equation*}
\label{cor:guarantee1_variant}
\end{corollary}
\begin{corollary}
We follow Algorithm \ref{alg:eligibilityset_variant} to obtain $\hat{\mathcal E}_m^v$. If $m=o(N)$ and $n=\Omega(N)$ as $N\rightarrow \infty$, then
\begin{equation*}
    \lim_{m,n,N\rightarrow\infty}\mathbb{P}\left(\exists i=1,\cdots,m \text{ s.t. } \sup_{x\in\R}|F^{\theta_i}(x)-F^{\theta_0}(x)|>\sqrt{\frac{\log m}{m}},\ \theta_i\in \hat{\mathcal E}_m^v\right)=0.
\end{equation*}
\label{cor:guarantee2_variant}
\end{corollary}

Generally, we would recommend using the two-sample variant compared to one-sample since the former does not require that the simulation size $n$ grows faster than the real  size $N$. Instead, $n$ could be equal to $N$, which reduces the computational load. However, when we already have a large number of simulation runs ($n\gg N$) (e.g., used to train the ``features" for high-dimensional outputs discussed in Section \ref{sec:construction}), we are contented with the one-sample version.

\section{Elementary Numerics}\label{sec:elementary}
In this section, we apply our calibration framework to calibrate simple queueing models. We will show how our methodology recovers the parameters which are nearly non-identifiable from the truth. 
% Then we apply the calibration framework to calibrate a well known financial trading simulation platform called ABIDES. In ABIDES model calibration, we compare the combinations of different feature extraction techniques and different aggregation statistics, test the robustness of the calibration framework and perform an ablation study for different feature extraction techniques with different network structures.

% \subsection{Queueing Model Calibration}
%\subsection{Queueing Model Examples}
\subsection{An M/M/1 Example}
First, we apply our methodology on a simple M/M/1 queueing model example. The interarrival time distribution is $Exp(\lambda)$ and the service time distribution is $Exp(\mu)$. Moreover, we follow the first-come first-served (FCFS) discipline. Our goal is to calibrate $(\lambda,\mu)$ from the average sojourn time of the first 100 customers. The parameter space is set as $(0,2)^2$ and the true value is $(0.5,1)$. 

\begin{figure}[htbp]
    \centering
    \includegraphics[width=0.5\textwidth]{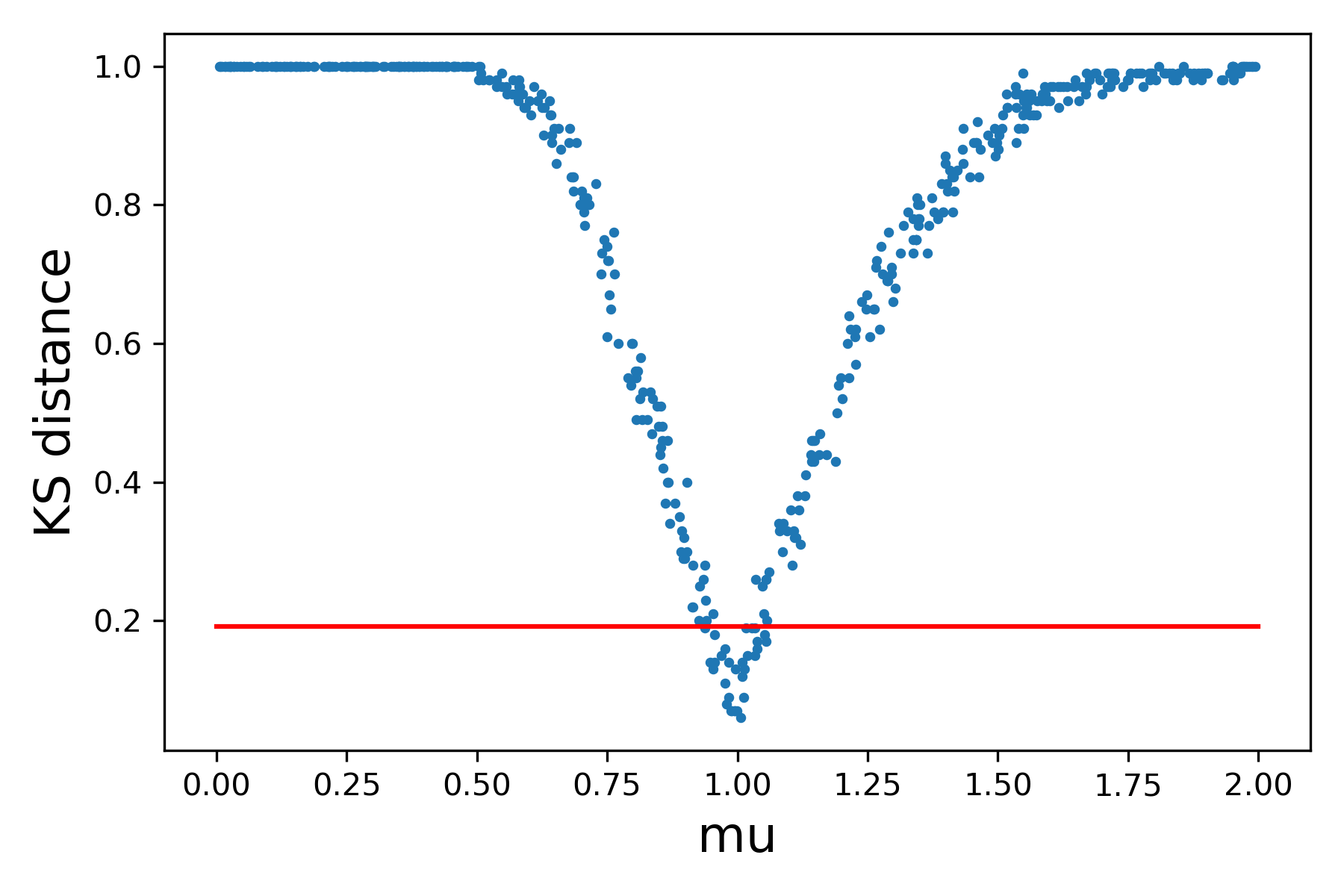}
    \caption{Numerical result for calibrating $\mu$ in the M/M/1 example: KS distance against $\mu$.}
    \label{fig:MM1_mu}
\end{figure}

First of all, to show that our framework could recover the true parameter when the problem is identifiable, we suppose that the true value of $\lambda$ is known and we only calibrate $\mu$. We use the two-sample version of Algorithm \ref{alg:eligibilityset} to obtain the eligibility set with $m=500$, $n=N=100$ and $\alpha=0.5$. Figure \ref{fig:MM1_mu} shows the approximated KS distance against the value of $\mu$. The red line represents the threshold $\eta$, and thus the points below the red line form the eligibility set. From the figure, we see that the eligibility set centers around the true value 1.

\begin{figure}[htbp]
     \centering
     \begin{subfigure}[b]{0.48\textwidth}
         \centering
         \includegraphics[width=\textwidth]{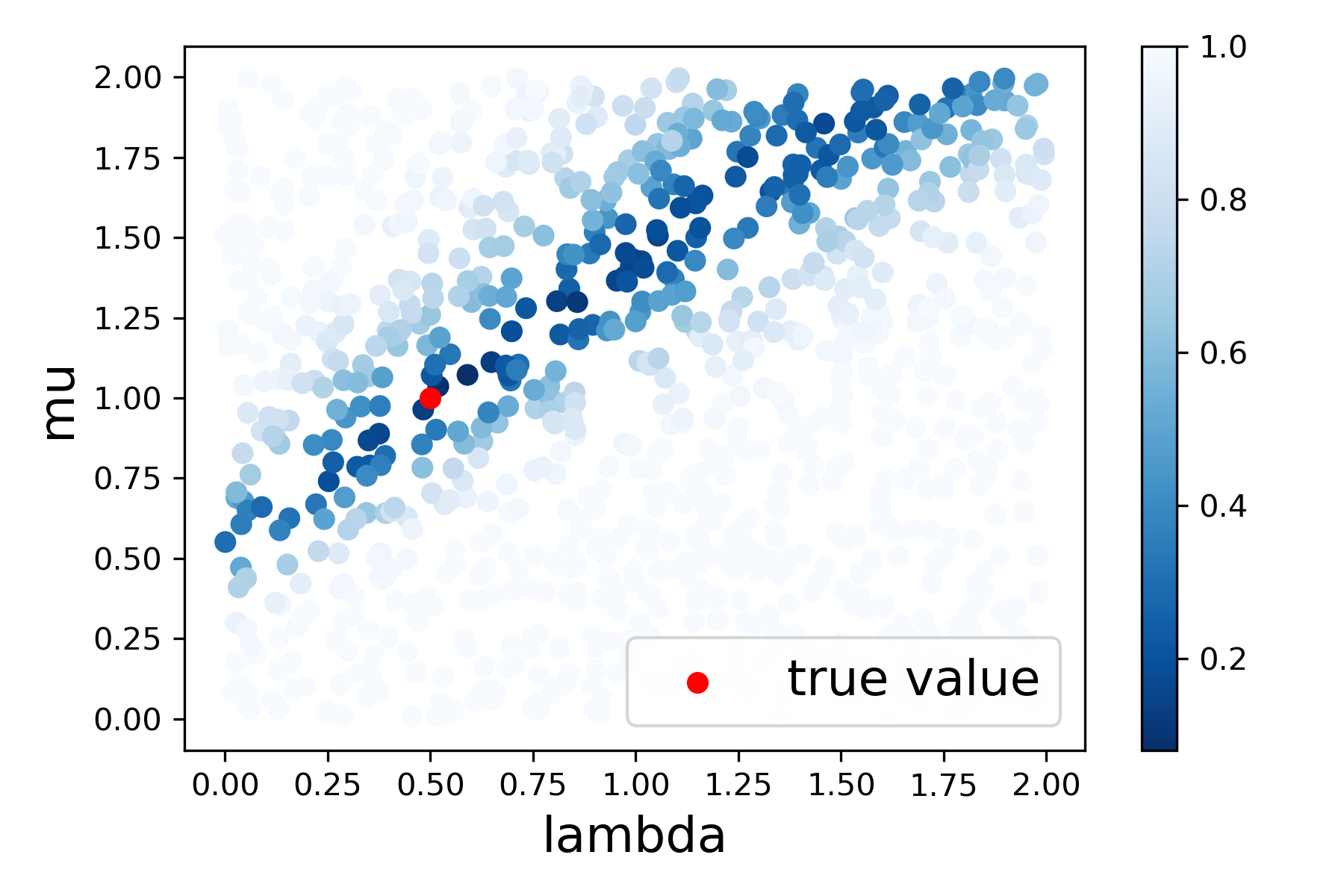}
         \caption{KS distance against $\lambda$ and $\mu$.}
         \label{fig:MM1_3dim}
     \end{subfigure}
     \hfill
     \begin{subfigure}[b]{0.48\textwidth}
         \centering
         \includegraphics[width=\textwidth]{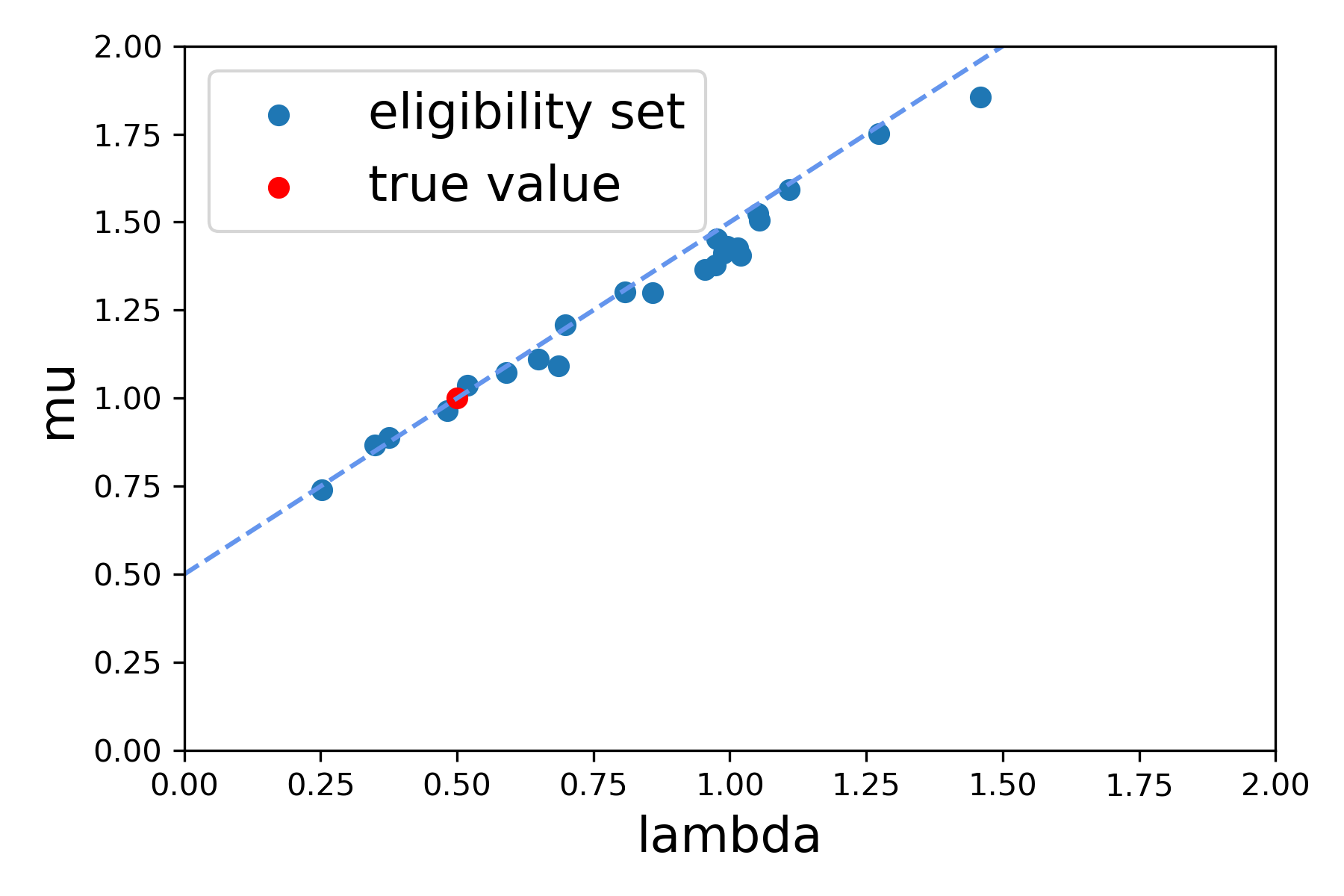}
         \caption{Eligibility set.}
         \label{fig:MM1_2dim}
     \end{subfigure}
     \caption{Numerical results for calibrating $(\lambda,\mu)$ in the M/M/1 example.}
     \label{fig:MM1}
\end{figure}

Now we calibrate $(\lambda,\mu)$ simultaneously with $m=1000$, $n=N=100$ and $\alpha=0.05$. Figure \ref{fig:MM1_3dim} shows how the KS distance changes with $\lambda$ and $\mu$. Figure \ref{fig:MM1_2dim} visualizes the resulting eligibility set.

Judging from Figure \ref{fig:MM1}, we observe that this problem is close to non-identifiable. In fact, it is well known that when $\lambda/\mu<1$, the steady-state sojourn time distribution is $Exp(\mu-\lambda)$. Our simulation output is the average sojourn time of the first 100 customers, which serves as an approximation to the steady-state. Thus, intuitively, the distance between the simulated data and the true data mainly depends on the distance between $\mu-\lambda$ and the truth. In Figure \ref{fig:MM1_2dim}, the dashed line represents $\mu-\lambda=0.5$. The points in the eligibility set indeed lie around this line.

In such a case where the problem is close to non-identifiable while the data size and the simulation size is not sufficiently large, if we carry out a stochastic gradient descent (SGD) or other optimization approaches attempting to minimize the distance between the true data and the simulated data, then it is possible that we will arrive at a misleading point which is actually far from the true value. However, with the concept of eligibility set, we could construct a region that covers the true parameter with high confidence. This demonstrates the key motivation of our approach in addressing non-identiafiability by relaxing point to set-level estimation. 

% and hence obtain confidence bounds for target quantities by solving robust optimization problems in the downstream tasks. 

\subsection{A G/G/1 Example}
\begin{figure}[h]
    \centering
    \includegraphics[width=\textwidth]{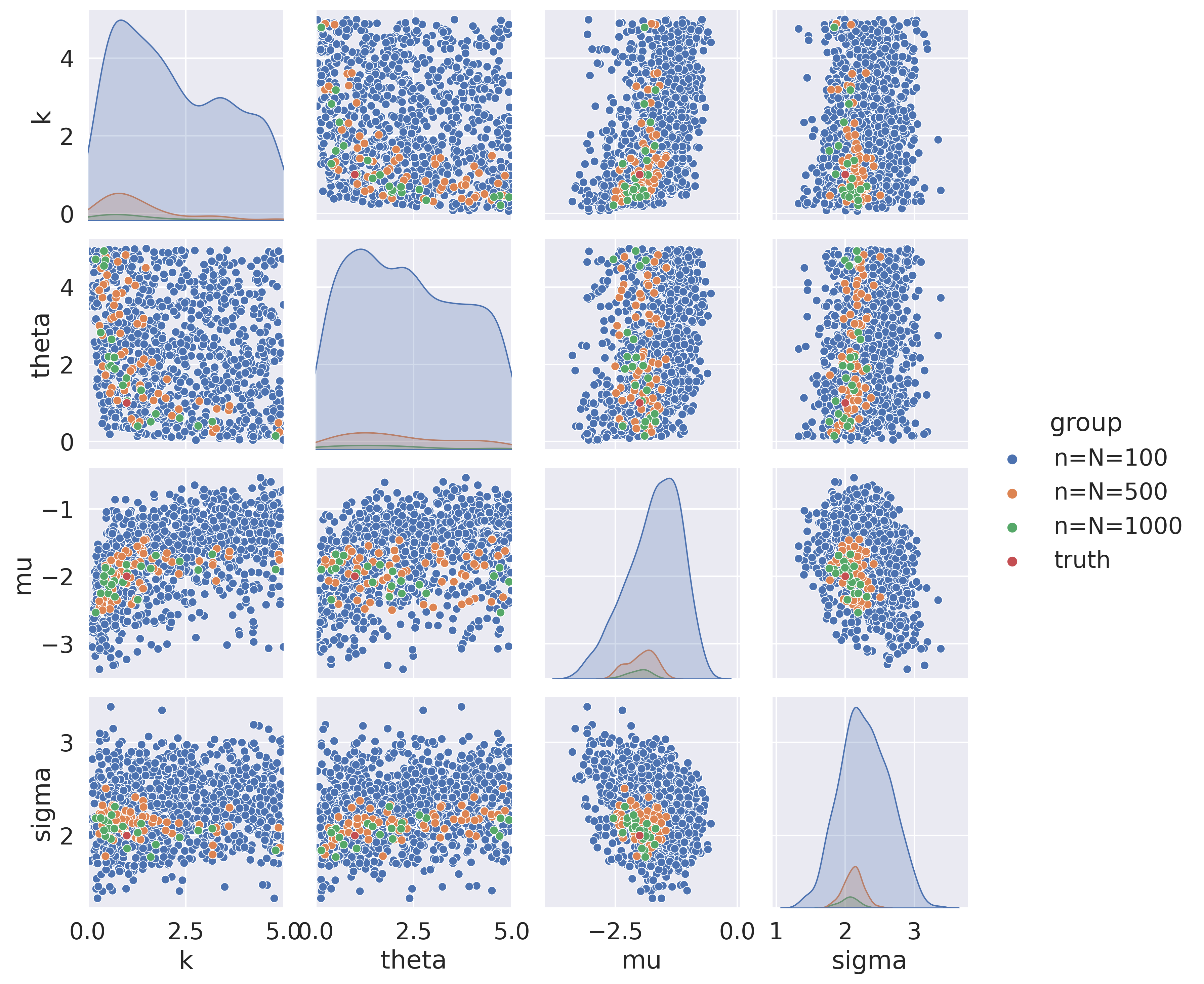}
    \caption{Evolution of the eligibility set as the data size and the simulation size increase.}
    \label{fig:eligibility_set}
\end{figure}

Now we generalize the problem setting to a G/G/1 queue with interarrival time distribution $Gamma(k, \theta)$ ($k$ is the shape parameter and $\theta$ is the scale parameter) and service time distribution $Lognormal(\mu,\sigma^2)$. Suppose that the output of the simulation model is the sojourn time of the first $K=10$ customers. We aim to calibrate the parameter $(k,\theta,\mu,\sigma)$ from the output data. The true value is set as $(1,1,-2,2)$ and the parameter space is chosen as $(0,5)\times(0,5)\times(-5,5)\times(0,5)$. We construct eligibility sets with $m=100,000$ and $n=N=100,500,1000$. Figure \ref{fig:eligibility_set} shows the evolution of the eligibility set as the data size and the simulation size grow. The diagonal graphs present the histograms of the points in the eligibility set in each dimension while the off-diagonal ones visualize the eligibility set via pairwise scatter plots.

From the figure, we find that as the data size and the simulation size grow, the eligibility set gradually becomes more concentrated around the true value. Moreover, we observe that though $\mu$ and $\sigma$ could be tuned relatively accurately, $(k,\theta)$ is close to non-identifiable. Even when $n=N=1000$, the relationship between the first two dimensions of points in the eligibility set still looks like a reciprocal curve. Intuitionally, the mean value of $Gamma(k,\theta)$ is $k\theta$, and thus as long as $k\theta$ is close to the true value 1, it is hard to well distinguish whether $(k,\theta)$ is also close to the truth only judging from the customers' sojourn time. 

\begin{figure}[htbp]
     \centering
     \begin{subfigure}[b]{0.4\textwidth}
         \centering
         \includegraphics[width=\textwidth]{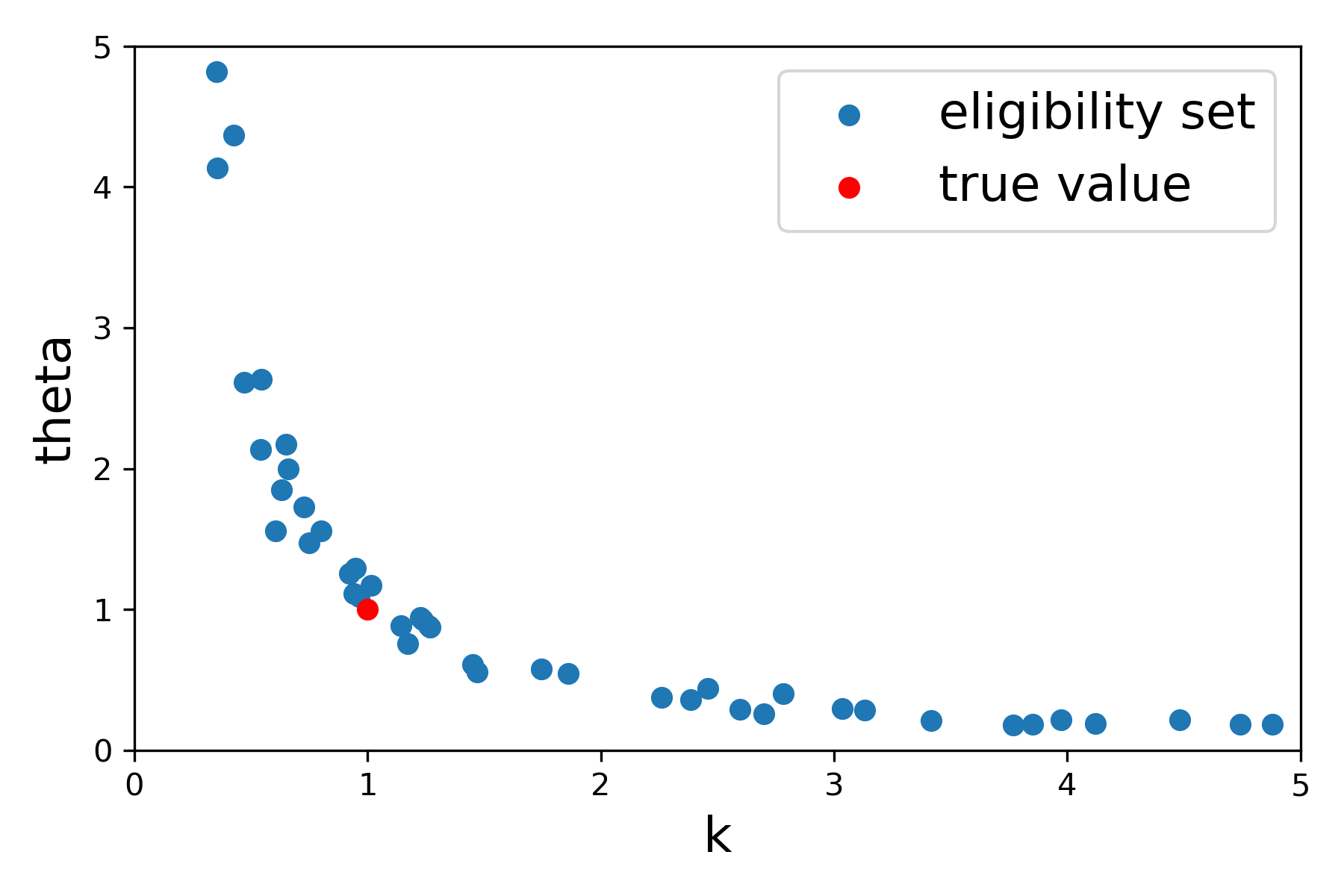}
         \caption{$n=N=1000$.}
         \label{fig:GG1_1000}
     \end{subfigure}
     \begin{subfigure}[b]{0.4\textwidth}
         \centering
         \includegraphics[width=\textwidth]{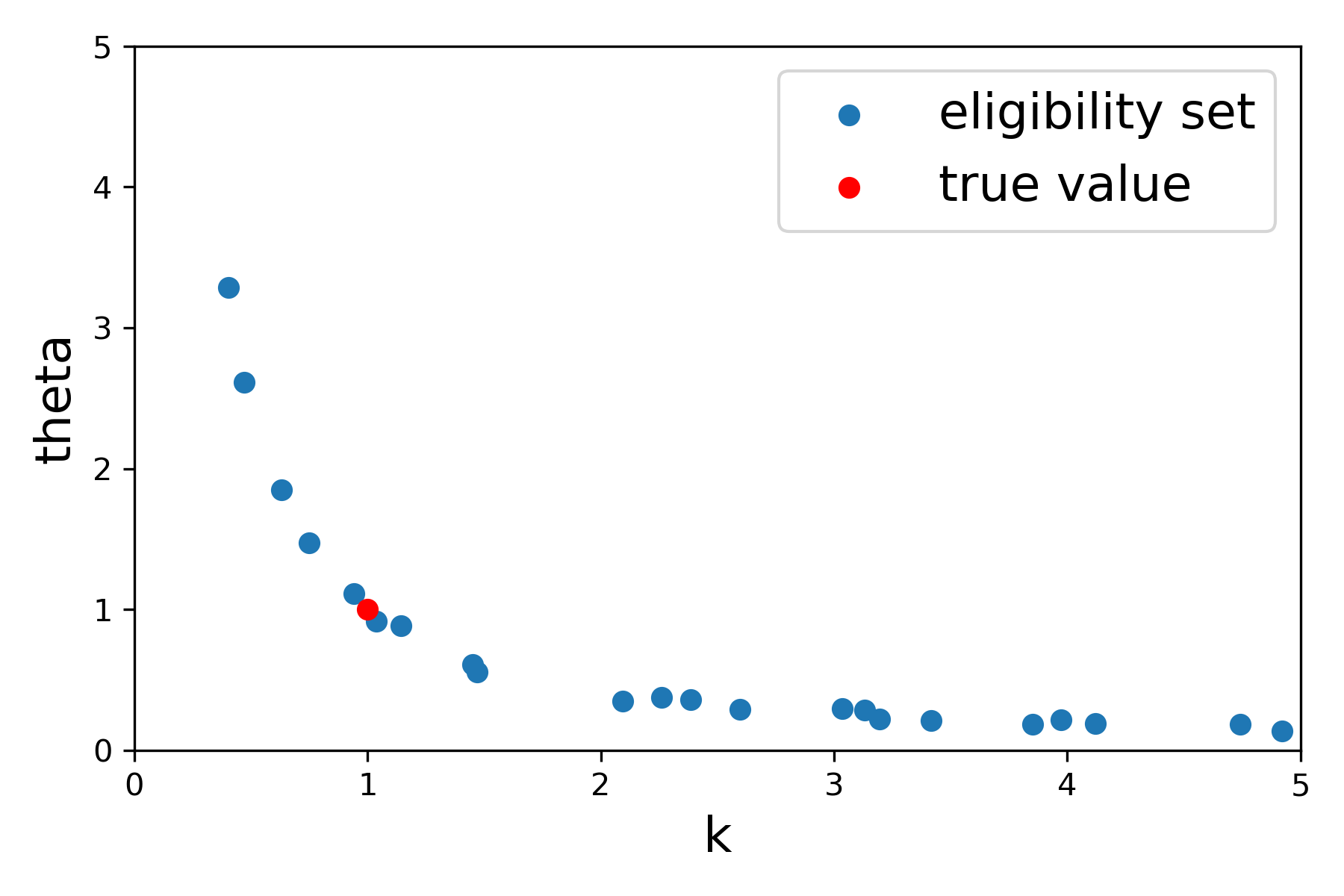}
         \caption{$n=N=2000$.}
         \label{fig:GG1_2000}
     \end{subfigure}
     \\
     \begin{subfigure}[b]{0.4\textwidth}
         \centering
         \includegraphics[width=\textwidth]{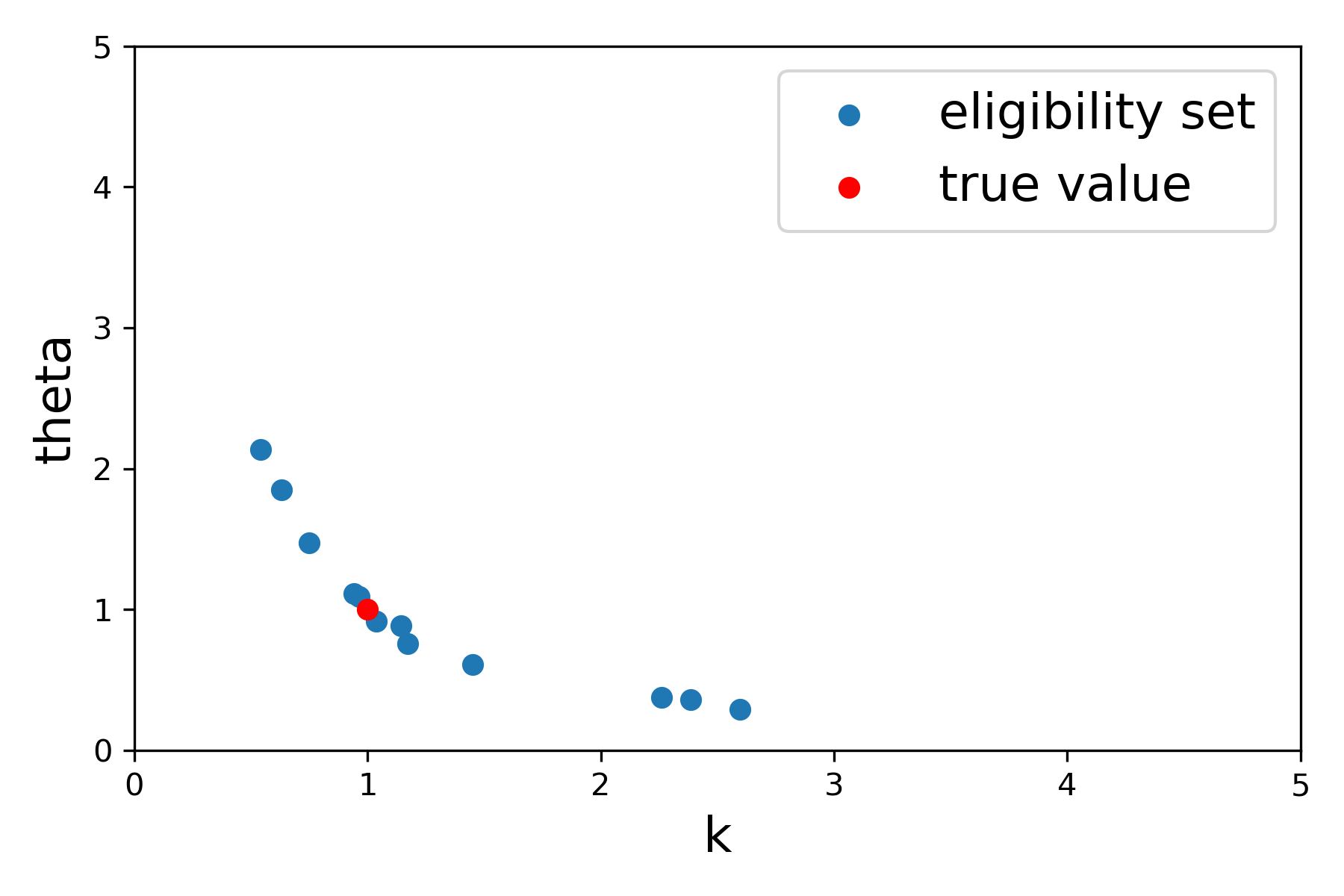}
         \caption{$n=N=5000$.}
         \label{fig:GG1_5000}
     \end{subfigure}
     \begin{subfigure}[b]{0.4\textwidth}
         \centering
         \includegraphics[width=\textwidth]{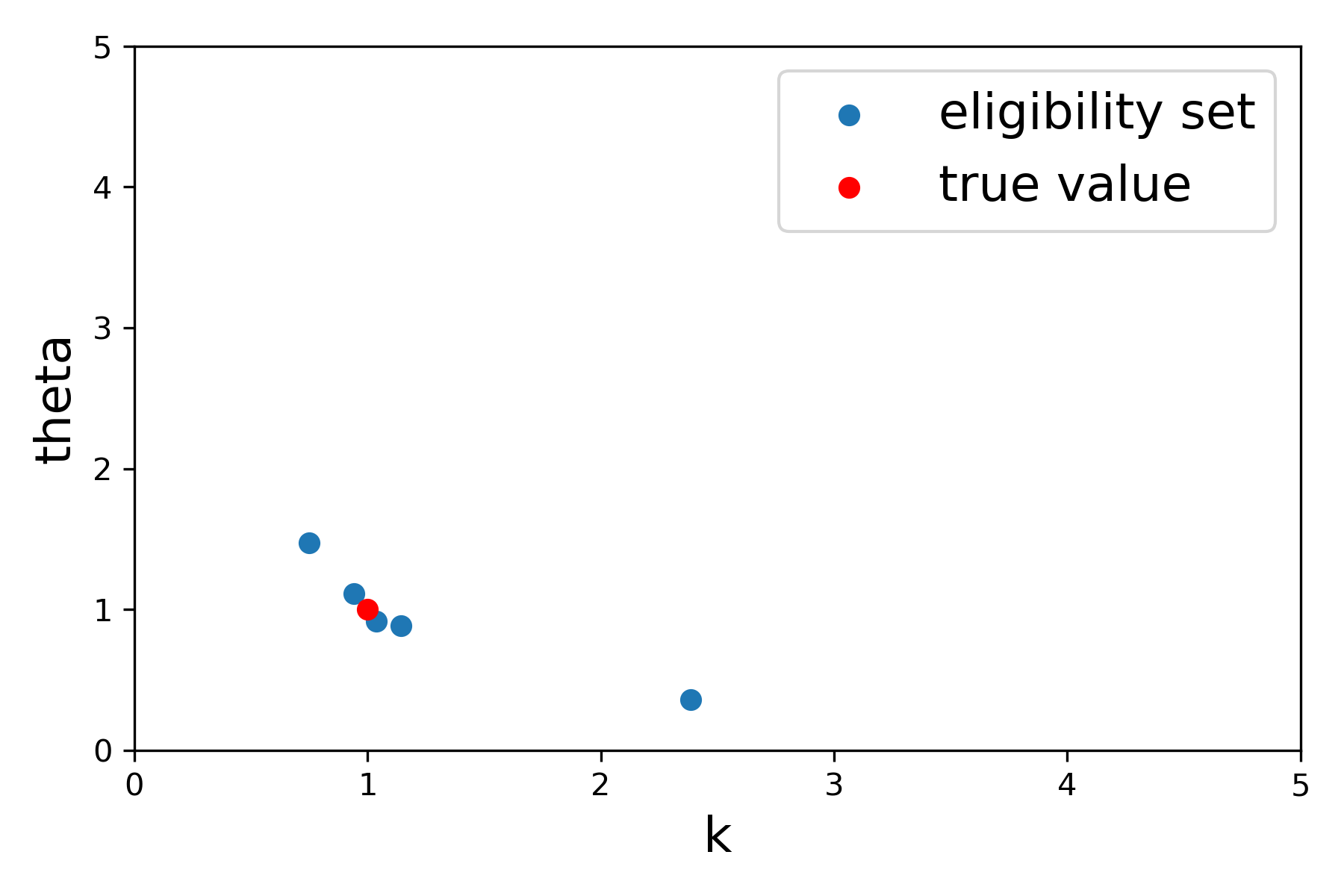}
         \caption{$n=N=10000$.}
         \label{fig:GG1_10000}
     \end{subfigure}
     \caption{Numerical results for the G/G/1 example.}
     \label{fig:GG1}
\end{figure}

Now we fix $\mu=-2$ and $\sigma=2$ and focus on calibrating $(k,\theta)$. We randomly sample $m=1000$ points in the parameter space $(0,5)^2$ and for each point, we test whether it is in the eligibility set with $n=N=1000,2000,5000,10000$. Figure \ref{fig:GG1} shows how the eligibility set shrinks as $n$ and $N$ grow. Though the problem is not structurally non-identifiable, it indeed requires a large number of samples in order to accurately locate the truth.

\section{Feature-Based Construction of Eligibility Sets}\label{sec:construction}
\begin{figure}[h]
\centering
\includegraphics[width=130mm,scale=0.5]{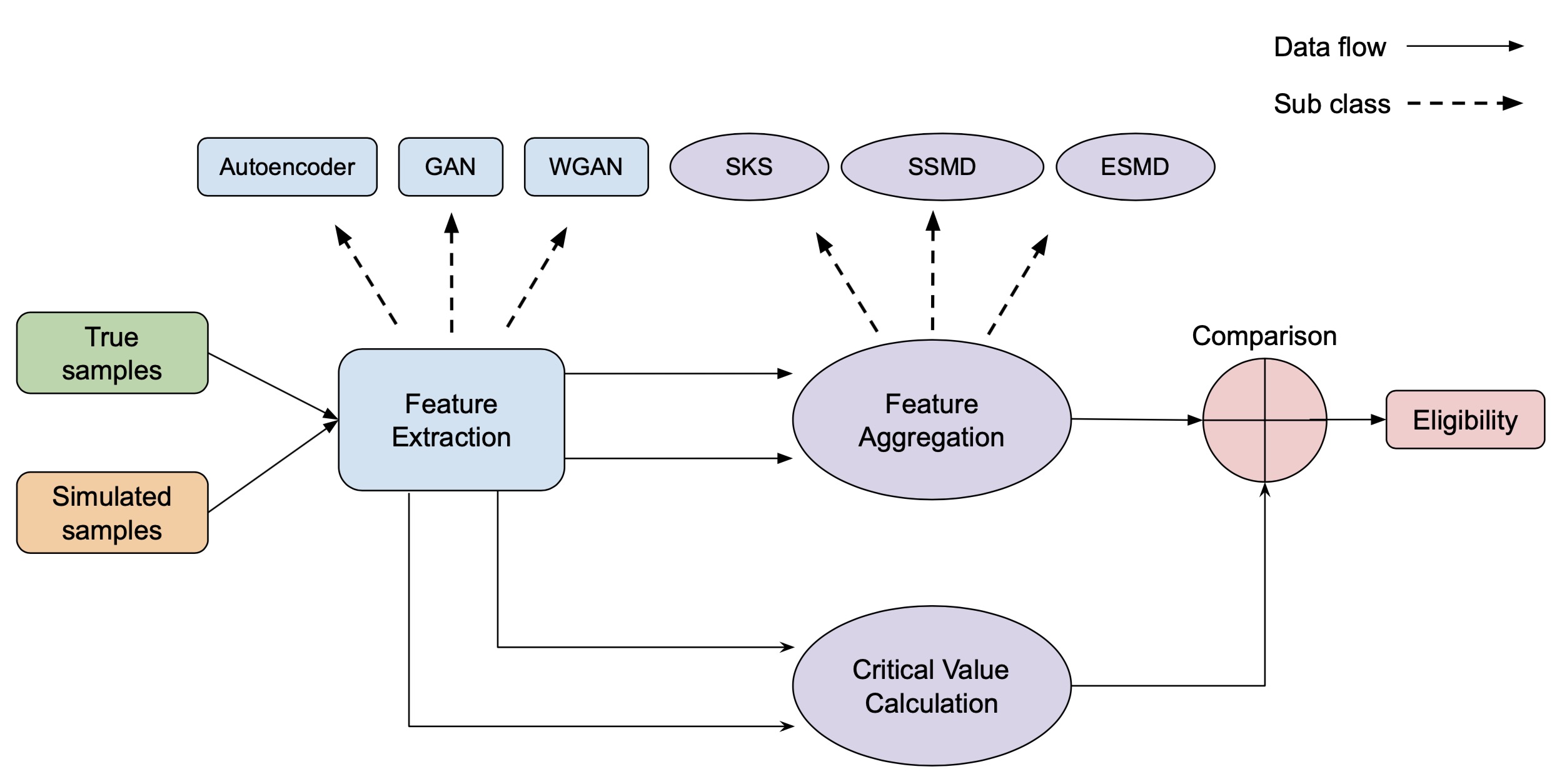}
\caption{\small Feature-based calibration framework diagram.}
\label{Fig:1}
\end{figure}

We present our recipe to calibrate simulation models beyond the simple cases considered in Sections \ref{sec:simple} and \ref{sec:elementary}. This requires suitable construction of the distance measure $d$. We propose a two-stage procedure for this construction, first a \emph{feature extraction} stage that reduces the dimension of the outputs to a manageable number, and second a \emph{feature aggregation} stage to put together the extracted features into a statistical distance $d$ that satisfy the three attractive properties discussed in Section \ref{sec:simple}. The latter requires analyzing the statistical properties of the aggregation schemes as we will illustrate. For the first stage, we extract features using the penultimate layer of a neural network, trained using three approaches: auto-encoder, generative adversarial network (GAN) and its variant Wasserstein GAN (WGAN). For the second stage, we aggregate features using three methods: supremum of Kolmogorov–Smirnov statistics (SKS), supremum of sample mean differences (SSMD), and ellipsoidal sample mean differences (ESMD). These aggregated statistics give rise to a statistical distance, which we compare against computable critical values (at a given confidence level) to decide the eligibility of a parameter value. Figure \ref{Fig:1} gives an overview of our feature-based eligibility decision framework.

% Our construction framework is as follows. Given true samples and simulated samples generated by specific parameters, we use feature extraction techniques to summarize the output, potentially with lower dimension. In machine learning, common feature extraction techniques include as a representative. With the embedded representations of the samples, we apply the feature aggregation techniques to measure the statistical distance between the true and the simulated embedded distributions. 

\subsection{Feature Extraction}

Our feature-based calibration starts by extracting features from the output $Z$, which from now on are assumed to be potentially high-dimensional. These features are defined via a summary function $f$ acted on the original input space. That is, we find a proper summary function $f$ such that $f(Z)\in\R^K$, in which case we extract $K$ features from the output. 

In general, there are many possible methods to extract features and the choice of method does not affect the correctness. More specifically, with summary function $f$, the eligibility set is $\{\theta\in\Theta:d(P^{\theta}\circ f^{-1},P^{\theta_0}_N\circ f^{-1})\leq\eta\}$ (or approximately $\{\theta\in\Theta:d(P_n^{\theta}\circ f^{-1},P^{\theta_0}_N\circ f^{-1})\leq\eta\}$). As long as the threshold $\eta$ is calibrated properly according to the distance measure $d$, the eligibility set is guaranteed to cover the true parameter $\theta_0$ with high probability.  However, even though such a correctness guarantee can be readily ensured by properly choosing $\eta$, the choice of $f$ affects the conservativeness in terms of the size of the eligibility set. Consider the case that $Z_1\sim P^{\theta_1}$, $Z_2\sim P^{\theta_2}$ and $P^{\theta_1}\ne P^{\theta_2}$. If we choose the summary function $f$ such that $f(Z_1)$ and $f(Z_2)$ have the same distribution, then we cannot distinguish between $\theta_1$ and $\theta_2$ only with the extracted features, though they should have been distinguishable. Therefore, a good method should have the power to distinguish different parameters. 

Here we leverage several machine learning methods to extract features, including auto-encoder \citep{baldi2012autoencoders}, GAN \citep{goodfellow2014gan}, and WGAN \citep{arjovski2017wgan}. The details can be found in Appendix \ref{ML model details}, and here we provide some key explanation and how we use these methods. Auto-encoder is a bottleneck-like structure, where the information is compressed by the first half of the network and then reconstructed by the second half. By minimizing the mean squared error between the original and reconstructed inputs, the network is trained to learn the latent representation in the middle layer. GAN contains two competing neural networks, where one of them is a generative model, which learns how to generate samples similar to the input distribution, and the other one is a discriminative model, which learns how to differentiate the generated samples from the input ones. The training procedure of GAN can be framed as a minimax two-layer game, where the discriminator is trained to minimize the probability of incorrectly classifying the input data and the simulated data, and the generator is trained to maximize the probability of the discriminator to make a classification mistake. \citet{goodfellow2014gan} shows that when the discriminator is trained to be optimal for any given generator, then the training of GAN can be viewed as minimizing the Jensen–Shannon (JS) divergence. However, \citet{arjovsky2017towards} mentions that the instability of GAN's training is caused by vanishing gradient when JS divergence is applied to distributions supported by low dimensional manifolds. As an improvement, \citet{arjovski2017wgan} proposes WGAN, where it shares similar network architecture with GAN but the objective is about minimizing the Wasserstein-1 distance instead. Though GAN or WGAN were originally proposed to generate data with the same distribution as the true data, its generator or discriminator could serve as a feature extractor. Specifically, the discriminator-based feature extractor has been successfully applied in various fields \citep{zhang2019unsupervised,lin2017MARTA}. 
% More detailed introductions to those models are provided in Section \ref{ML model details}.

In our work, we consider five feature extraction techniques, including auto-encoder, GAN discriminator output, GAN hidden layer, WGAN critic output and WGAN hidden layer. With a pre-trained auto-encoder, one can apply its encoder part to extract hidden feature from the input data. For GAN discriminator output, a usual way is to directly use the trained discriminator for extraction, which can summarize the output with a single number. GAN hidden layer serves an alternative way to leverage the discriminator, that is, using the output from the last hidden layer instead of the final output. Similarly, since WGAN shares a similar network architecture, one can also use the direct output from the trained critic part (like the discriminator in GAN), or apply a similar trick to obtain output from the last hidden layer. Intuitively, using GAN discriminator output or WGAN critic output to summarize a high dimensional data into a single number might result in much loss of information and thus bring conservativeness. However, using the output of the last hidden layer not only helps in dimensionality reduction without losing too much information but also introduces less conservativeness. We will make a comparison and discuss more in our experimental results in Section \ref{sec:abides}.

Finally, we note that machine learning techniques have been applied to solve calibration problems, but they focus on building surrogate models and minimizing the distance instead of the eligibility set construction studied in our paper. More specifically, when the simulation model is computationally costly, a surrogate model could be built as a proxy in order to speed up the evaluation and calibration. For instance, \citet{alonso2020image} proposes a model-assisted GAN where a convolutional neural network (CNN) is trained as an emulator to mimic the simulator. The similarity between the simulated data and the emulated data is determined by encoding the data with the same CNN and comparing the $L_2$ norm of the difference. Alternatively, \citet{lamperti2018agent} suggests to use extreme gradient boosted trees (XGBoost) to learn an ensemble of classification and regression trees (CART) which ultimately provides an approximation for agent-based models. After a surrogate is built, the computational load could potentially be greatly reduced. We could also integrate such surrogate models in our framework. That is, if it is time-consuming to finish a simulation run, then we could build a surrogate model for the original simulation model. 

\subsection{Feature Aggregation}
\label{sec:aggregation}
From now on, we suppose that we have found a way to extract $K$ features from the output. To facilitate presentation, we abuse the notation $X_i$ in this subsection to refer to the extracted features, i.e., the summary function $f$ applied to the raw real or simulated samples. Suppose that we have the true sample features $X_1,\dots,X_N\in\R^K$ with parameter $\theta_0$ and simulated  sample features $Y_1^{\theta},\dots,Y_n^{\theta}\in\R^K$ with parameter $\theta$. We need a way to aggregate the features to judge whether $\theta=\theta_0$. 

We will introduce three methods and analyze their correctness and conservativeness. The first method, SKS, is a generalization of our approach in Section \ref{sec:simple} to multiple output dimensions by using the Bonferroni correction. We will provide theoretical guarantees for its errors. The second method is SSMD that applies the Bonferroni correction to sample mean differences, and the third method ESMD uses the sum of squares of sample mean differences. For the last two methods, we calibrate the thresholds or critical values with normal approximations. However, it is challenging to analyze the Type II error probabilities for them, so we will investigate them via examples and draw insights therein. 

Intuitively, using more features usually implies more information, and it is easier to distinguish different parameters. However, with more features the threshold is also higher, which in turn makes it more difficult to detect wrong parameters. Generally, for each method, there is no certain conclusion regarding whether using more or less features would be better in terms of conservativeness. Overall, we would recommend using SKS with all the features since theoretically it does not introduce much conservativeness, whereas SSMD and ESMD do not, at least currently, have the same level of theoretical guarantees. Moreover, SSMD and ESMD only compare the mean and the covariance matrix while SKS compares the whole distribution, and hence accounts for more information.

\subsubsection{Supremum of Kolmogorov-Smirnov Statistics (SKS).}
% In the one dimensional case (i.e. $K=1$), we can use the max difference in empirical distribution functions to measure the distance between two samples. That is, we say $\theta$ is eligible if 
% $$
% \sup_{x\in\R}|F_n^{\theta}(x)-F_N^{\theta_0}(x)|\leq q_{1-\alpha}/\sqrt{N}
% $$
% where $F_N^{\theta_0},F_n^{\theta}$ are respectively the empirical distribution functions for $X_1,\dots,X_N$ and $Y_1^{\theta},\dots,Y_n^{\theta}$, $q_{1-\alpha}$ is the ($1-\alpha$)-quantile of $\sup_{t\in[0,1]}|BB(t)|$.

This is a generalization of the calibration approach in Section \ref{sec:simple} from $K=1$ to $K>1$, by using the Bonferroni correction. We focus on the one-sample version, as the results for the two-sample version could be adapted similarly. In this case, we say $\theta$ is eligible (i.e., inside the eligibility set) if 
$$
\sup_{x\in\R}|F_{n,k}^{\theta}(x)-F_{N,k}^{\theta_0}(x)|\leq\frac{q_{1-\alpha/K}}{\sqrt{N}},\forall k=1,\dots,K
$$
where $F_{N,k}^{\theta_0}$ and $F_{n,k}^{\theta}$ are the empirical distribution functions for $X_{1,k},\dots,X_{N,k}$ and $Y_{1,k}^{\theta},\dots,Y_{n,k}^{\theta}$. 
% Similarly, in the two-sample version, $\theta$ is eligible if 
% $$
% \sup_{x\in\R}|F_{n,k}^{\theta}(x)-F_{N,k}^{\theta_0}(x)|\leq \sqrt{\frac{n+N}{nN}}\sqrt{-\frac{1}{2}\log(\frac{\alpha}{2K})},\forall k=1,\dots,K.
% $$

To justify the correctness of our methods, we analyze the probability of Type I error, i.e., $\theta_0$ is eligible according to our criterion.
% If this probability is asymptotically no less than the confidence level $1-\alpha$, then the eligibility set serves as a valid confidence region for $\theta_0$. 
Special focus will be on how the required simulation size depends on the number of features $K$, that is, whether the requirement is substantial when facing high-dimensionality. In the one-dimensional case, we have stated Theorem \ref{thm:confidence}. Now suppose that we jointly consider $K$ features and we use the Bonferroni correction. Then we have the following theorem.
\begin{theorem}
	Suppose that $X_1,\cdots,X_N\in\R^K$ is an i.i.d. true sample from $P^{\theta_0}$ and $Y_1^{\theta_0},\cdots,Y_n^{\theta_0}\in\R^K$ is an i.i.d. simulated sample from $P^{\theta_0}$. $F_{N,k}^{\theta_0}$ and $F_{n,k}^{\theta_0}$ are respectively the empirical distribution functions of the $k$-th component of the two random samples. If $n=\omega(N)$ as $N\rightarrow\infty$, then 
	\begin{equation*}
	\liminf_{n,N\rightarrow\infty}\mathbb{P}\left(\sup_{x\in\R}|F_{n,k}^{\theta_0}(x)-F_{N,k}^{\theta_0}(x)|\leq\frac{q_{1-\alpha/K}}{\sqrt{N}},\forall 1\leq k\leq K\right)\geq 1-\alpha.
	\end{equation*}
	\label{thm:confidence_bonferroni}
\end{theorem}

Similar to Theorem \ref{thm:confidence}, Theorem \ref{thm:confidence_bonferroni} justifies that the eligibility set is an asymptotically valid confidence region for any fixed $K$ as long as $n$ is of higher order than $N$. Next, to see how $n$ should scale with a growing $K$, we state Theorem \ref{thm:confidence_bonferroni_generalized}, which shows that the asymptotic guarantee still holds as long as $n/N$ grows in a higher order than $\log^2K$, which is a relatively slow rate.
\begin{theorem}
	Suppose that $X_1,\cdots,X_N\in\R^K$ is an i.i.d. true sample from $P^{\theta_0}$ and $Y_1^{\theta_0},\cdots,Y_n^{\theta_0}\in\R^k$ is an i.i.d. simulated sample from $P^{\theta_0}$. $F_{N,k}^{\theta_0}$ and $F_{n,k}^{\theta_0}$ are respectively the empirical distribution functions of the $k$-th component of the two random samples. If $n/N=\omega(\log^2 K)$ as $K\rightarrow\infty$, then 
	\begin{equation*}
	\liminf_{K\rightarrow\infty}\mathbb{P}\left(\sup_{x\in\R}|F_{n,k}^{\theta_0}(x)-F_{N,k}^{\theta_0}(x)|\leq\frac{q_{1-\alpha/K}}{\sqrt{N}},\forall 1\leq k\leq K\right)\geq 1-\alpha.
	\end{equation*}
	\label{thm:confidence_bonferroni_generalized}
\end{theorem}

The above theorems provide us some guidance on how to choose the simulation size $n$ according to the data size $N$ and number of features $K$ in order to ensure the correctness. Note that in the high-dimensional case, by replacing $q_{1-\alpha}$ with $q_{1-\alpha/K}$, we are increasing the threshold. However, this increase is not that substantial. Indeed, from \citet{feller1948}, we know that $\mathbb{P}(\sup_{0\leq t\leq 1}|BB(t)|\leq z)=1-2\sum_{v=1}^{\infty}(-1)^{v-1}{e^{-2v^2z^2}}$ for $z>0$, and thus $\alpha=2\sum_{v=1}^{\infty}(-1)^{v-1}{e^{-2v^2q_{1-\alpha}^2}}\leq 2e^{-2q_{1-\alpha}^2}$. Hence we get that $q_{1-\alpha}^2\leq -\log(\alpha/2)/2$. Therefore, $q_{1-\alpha/K}^2=O(\log K)$ as $K\rightarrow\infty$. That is to say, when we simultaneously consider a large number of features, using Bonferroni correction will not bring much conservativeness.

Parallel to the one-dimensional case, we are also interested in whether our methods can efficiently detect the wrong parameters. Thus, we analyze the probability of Type II error, which is characterized by the scenario that $\theta\ne\theta_0$ is eligible. Intuitively, the difficulty still depends on the discrepancy between the distribution functions. If $\theta\ne\theta_0$ but they result in the same output distribution, then we cannot identify which is the true parameter only by comparing the output data. The following theorem, which generalizes Theorem \ref{thm:error}, shows how the conservativeness depends on this discrepancy as well as $n,N,K$.

% In the one-dimensional case, we have already developed Theorem \ref{thm:error}. If we jointly consider $K$ features, then we have the following theorem.
\begin{theorem}
	Suppose that $X_1,\cdots,X_N\in\R^K$ is an i.i.d. true sample from $P^{\theta_0}$ and $Y_1^{\theta},\cdots,Y_n^{\theta}\in\R^K$ is an i.i.d. simulated sample from $P^{\theta}$. $F_{N,k}^{\theta_0}$ and $F_{n,k}^{\theta}$ are respectively the empirical distribution functions of the $k$-th component of the two random samples. $F_k^{\theta_0}$ and $F_k^{\theta}$ denote the cumulative distribution functions of the $k$-th component under $P^{\theta_0}$ and $P^{\theta}$. Suppose that $\max_{1\leq k\leq K}\sup_{x\in\R}|F_k^{\theta}(x)-F_k^{\theta_0}(x)|>0$. For any $\varepsilon_1,\varepsilon_2>0$ such that $\varepsilon_1+\varepsilon_2<\max_{1\leq k\leq K}\sup_{x\in\R}|F_k^{\theta}(x)-F_k^{\theta_0}(x)|$, if 
	\begin{equation*}
	N>\left(\frac{q_{1-\alpha/K}}{\max_{1\leq k\leq K}\sup_{x\in\R}|F_k^{\theta}(x)-F_k^{\theta_0}(x)|-\varepsilon_1-\varepsilon_2}\right)^2,  
	\end{equation*}
	then 
	\begin{equation*}
	\mathbb{P}\left(\sup_{x\in\R}|F_{n,k}^{\theta}(x)-F_{N,k}^{\theta_0}(x)|\leq\frac{q_{1-\alpha/K}}{\sqrt{N}},\forall 1\leq k\leq K\right)\leq 2\left(e^{-2n\varepsilon_1^2}+e^{-2N\varepsilon_2^2}\right).
	\end{equation*}
	\label{thm:error_bonferroni}
\end{theorem}

Comparing with Theorem \ref{thm:error}, we note that $q_{1-\alpha/K}$ grows only logarithmically in $K$ as $K\rightarrow\infty$, so the minimum required $N$ does not increase much. Overall, using the Bonferroni correction does not bring much conservativeness compared to only using one feature, and thus appears superior as it accounts for more information. In fact, numerical experiments in Section \ref{sec:abides} shows that the Bonferroni correction generally works well.

\subsubsection{Supremum of Sample Mean Differences (SSMD).}
Another natural idea to aggregate features is to directly compare their sample means. We denote $\bar{X}_k=\frac{1}{N}\sum_{j=1}^N X_{j,k}$ and $\bar{Y}_k=\frac{1}{n}\sum_{i=1}^n Y_{i,k}^{\theta}$. Assume that $var(X_{1,k}),var(Y_{1,k}^{\theta})<\infty$ for any $k$. We say $\theta$ is eligible if 
$$
|\bar{X}_k-\bar{Y}_k|\leq \eta_k,\ \forall k=1,\dots,K
$$
where 
$$
\eta_k=z_{1-\alpha/(2K)}\sqrt{\left(\frac1N+\frac1n\right)var(X_{1,k})}
$$
with $z_{1-\alpha/(2K)}$ being the $(1-\alpha/(2K))$-quantile of standard normal distribution. In practice, $var(X_{1,k})$ could be estimated using the sample variance. 

Indeed, by the central limit theorem (CLT), we know that for sufficiently large $N$ and $n$, $\bar{X}_k-\bar{Y}_k$ is approximately distributed as $N\left(E(X_{1,k})-E(Y_{1,k}),\frac{var(X_{1,k})}{N}+\frac{var(Y_{1,k}^{\theta})}{n}\right)$. If $\theta=\theta_0$, then the approximate distribution is $N\left(0,\left(\frac1N+\frac1n\right)var(X_{1,k})\right)$. We combine this normal approximation with the Bonferroni correction. Note that the idea is similar to the two-sample hypothesis testing, and thus when $n$ and $N$ are large, the approximate correctness is implied. More concretely, we have the following theorem:
\begin{theorem}
Suppose that $X_1,\cdots,X_N\in\R^K$ is an i.i.d. true sample from $P^{\theta_0}$ and $Y_1^{\theta_0},\cdots,Y_n^{\theta_0}\in\R^K$ is an i.i.d. simulated sample from $P^{\theta_0}$. For any $k=1,\dots,K$, we denote $\bar{X}_k=\frac{1}{N}\sum_{j=1}^N X_{j,k},\bar{Y}_k=\frac{1}{n}\sum_{i=1}^n Y_{i,k}^{\theta_0}$ and assume that $var(X_{1,k})=var(Y_{1,k}^{\theta_0})<\infty$. Use $\widehat{var}_k$ to denote the sample variance of $X_{j,k}$'s. If $n/(n+N)\to \rho$ where $0\leq \rho\leq 1$ as $n,N\to\infty$, then 
\begin{equation*}
    \liminf_{n,N\to\infty} \mathbb P\left(|\bar{X}_k-\bar{Y}_k|\leq z_{1-\alpha/(2K)}\sqrt{\left(\frac1N+\frac1n\right)\widehat{var}_k},\ \forall k=1,\dots,K\right)\geq 1-\alpha.
\end{equation*}
\label{thm:ssmd_correctness}
\end{theorem}

Like Theorem \ref{thm:confidence_bonferroni} for SKS, Theorem \ref{thm:ssmd_correctness} controls the Type I error of SSMD. Regarding Type II error that drives the level of conservativeness, SSMD appears challenging to fully analyze. Nonetheless, we will look at special cases to draw insights. First, we give an example to show that using Bonferroni correction can be conservative in terms of relative difference in the Type II error probability. That is, the ratio of the Type II error probability of using all the $K$ features to the error of using only one feature could be pretty large. Suppose that $X_1,\cdots,X_N\in\R^K$ is an i.i.d. random sample from $N(\theta_0,I_K)$ and $Y_1^{\theta},\cdots,Y_n^{\theta}\in\R^K$ is an i.i.d. random sample from $N(\theta,I_K)$. Given a confidence level $1-\alpha$, we define $\eta$ and $\eta'$ by 
$$
\mathbb P\left(\left|N\left(0,\frac{1}{N}+\frac{1}{n}\right)\right|\leq \eta\right)=1-\alpha, \mathbb P\left(\left|N\left(0,\frac{1}{N}+\frac{1}{n}\right)\right|\leq \eta'\right)=1-\frac{\alpha}{K}.
$$
 We suppose that $\Delta_1=\theta_{01}-\theta_1>0$ and $\theta_{0k}=\theta_k$ for $k=2,\dots,K$ ($\theta_{0k}$ and $\theta_k$ respectively denote the $k$-th component of $\theta_0$ and $\theta$). If we only use the first feature, that is, we reject $\theta=\theta_0$ if $|\bar{X}_1-\bar{Y}_1|>\eta$, then the Type II error probability is 
$$
p_1:=\mathbb P(|\bar{X}_1-\bar{Y}_1|\leq \eta)=\mathbb P\left(\left|N\left(\Delta_1,\frac{1}{N}+\frac{1}{n}\right)\right|\leq\eta\right).
$$
If we use all the $K$ features, that is, we reject $\theta=\theta_0$ if $\exists k, |\bar{X}_k-\bar{Y}_k|>\eta'$, then the Type II error probability is 
$$
p_2:=\mathbb P(\forall k,|\bar{X}_k-\bar{Y}_k|\leq \eta')=\left(1-\frac{\alpha}{K}\right)^{K-1}\mathbb P\left(\left|N\left(\Delta_1,\frac{1}{N}+\frac{1}{n}\right)\right|\leq\eta'\right).
$$
The following theorem shows that in this setting, $p_2/p_1$ grows exponentially:
\begin{theorem}
    Suppose that $X_1,\cdots,X_N\in\R^K$ is an i.i.d. true sample from $N(\theta_0,I_K)$ and $Y_1^{\theta},\cdots,Y_n^{\theta}\in\R^K$ is an i.i.d. simulated sample from $N(\theta,I_K)$. We suppose that $\Delta_1=\theta_{01}-\theta_1>0$ and $\theta_{0k}=\theta_k$ for $k=2,\dots,K$. $p_1$ and $p_2$ are respectively the Type II error probability of only using the first feature and using all the $K$ features. For fixed $K>1$, as $N,n\rightarrow\infty$, we have that $p_2/p_1$ grows exponentially in $N$ and $n$.
    \label{thm:sos_1}
\end{theorem}

Next, we show that using all the features with Bonferroni correction is not too conservative in terms of the absolute difference in the Type II error probability. Here we consider a more general setting. Suppose that $X_1,\cdots,X_N\in\R^K$ is an i.i.d. random sample from $N(\theta_0,\Sigma)$ and $Y_1^{\theta},\cdots,Y_n^{\theta}\in\R^K$ is an i.i.d. random sample from $N(\theta,\Sigma)$. Given a confidence level $1-\alpha$, we define $\eta_k$ and $\eta_k'$ by 
$$
\mathbb P\left(\left|N\left(0,\left(\frac{1}{N}+\frac{1}{n}\right)\Sigma_{kk}\right)\right|\leq \eta_k\right)=1-\alpha, \mathbb P\left(\left|N\left(0,\left(\frac{1}{N}+\frac{1}{n}\right)\Sigma_{kk}\right)\right|\leq \eta_k'\right)=1-\frac{\alpha}{K}.
$$
We suppose that $\theta\ne\theta_0$. We denote $\Delta_k=\theta_{0k}-\theta_k$. If we only use the $k$-th feature, that is, we reject $\theta=\theta_0$ if $|\bar{X}_k-\bar{Y}_k|>\eta_k$, then the Type II error probability is 
$$
p_1:=\mathbb P(|\bar{X}_k-\bar{Y}_k|\leq \eta_k)=\mathbb P\left(\left|N\left(\Delta_k,\left(\frac{1}{N}+\frac{1}{n}\right)\Sigma_{kk}\right)\right|\leq\eta_k\right).
$$
If we use all the $K$ features, that is, we reject $\theta=\theta_0$ if $\exists k, |\bar{X}_k-\bar{Y}_k|>\eta_k'$, then the Type II error probability is
$$
p_2:=\mathbb P(\forall k,|\bar{X}_k-\bar{Y}_k|\leq \eta_k').
$$
\begin{theorem}
    Suppose that $X_1,\cdots,X_N\in\R^K$ is an i.i.d. true sample from $N(\theta_0,\Sigma)$ and $Y_1^{\theta},\cdots,Y_n^{\theta}\in\R^K$ is an i.i.d. simulated sample from $N(\theta,\Sigma)$. $p_1$ and $p_2$ are respectively the Type II error probability of only using the $k$-th feature and using all the $K$ features. If $\Delta_k\ne 0$, then for fixed $K>1$, as $N,n\rightarrow\infty$, both $p_1$ and $p_2$ converge to 0 exponentially in $N$ and $n$. If further we have $N=\omega(\log K)$ and $n=\omega(\log K)$ as $K\rightarrow\infty$, then $p_2\rightarrow 0$.
    \label{thm:sos_2}
\end{theorem}

Theorem \ref{thm:sos_2} shows that in this more general setting, the Type II error probability of either using only one feature or using all the features decays exponentially in $N$ and $n$. Moreover, if the number of features $K$ is also growing, the Type II error probability of using all the features still converges to 0 as long as $N$ and $n$ grow in a higher order than $\log K$. 

Though the above theorems are developed for the special case of Gaussian distribution, they convey some information regarding the conservativeness of using more features. Intuitively, in the general case we may apply CLT and then the sample means are approximately Gaussian for sufficiently large sample sizes. Our conclusion is that even though in some cases using more features makes the method more conservative, overall we can still get an acceptable Type II error probability.

\subsubsection{Ellipsoidal Sample Mean Difference (ESMD).}
Now we consider further aggregating the sample mean difference of each feature with sum of squares. We say that $\theta$ is eligible if 
$$
\sum_{k=1}^K(\bar{X}_k-\bar{Y}_k)^2\leq \eta
$$
where $\eta$ is the $(1-\alpha)$-quantile of the generalized chi-square distribution given by $\left(\frac{1}{N}+\frac{1}{n}\right)Z^T\Sigma_X Z$ with $Z\sim N(0,I_K)$ and $\Sigma_X$ being the covariance matrix of $X_1$. In practice, $\Sigma_X$ could be estimated using the samples and the quantile could be numerically evaluated. Again, the approximate correctness is implied by CLT. We have the following theorem:

\begin{theorem}
Suppose that $X_1,\cdots,X_N\in\R^K$ is an i.i.d. true sample from $P^{\theta_0}$ and $Y_1^{\theta_0},\cdots,Y_n^{\theta_0}\in\R^K$ is an i.i.d. simulated sample from $P^{\theta_0}$. For any $k=1,\dots,K$, we denote $\bar{X}_k=\frac{1}{N}\sum_{j=1}^N X_{j,k},\bar{Y}_k=\frac{1}{n}\sum_{i=1}^n Y_{i,k}^{\theta_0}$. Assume that the covariance matrix of $X_1$ exists and denote it as $\Sigma_X$. Use $\hat{\Sigma}$ to denote the sample covariance matrix of $X_j$'s. If $n/(n+N)\to \rho$ where $0\leq \rho\leq 1$ as $n,N\to\infty$, then 
\begin{equation*}
    \liminf_{n,N\to\infty} \mathbb P\left(\sum_{k=1}^K(\bar{X}_k-\bar{Y}_k)^2\leq\eta\right)\geq 1-\alpha
\end{equation*}
where $\eta$ is the $(1-\alpha)$-quantile of the generalized chi-square distribution given by $\left(\frac{1}{N}+\frac{1}{n}\right)Z^T\hat\Sigma Z$ with $Z\sim N(0,I_K)$.
\label{thm:esmd_correctness}
\end{theorem}

Theorem \ref{thm:esmd_correctness} gives a guarantee on Type I error. Regarding Type II error or conservativeness, we give an example to show that including or excluding more features do not lead to generally dominant results. 

Suppose that we have a random sample $X_1,\dots,X_N\sim N(\theta_0,I_K)$ and a random sample $Y_1^{\theta},\dots,Y_n^{\theta}\sim N(\theta,I_K)$ where $\theta_0,\theta\in\R^K$. If $\theta=\theta_0$, then $\bar{X}_k-\bar{Y}_k\sim N(0,\frac{1}{N}+\frac{1}{n})$ and $\sum_{k=1}^K (\bar{X}_k-\bar{Y}_k)^2/(\frac{1}{N}+\frac{1}{n})\sim \chi_K^2$. Given a confidence level $1-\alpha$, we define $\eta$ and $\eta'$ by
$$
\mathbb P\left(\left|N\left(0,\frac{1}{N}+\frac{1}{n}\right)\right|\leq \eta\right)=1-\alpha, \mathbb P\left(\chi_K^2\leq\frac{\eta'}{\frac{1}{N}+\frac{1}{n}}\right)=1-\alpha.
$$

Now we suppose that $\theta\ne\theta_0$. For any $k=1,\dots,K$, we denote $\Delta_k=\theta_{0k}-\theta_k$. If we only use the $k$-th feature, that is, we reject $\theta=\theta_0$ if $|\bar{X}_k-\bar{Y}_k|>\eta$, then the Type II error probability is 
$$
p_1:=\mathbb P(|\bar{X}_k-\bar{Y}_k|\leq \eta)=\mathbb P\left(\left|N\left(\Delta_k,\frac{1}{N}+\frac{1}{n}\right)\right|\leq\eta\right).
$$
If we use all the features, that is, we reject $\theta=\theta_0$ if $\sum_{k=1}^K (\bar{X}_k-\bar{Y}_k)^2>\eta'$, then the Type II error probability is 
$$
p_2:=\mathbb P\left(\sum_{k=1}^K (\bar{X}_k-\bar{Y}_k)^2\leq \eta'\right)=\mathbb P\left(\chi_{K,\nu}^2\leq \frac{\eta'}{\frac{1}{N}+\frac{1}{n}}\right)
$$
where 
$$
\nu=\frac{\sum_{k=1}^K\Delta_k^2}{\frac{1}{N}+\frac{1}{n}}.
$$

Given the value of $\alpha,k,K,n,N$ and $\Delta_k$'s, we are able to numerically compute $p_1$ and $p_2$. Below are some results from the numerical computation.

\begin{figure}[htbp]
    \centering
    \begin{subfigure}[b]{0.45\textwidth}
        \centering
        \includegraphics[width=\textwidth]{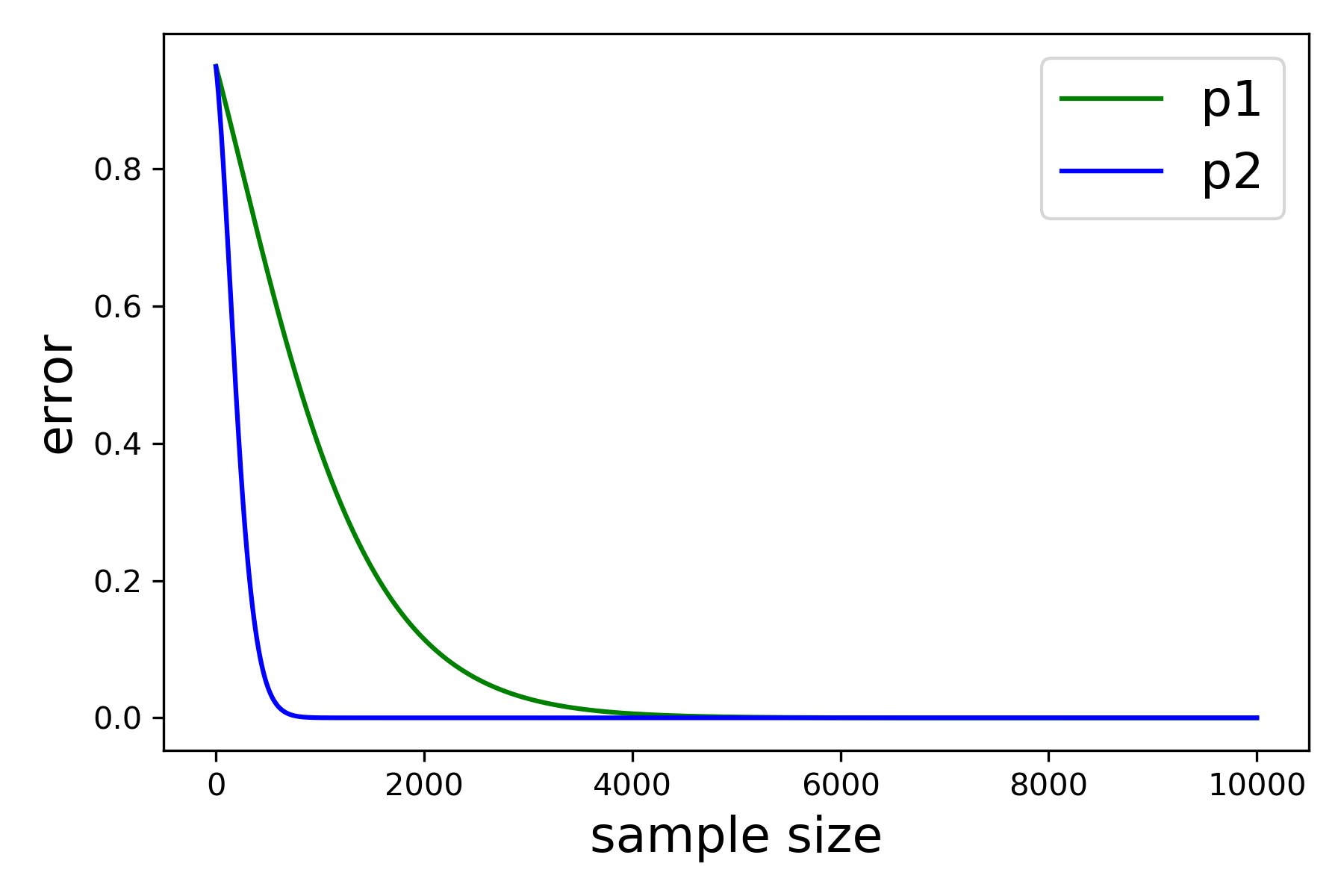}
        \caption{$\Delta=0.1*(1,1,\dots,1)$}
        \label{fig:1}
    \end{subfigure}
    \hfill
    \begin{subfigure}[b]{0.45\textwidth}
        \centering
        \includegraphics[width=\textwidth]{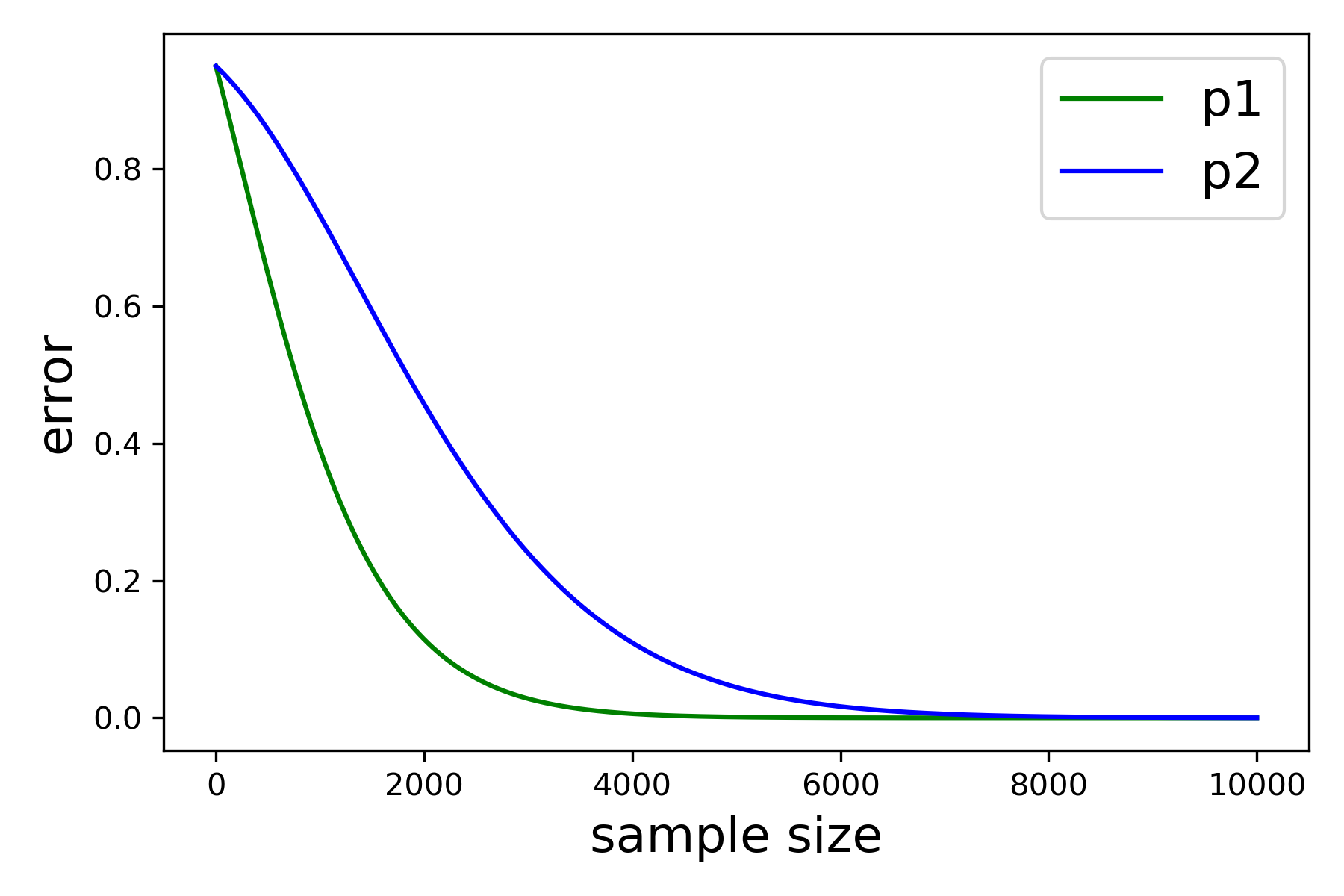}
        \caption{$\Delta=0.1*(1,0,\dots,0)$}
        \label{fig:2}
        \end{subfigure}
    \caption{Change of Type II error probabilities with the sample size.}
\end{figure}

First, we set $\alpha=0.05,k=1,K=10$ and $\Delta=0.1*(1,1,\dots,1)$. We also let $n=N$. Then the change of Type II error probabilities with the sample size is shown in Figure \ref{fig:1}. We find that it is better to use all the features in this case. Next, we set $\Delta=0.1*(1,0,\dots,0)$. The result is shown in Figure \ref{fig:2}. We find that in this case it is better to only use the first feature. The results are interpretable. In the first case, the marginal distributions in each dimension are different for the two samples, so using more features help us detect the wrong parameter. In the second case, however, the distributions are the same except for the first dimension, so including more features in turn makes it more difficult to distinguish the parameters. Therefore, there is no certain conclusion that one choice is better than the other.

\begin{figure}[htbp]
   \centering
    \begin{subfigure}[b]{0.45\textwidth}
        \centering
        \includegraphics[width=\textwidth]{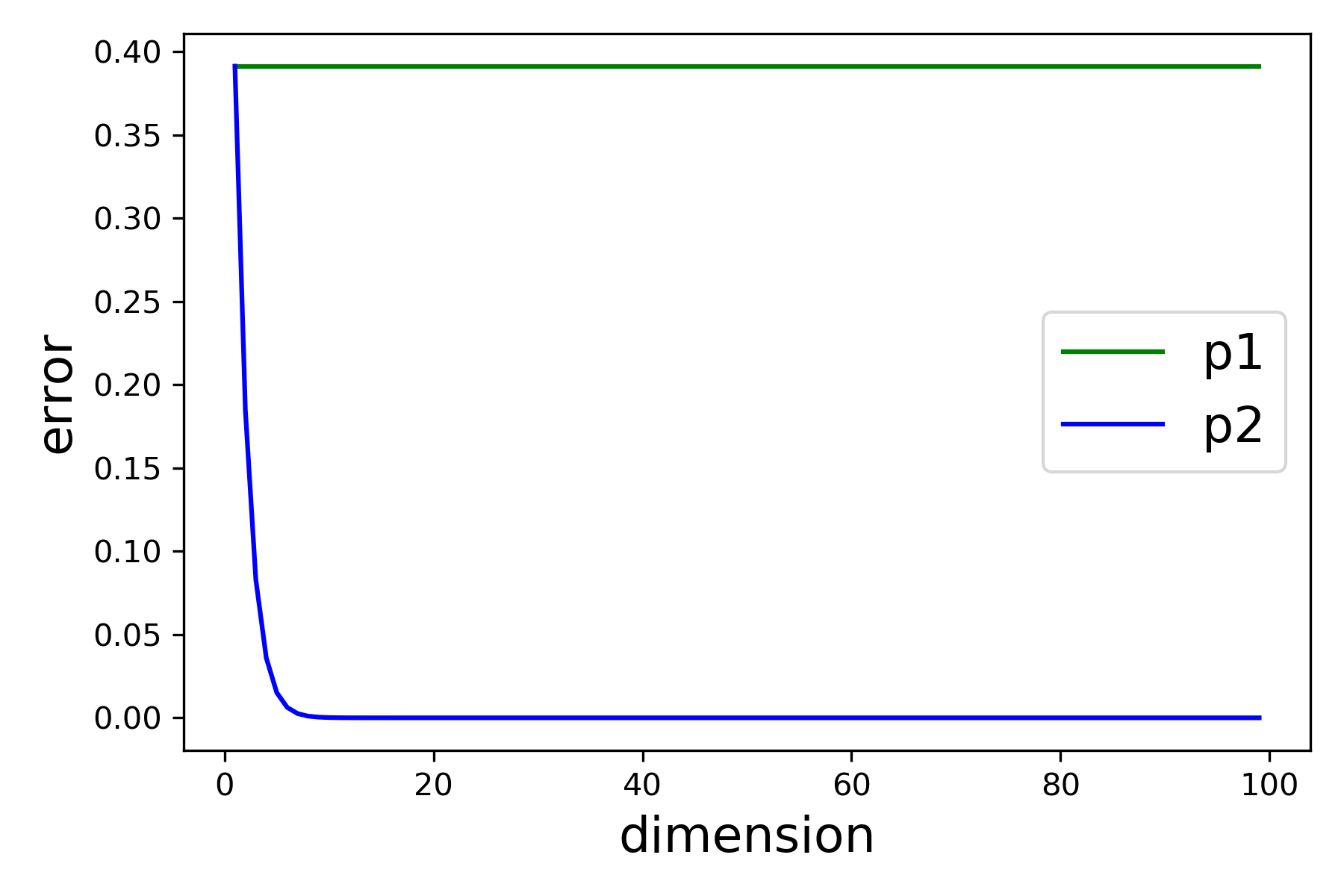}
        \caption{$\Delta=0.1*(1,1,\dots,1)$}
        \label{fig:3}
    \end{subfigure}
    \hfill
    \begin{subfigure}[b]{0.45\textwidth}
        \centering
        \includegraphics[width=\textwidth]{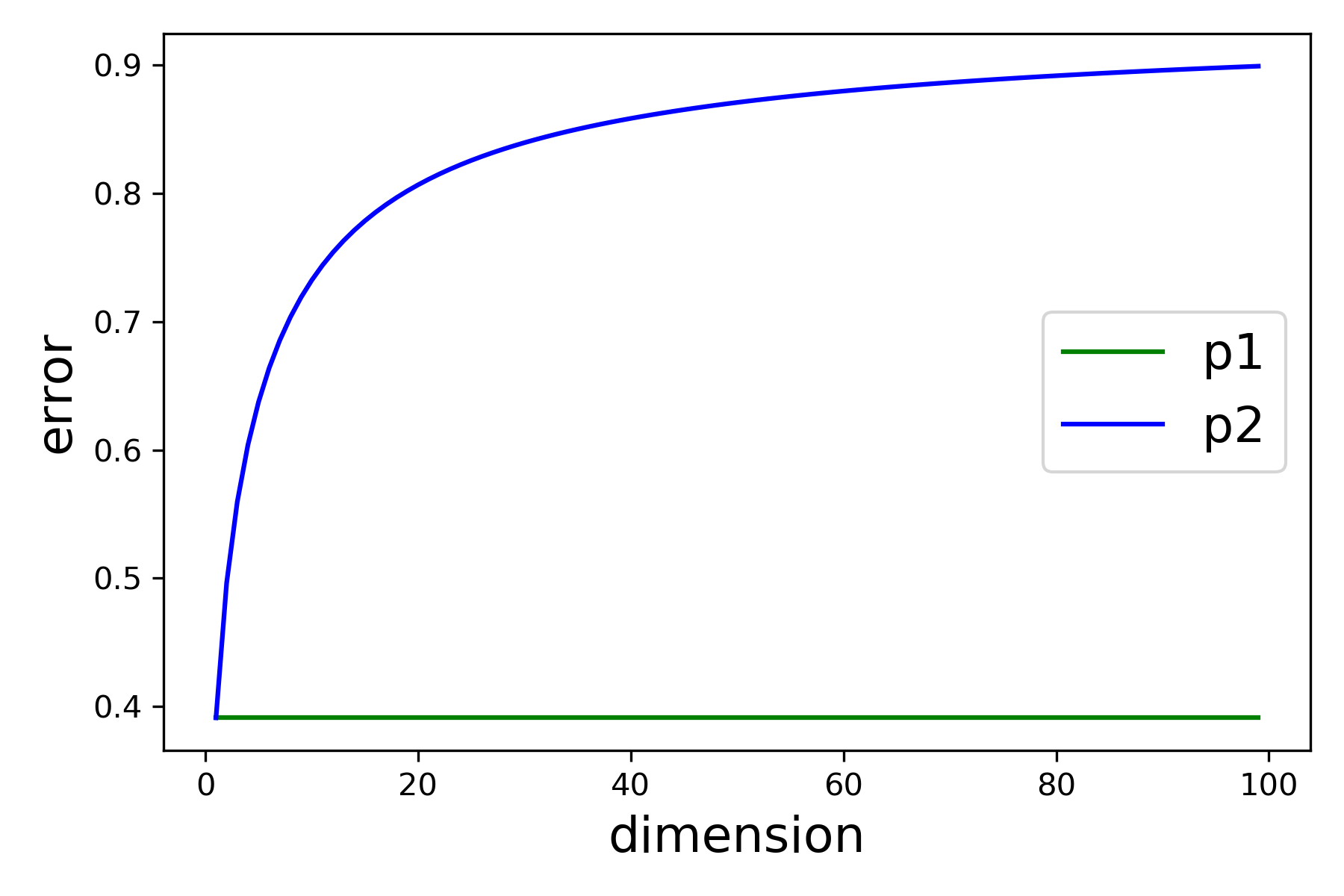}
        \caption{$\Delta=0.1*(1,0,\dots,0)$}
        \label{fig:4}
        \end{subfigure}
    \caption{Change of Type II error probabilities with the number of features.}
\end{figure}

We are also interested in how the Type II error probability will change with the dimension $K$. We fix $n=N=1000$ and the results are shown in Figure \ref{fig:3} and \ref{fig:4}. From the figures, we see that in both cases, the Type I error probability of only using the first feature does not change with the number of features. In the first case, the Type II error probability of using all the features decreases as the number of features grows. As explained, each dimension contributes to detecting the wrong parameter, so it is easier with more features. By comparison, in the second case, the Type II error probability increases as the number of features grows, as the additional features do not help us distinguish the parameters but nullify the difference in the first dimension. This result further supports that neither of the methods dominates the other one.

\section{{Multi-Agent Simulator Calibration}}
\label{sec:abides}

In this section, we apply our feature-based calibration framework in Section \ref{sec:construction} to the Agent-Based Interactive Discrete-Event Simulation (ABIDES) environment \citep{byrd2020abides,vyetrenko2019get} which, among other uses, can be used to simulate limit order book exchange markets such as NASDAQ or New York Stock Exchange. 
% . The output of the ABIDES model is a time series. We will show how our approach is able to construct the eligibility set and rediscover the true parameter in a high dimensional output setting.
% The ABIDES model is an agent-based large-scale discrete event simulation model. 
 ABIDES provides a selection of background agent types (such as market makers, noise agents, value agents, etc.), a NASDAQ-like exchange agent which lists any number of securities for trade against a limit order book with price-then-FIFO matching rules, and a simulation kernel which manages the flow of time and handles all inter-agent communication. 
 
We use background agent implementation introduced in \citet{vyetrenko2019get}. Value agents are designed to simulate
the actions of fundamental traders that trade according
to their belief of the exogenous value of a stock (also called fundamental price -- which in our model follows
 a mean reverting price time series). Value traders choose to buy or sell a stock
depending on whether it is cheap or expensive relative to their
noisy observation of a fundamental price. The fundamental follows a discrete-time mean-reverting Ornstein-Uhlenbeck process~\citep{byrd2019explaining}. Noise agents are designed to emulate the
actions of consumer agents who trade on demand without
any other considerations; they
arrive to the market at times that are uniformly distributed
throughout the trading day and place an order of random size in random direction. Market makers act as liquidity providers by placing
limit orders on both sides of the limit order book with a constant
arrival rate.

We test our calibration algorithm on ABIDES against a synthetic configuration that is designated as a ground truth. Our experimental analysis consists of two parts. The first is comparison of different feature extraction and aggregation techniques in deciding the acceptance/rejection of the eligibility of  configurations. The second part is the study of how the eligible configurations conform with the stylized facts, which are the important realism metrics in financial markets.
Stylized fact similarity to the ground truth for the accepted configurations indicates that they are indeed more ``realistic" than the rejected models.

% , aiming at finding out a set of eligible parameters given the ground truth. 
%\indent We implemented our calibration algorithm on the ABIDES model, aiming at finding out a set of eligible parameters given the ground truth. The first part is about the calibration on synthetic ground truth, the second part is about applying variational optimization method to search for eligible parameters, and the last part is about the application of Distributionally Robust Optimization.

\subsection{Comparisons of Feature Extraction and Aggregation Techniques}
We first test the various approaches in Section \ref{sec:construction} on a specific example. Then we vary the number of input parameters and conduct ablation studies on our training methods.

\subsubsection{A Basic Example.}\label{sec:basic_example}
We generate limit order book time series that result from simulating the market from 9:30 to 10:00 AM with configurations that consists of an exchange agent, a market maker agent, $n$ noise agents and $m$ value agents that each follow a nosy observation of an Orstein-Uhlenbeck fundamental with the mean reversion rate $\kappa$, mean fundamental value $\overline{r}$ and arrival rate $\lambda_a$ (defined as in \citealt{byrd2019explaining}). The representative time series is taken as the log returns of the resultant mid-price sampled at every second, $r_t := \operatorname{log}(m_{t+1} / m_t)$, where $m_t$ is the mid price at second $t$ -- hence, in our experiment, each data point from which to extract relevant features is a time series of length 1799. 

We run $N = 100,000$ simulations for each configuration. The feature extraction network architectures are shown in Tables \ref{Tab:autoencoder_architecture}, \ref{Tab:GAN_architecture} and \ref{Tab:WGAN_architecture} in Appendix \ref{sec:details}. For the training procedure, autoencoder is trained with ADAM \citep{kingma2014adam} with learning rate $10^{-5}$, batch size $1000$, and number of epochs $20$. GAN and WGAN are trained with RMSprop \citep{hinton2012neural} with learning rate $5\times10^{-5}$, batch size $1000$, and number of epochs $300$. The parameters that we aim to calibrate are $m$, $n$, $\overline{r}$, $\kappa$ and $\lambda_a$.

\begin{figure}[h]
\centering
\includegraphics[width=145mm,scale=0.5]{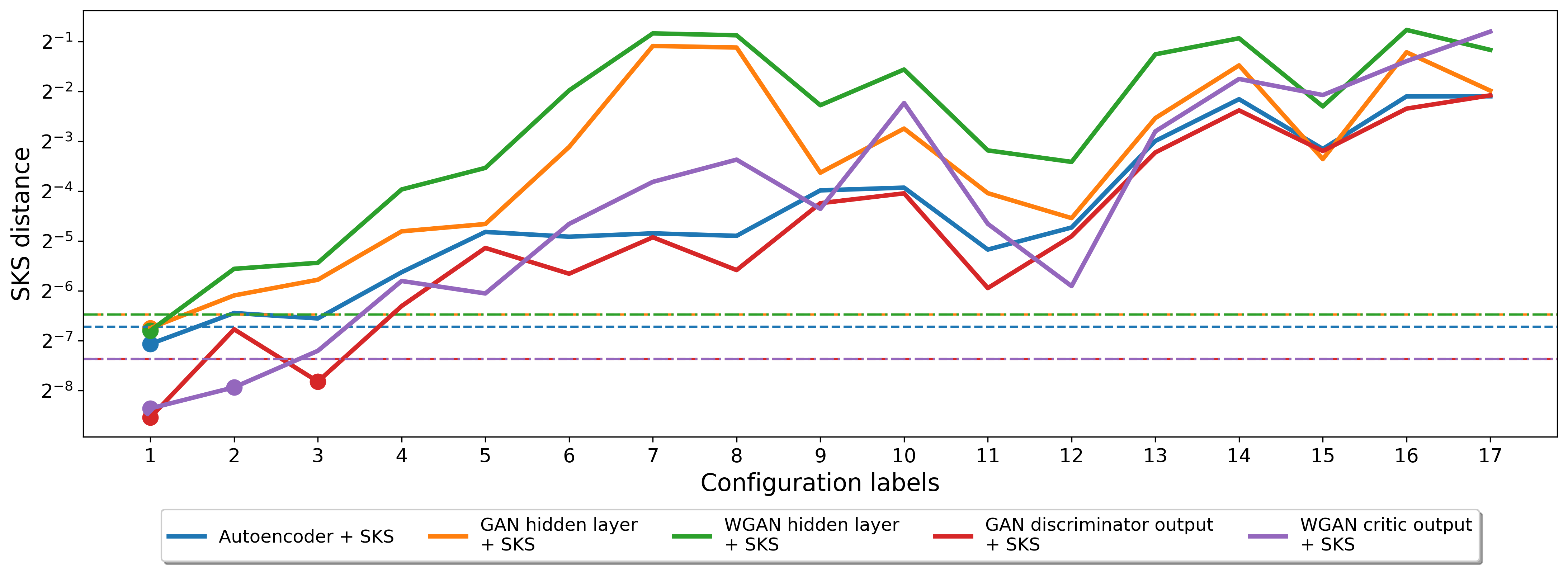}
\caption{\small Comparisons of different feature extraction methods with SKS. Eligible configurations are the ones below the thresholds (shown in dash lines) and plotted with dots. }
\label{Fig:KS}
\end{figure}

\begin{figure}[h]
\centering
\includegraphics[width=145mm,scale=0.5]{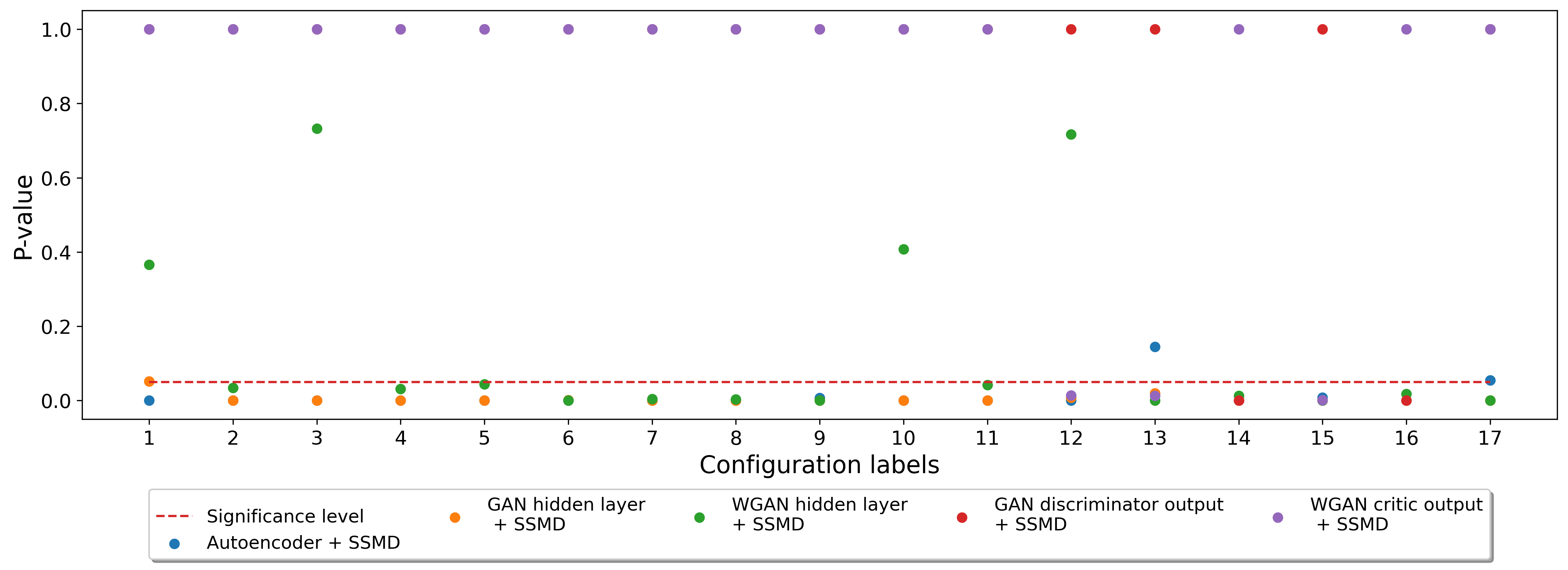}
\caption{\small Comparisons of different feature extraction methods with SSMD. Eligible configurations are the ones above the thresholds (shown in dash lines). }
\label{Fig:SSMD}
\end{figure}

\begin{figure}[h]
\centering
\includegraphics[width=145mm,scale=0.5]{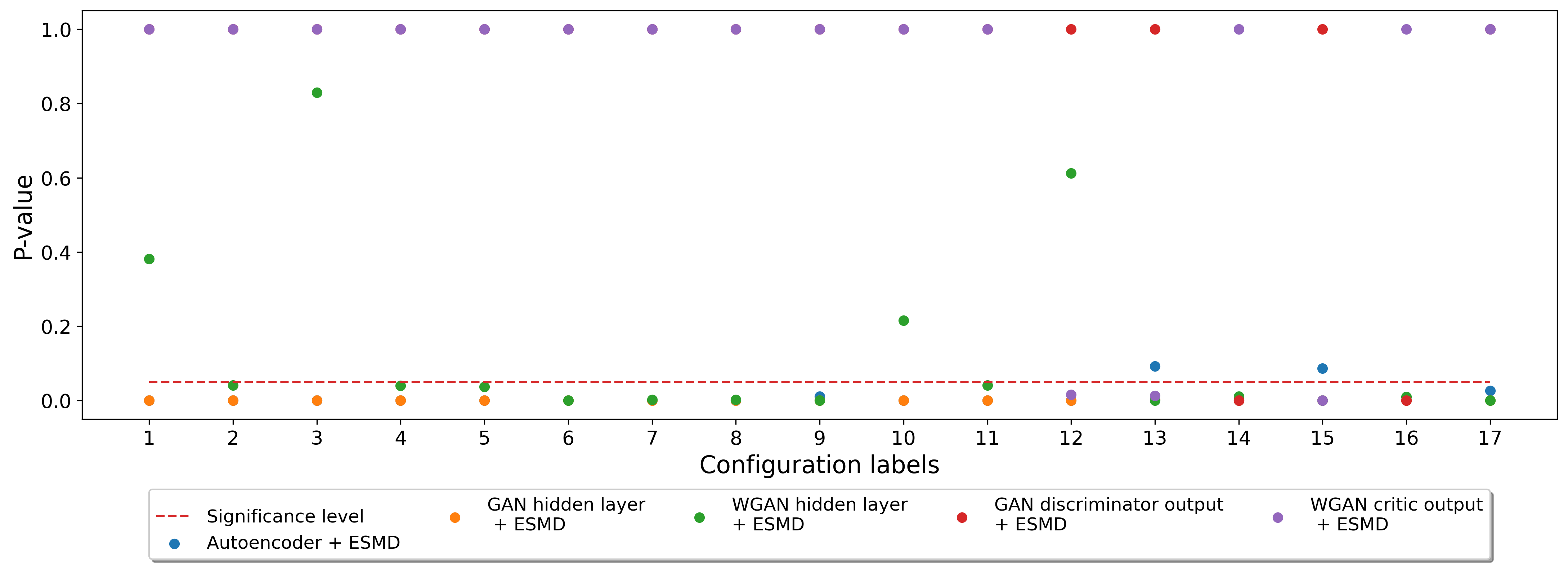}
\caption{\small Comparisons of different feature extraction methods with ESMD. Eligible configurations are the ones above the thresholds (shown in dash lines). }
\label{Fig:ESMD}
\end{figure}

We compare combinations of the five feature extraction methods: autoencoder hidden layer, GAN hidden layer, WGAN hidden layer, GAN discriminator output and WGAN critic output, and three aggregation methods: SKS, SSMD and ESMD. We set the confidence level $1-\alpha=0.95$. Figures~\ref{Fig:KS},~\ref{Fig:SSMD}~and~\ref{Fig:ESMD} summarize the calibration results of SKS, SSMD and ESMD respectively. In all figures, Models 1 to 17 in the $x$-axis refer to 17 different configurations on ABIDES (see Appendix~\ref{sec:details} for the details). Model 1 represents the ground truth configuration. Models 2 to 5 are the configurations close to the truth, because their parameters are chosen with smallest perturbations from the ground truth. The remaining models refer to configurations with parameters that differ significantly. In Figure~\ref{Fig:KS}, we use the value of the SKS distance as the $y$-axis, so that an SKS value below the cutoff line of the corresponding aggregation method indicates eligibility. The accepted configurations are marked as a circle, and the rejected configurations are shown without any markers. In Figures \ref{Fig:SSMD} and \ref{Fig:ESMD}, we convert the scale of SSMD and ESMD into a ``$p$-value" on the $y$-axis under a normal approximation, as discussed in Section~\ref{sec:aggregation}. The ``$p$-value" here measures how often the distance is smaller than the critical value. Correspondingly, a dot value above the cutoff $0.05$ indicates eligibility. This $p$-value scaling for SSMD and ESMD serves to facilitate easier visual comparison than in the original scale. 

From these results, we see that SKS exhibits the most reasonable trend across the board, where configurations close to the ground truth are classified as eligible while others are not. In comparison, with SSMD and ESMD, some methods either cannot recover the truth or misclassify ineligible configurations as eligible. Even though some extraction methods with SSMD or ESMD are able to recover the truth, they also classify many of the other configurations as eligible. Thus, with the same feature extraction techniques, SSMD and ESMD appear to underperform compared to SKS in terms of conservativeness.

% In detail, Figure \ref{Fig:KS} is a line plot, where the x-axis shows the configurations, and y-axis shows the Sup KS distance. Figure \ref{Fig:SSMD} and Figure \ref{Fig:ESMD} are scatter plots, where the x-axes still show the configurations, but the y-axes are about $p$-value, measuring how often the distance is less than the critical value. If the $p$-value is less than the significance level, the corresponding configuration is then rejected. 

%(I DON'T SEE THE LOGIC FROM THE ABOVE SENTENCE TO THE FOLLOWING SENTENCE HERE - A SEPARATE OBSERVATION OR TOGETHER AS A WHOLE CRITISICM ON SUPSMD AND ESMD?; ALSO USE SHORTHAND OF METHODS THAT WE INTRODUCE, INCLUDING IN THE CAPTIONS OF FIGURES) On the other hand, some extraction methods with the supremum of sampled mean difference or the ellipsoidal sampled mean difference are able to recover the truth; nevertheless, they also classify many of the other configurations to be eligible, representing that fact that the supremum of KS statistic can provide with a much less conservative calibration than the other two aggregation statistics. (SAMPLE SIZE ETC? OR ANY OTHER MISSING EXPERIMENT DETAILS? SHOULD BE PUT AT THE BEGINNING)

\subsubsection{Calibration with Different Numbers of Model Parameters.}
Section \ref{sec:basic_example} indicates that combining SKS with different feature extraction methods outperforms other aggregation approaches, not only in recovering the truth but also in eliminating false candidates. This subsection aims to investigate the robustness of this observation. In particular, we investigate the performance of SKS in calibrating different numbers of model parameters.

Table \ref{Tab:1} shows five sets of parameters to be calibrated. For each set, we investigate a total of $20$ configuration candidates. Here, the data we calibrate against is the trading mid-price -- the setting is otherwise the same as in Section \ref{sec:basic_example}.

\begin{table}[h]
\centering
%\begin{tabular}{ |c|c|c|c|c|c| } 
\begin{tabular}{|P{2.2cm}|P{2.2cm}|P{2.2cm}|P{2.2cm}|P{2.2cm}|P{2.2cm}|}
 \hline
 Case No. & $1$ & $2$ & $3$ & $4$ & $5$\\ 
 \hline
 Parameters & $m$ & $\kappa, \lambda_a$ & $m, \kappa, \lambda_a$ & $m, \overline{r}, \kappa, \lambda_a$ & $m, n, \overline{r}, \kappa, \lambda_a$\\
 \hline
\end{tabular}
 \caption{Parameters to calibrate in each case}
 \label{Tab:1}
\end{table}

\begin{figure}[h]
\centering
\includegraphics[width=145mm,scale=0.4]{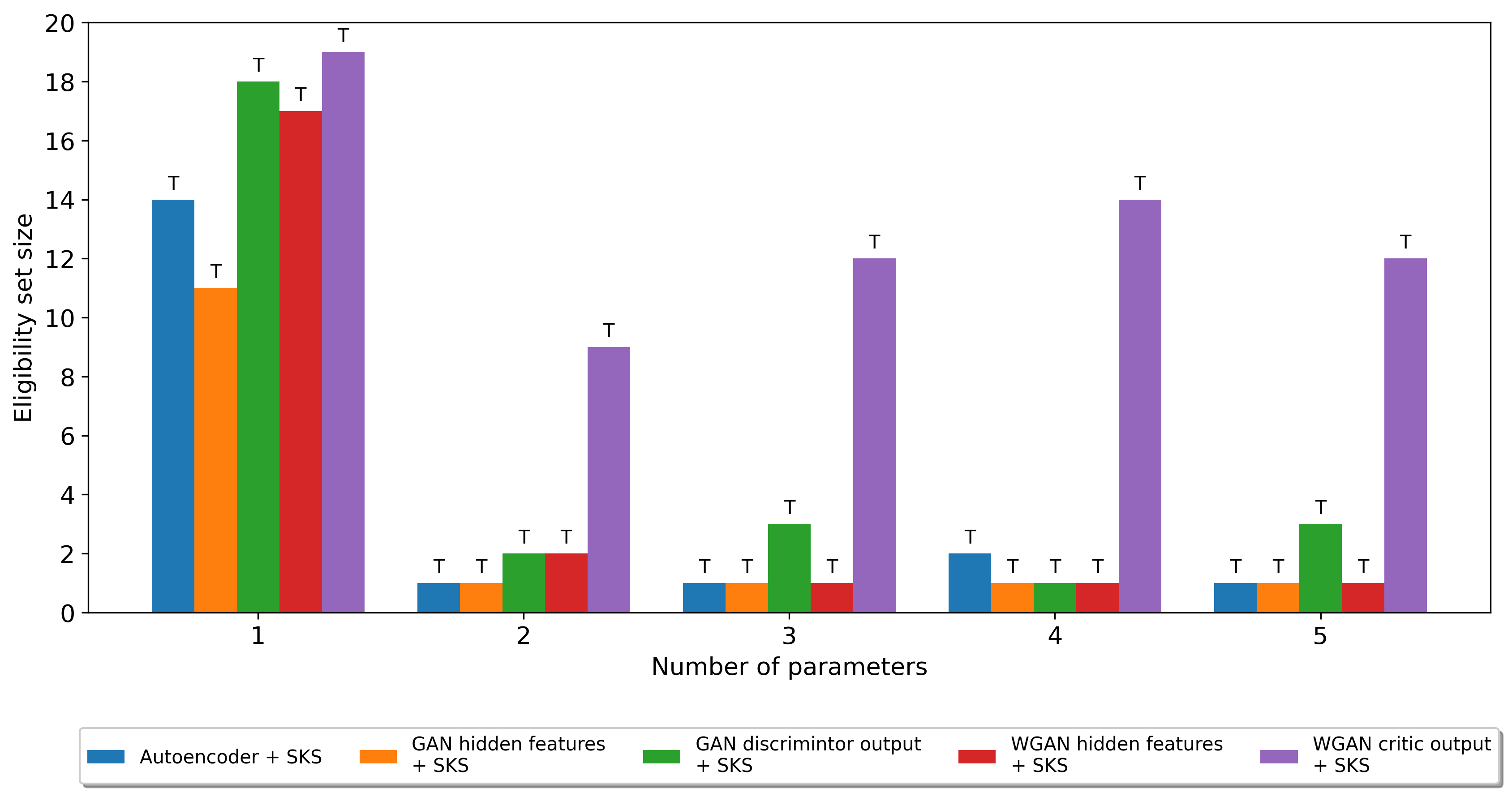}
\caption{\small Comparisons of different feature extraction method with SKS statistics. The annotations over the bars denote whether the true case can be recovered. ``T" means true and ``F" means false.}
\label{Fig:five_cases}
\end{figure}

Figure \ref{Fig:five_cases} shows the calibration results on $1$ to $5$ parameter sets. If a method recovers the true parameter and simultaneously maintains a small eligibility set (among the 20 configurations), then it is viewed as superior. In terms of the first metric, correctness, all of the methods are able to recover the truth. In terms of the second metric, conservativeness, GAN hidden features appears to be the best, since it consistently achieves the minimal eligibility set size while able to uncover the truth. Autoencoder follows next by achieving the smallest set size in $3$ cases and the second smallest set size in $2$ cases. WGAN hidden features performs a bit more conservatively than autoencoder because in almost all cases, WGAN hidden feature is able to attain the smallest or the second smallest set size except for case $1$. GAN discriminator output is also slightly more conservative than those methods mentioned before, and WGAN critic output is the most conservative one among all methods. Note that the calibration performance comparisons in terms of correctness and conservativeness among different methods are similar within each case, which further showcases the robustness of our methodologies. 

Therefore, in order to achieve a low conservativeness, we would first recommend using GAN hidden layer as a feature extraction method, followed by the autoencoder then WGAN hidden layer. GAN discriminator output or WGAN critic output are less recommended because of their over-conservativeness as compared with the first three methods.
%Therefore, in order to achieve a high chance of correctness, we recommend using autoencoder or WGAN hidden layer. However, if one focuses more on conservativeness at the cost of correctness, GAN hidden layer could be a better choice. WGAN critic output is not recommended because of the over-conservativeness, and GAN discriminator output is not recommended either because of the low correctness.

%(ANY CONCLUSION ON WHAT EXTRACTION METHOD TO RECOMMEND?)

\subsubsection{Ablation Study.}\label{sec:ablation_study}

We further study the performances of different approaches when the number of extracted features, $K$, varies. The magnitude of $K$ is controlled by the size of the final hidden layer in each neural network. We vary the size of this layer to $15$, $29$, $57$, $113$, $225$, and $450$. The network architecture can be adjusted by adding or removing a block of $\langle$Convolutional, Leaky ReLU, Dropout$\rangle$ layers in order to increase or decrease the network size by a factor of $2$. In each setting, we measure the Type I error probability and $q - \text{SKS}$ over $500$ experiments, where $q$ is calculated as the critical value of SKS. Intuitively, the magnitude of the quantity $q - \text{SKS}$ measures the gap between the SKS statistic and the threshold $q$. The larger the gap is, the more configurations can be considered as eligible, meaning that the method is more conservative. 

\begin{figure}[h]
    \hspace*{1.6cm}
    \includegraphics[width=130mm,scale = 0.4]{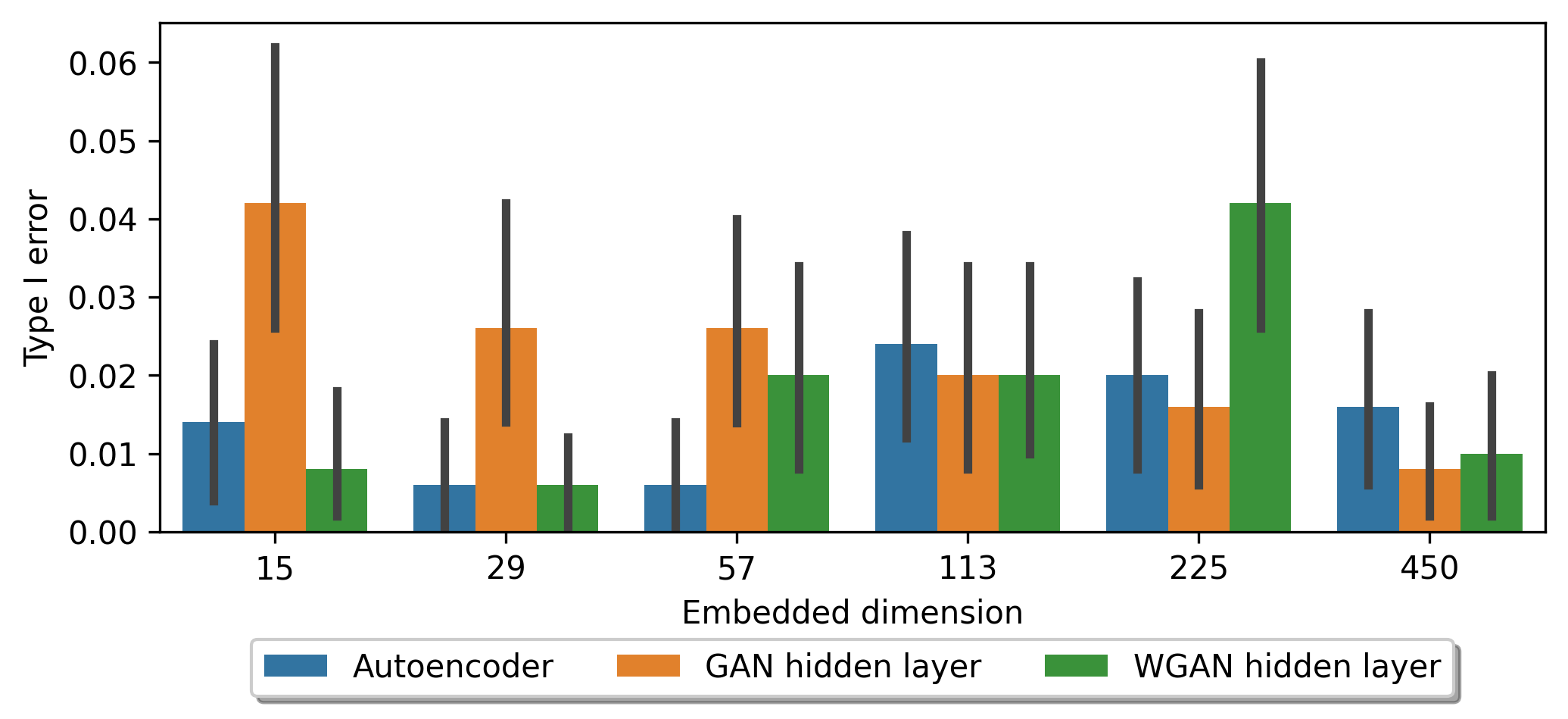}
    \caption{Type I error probability of Autoencoder, GAN hidden features extraction, WGAN hidden features extraction combined with SKS statistics as the dimension changes.}
    \label{fig:Type_I}
\end{figure}

\begin{figure}[h]
   \centering
    \includegraphics[width=110mm,scale = 0.4]{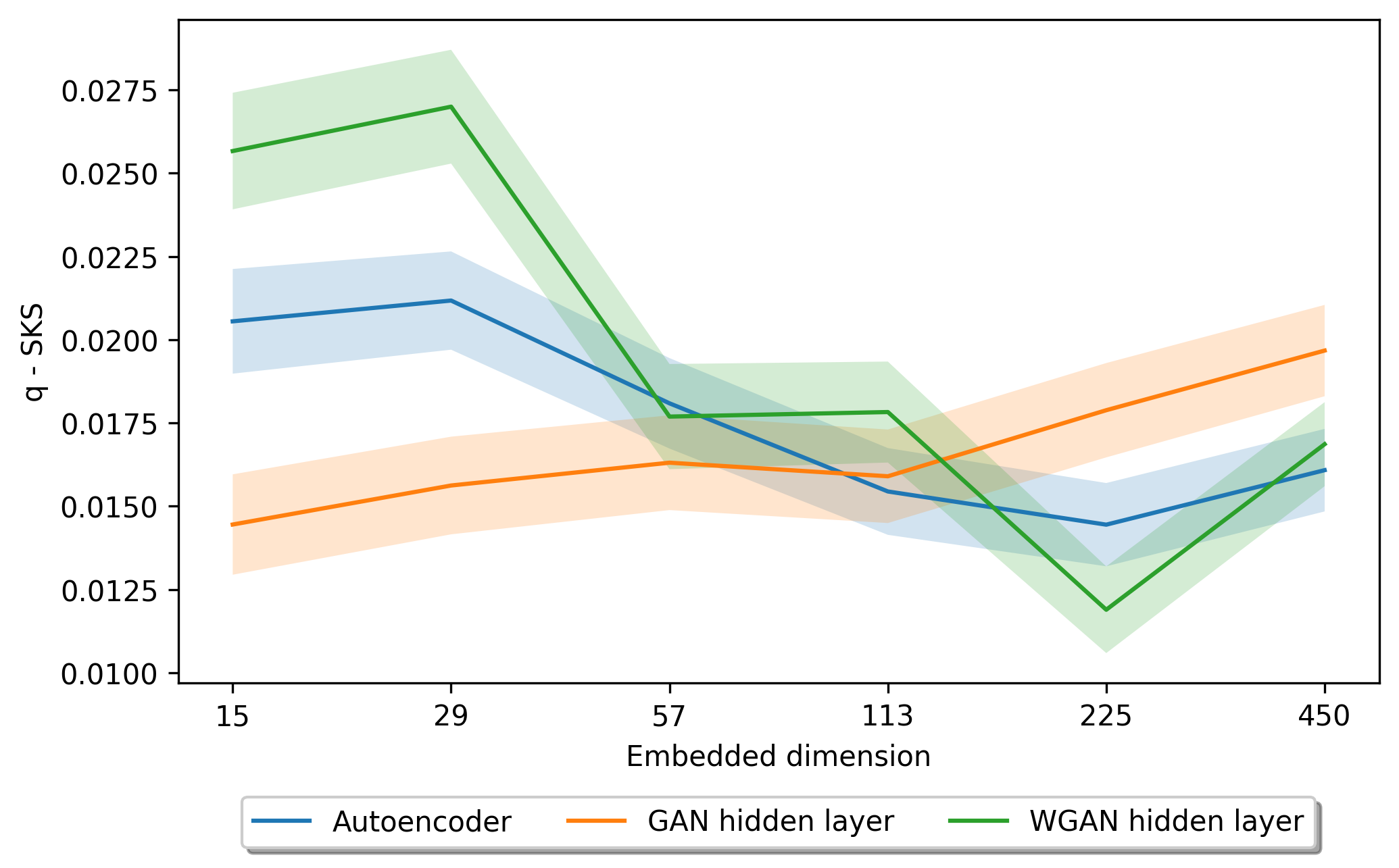}
    \caption{The difference between critical value and SKS of Autoencoder, GAN hidden features extraction, WGAN hidden features extraction combined with SKS statistics as the dimension changes. To get a sense of the estimation uncertainty, we also show shaded areas which have width $0.2$ times the standard deviation.}
    \label{fig:q_minus_ks}
\end{figure}

Figures \ref{fig:Type_I} and \ref{fig:q_minus_ks} present the results of the ablation study. First, we do not observe a clear trend of Type I error probability as the number of extracted features changes, and it is not obvious to see any methods consistently outperforming others. For $q - \text{SKS}$, there is no clear pattern to distinguish its performance either. Nonetheless, all methods have Type I error probabilities less than or equal to the nominal level $\alpha=0.05$. Furthermore, all methods constantly obtain a positive average value of $q - \text{SKS}$, meaning that the ground truth can still be recovered even with different hidden feature dimensions. These two observations once again verify the power of our methodology.

%  Even though it has been shown that our calibration framework is able to recover the true configuration along with some other configurations, the ability of those calibrated agent configurations to reproduce some stylized facts of the financial markets remains unclear. 
\subsection{Realism Metrics}
Properties of limit order book market behavior that are repeated across a wide range of instruments, markets, and time periods are referred to as stylized facts \citep{BouchaudStatistical,cont2001empirical,Bouchaud_book}. Evaluating the statistical properties of simulated asset returns, order volumes, order arrival times, order cancellations, etc and comparing them to those generated from real historical data allows us to infer the level of fidelity of a simulation; hence, stylized facts can be used as simulated realism metrics \citep{vyetrenko2019get}. In this subsection, we investigate how our eligible models can match the ground truth in terms of these metrics.

% Therefore, we leverage several related metrics to further 
% explore the reproducing ability of our calibrated configurations. 
We investigate the following metrics. Denote the midprice at time $t$ as $m_{t}$, and the log return at time scale $\Delta t$ as $r_{t, \Delta t} = \log m_{t+\Delta t} - \log m_{t}$:
\begin{itemize}
    \item Heavy tails and aggregational normality: The distribution of the asset prices exhibits a fat tail. However, as $\Delta t$ increases, the distribution tends to show a slimmer tail, and more like a normal distribution.
    \item Absence of autocorrelation: The autocorrelation function $corr(r_{t+\tau, \Delta t}, r_{t, \Delta t})$ becomes insignificant as $\tau$ gets longer than $20$ minutes.
    \item Volatility clustering: High-volatility events usually cluster in time. The autocorrelation function $corr(r_{t+\tau, \Delta t}^{2}, r_{t, \Delta t}^{2})$ is used to measure volatility clustering. Empirical results on various equities indicate that this quantity remains positive over several days.
\end{itemize}

We examine the realism metrics of the $17$ configurations studied in Section \ref{sec:basic_example}. The simulation sample size for each configuration is $100,000$. For the first metric, we examine minutely log return; for the second metric, we examine the autocorrelation of minutely log returns and take $\Delta t$ to be $25$ minutes; for the third metric, we examine the $10$-second log return and the lags of the autocorrelation function ranging from $1$ to $10$.

\begin{figure}[h]
    \centering
    \includegraphics[width=168mm,scale = 0.4]{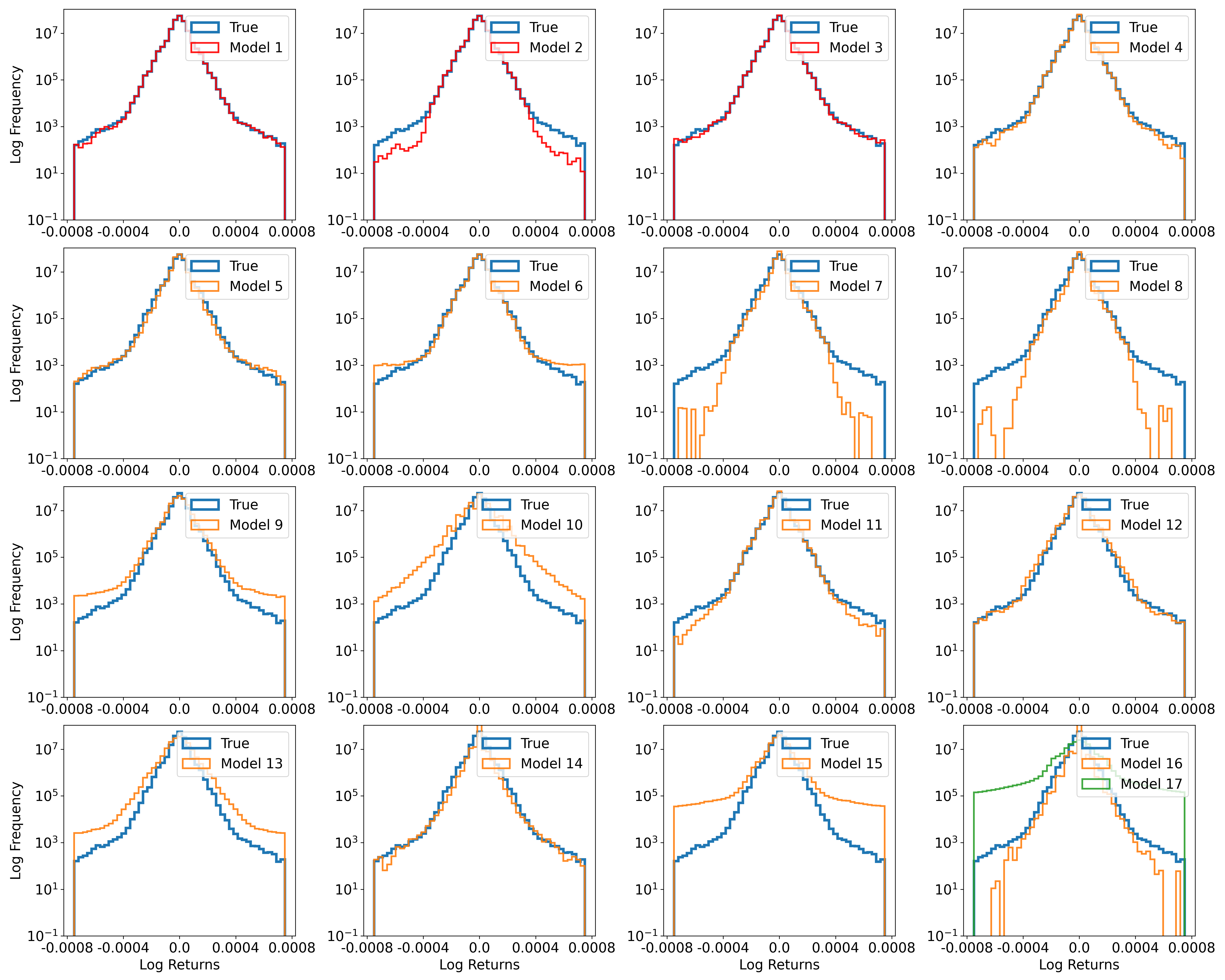}
    \caption{One-minute log return distributions. Accepted models are the first three in the first row, which are shown in red.}
    \label{Fig:minute_log_return}
\end{figure}

\begin{figure}[h]
    \centering
    \includegraphics[width=168mm,scale = 0.4]{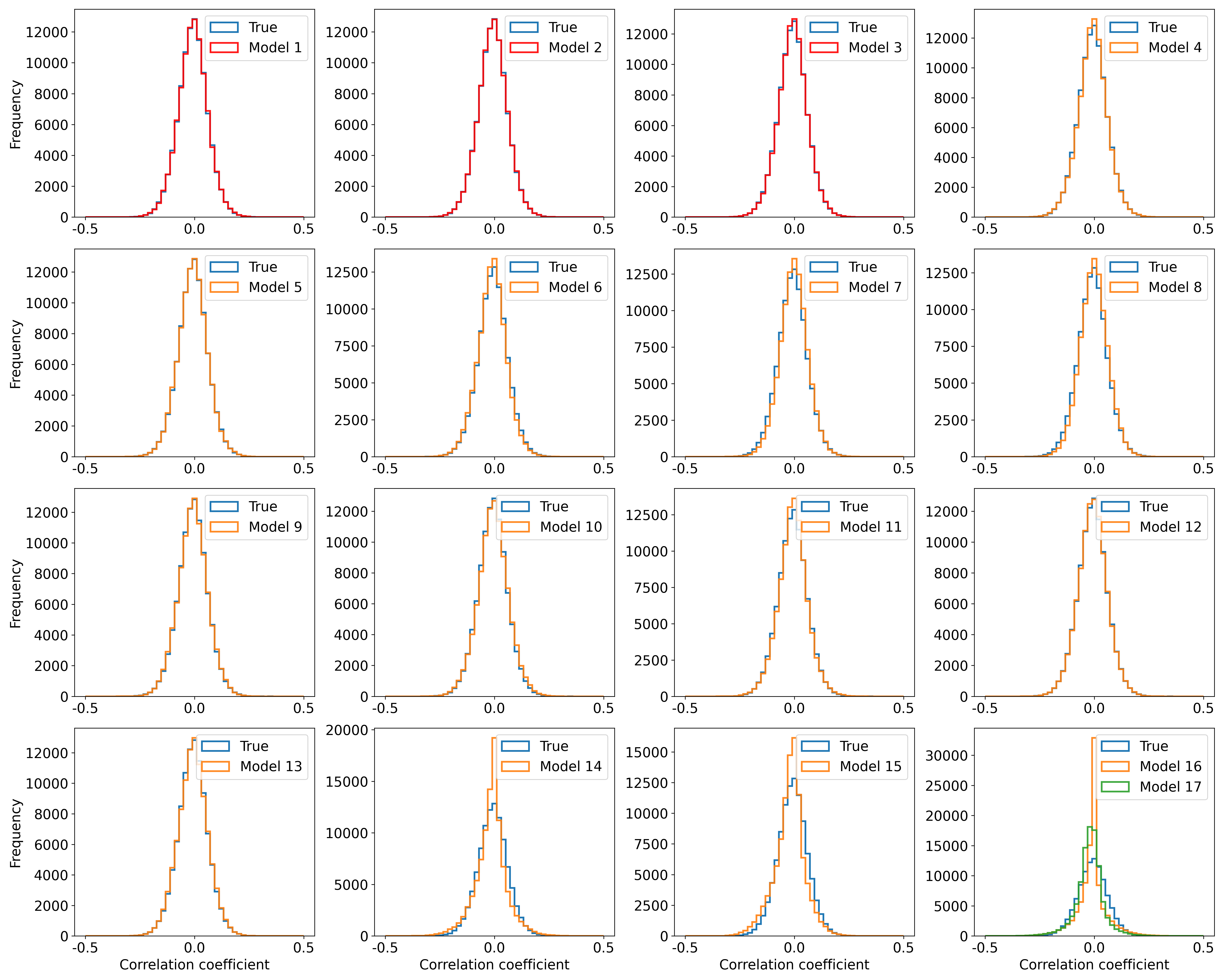}
    \caption{Distributions of return autocorrelation at 25 minutes. Accepted models are the first three in the first row, which are shown in red.}
    \label{Fig:log_return_autocorrelation}
\end{figure}

\begin{figure}[h]
    \centering
    \includegraphics[width=168mm,scale = 0.4]{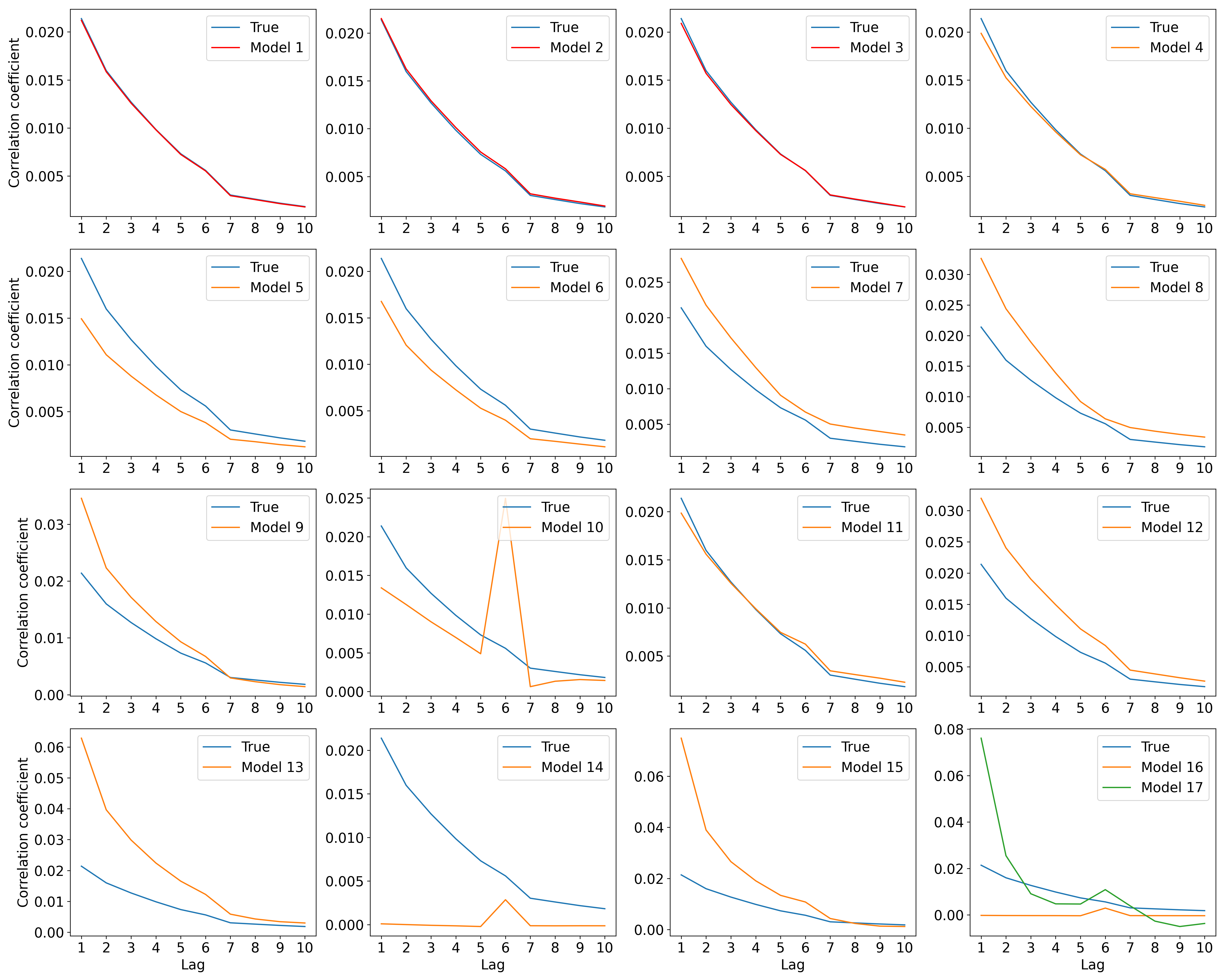}
    \caption{Average autocorrelation of squared returns as a function of time lag from 1 to 10. Squared returns are estimated for every 10 seconds. Accepted models are the first three in the first row, which are shown in red.}
    \label{Fig:volatility_clustering}
\end{figure}

Figures \ref{Fig:minute_log_return}, \ref{Fig:log_return_autocorrelation} and \ref{Fig:volatility_clustering} show the realism metrics exhibited by the 17 configurations. Recall that, according to Figure \ref{Fig:KS}, the three eligible configurations under our framework are Model 1 (accepted by all methods), Model 2 (accepted only by WGAN critic output with SKS), and Model 3 (accepted only by GAN discriminator output with SKS). %(LET'S MAKE THE TERMINOLOGY CONSISTENT THROUGHOUT THIS SECTION, IT SEEMS MORE NATURAL TO USE MODEL 1 INSTEAD OF LABEL 1 ETC.)
Model 1, in particular, is the true configuration. We observe that the distribution of the autocorrelation function value at $\Delta = 20$ minutes has most of the mass around $0$. Therefore, from this perspective, the true configuration follows the empirical rule of the market. For volatility clustering, the average autocorrelation function of squared returns among $100,000$ samples decays as the lag (based on $10$ seconds) increases but remains positive. However, we have also found that the autocorrelation function can become negative as we gradually increase the length of the lag, thus deviating from the empirical rule in this regime. Nonetheless, the true configuration still loosely follows the empirical rule when the lag is not too large.

%its log return distribution is heavy-tailed, with a shape looking similar to a normal distribution (THE NORMAL DISTRIBUTION DOES NOT HAVE HEAVY TAIL, SO THIS DESCRIPTION IS A BIT PROBLEMATIC; PERHAPS YOU HAVE A PARTICULAR DEFINITION OF HEAVY TAIL OR SIMILARITY TO NORMAL?), the distribution of the autocorrelation function value at $\Delta = 25$ minutes has most of the mass around $-0.25$ to $0$, and the average %(WHAT DOES IT MEAN TO BE AVERAGE HERE?) autocorrelation function of squared returns among $100000$ samples decays as the lag increases but remains positive. This true configuration behaves similar to the empirical rules of the market (ANY REFERENCE TO SUPPORT THIS POINT?). 

In terms of the similarity between the ground truth and the simulated configurations, we see that Models 1 and 3 behave almost the same as the truth in all three metrics. Model 2 also behaves similarly in log return autocorrelation and volatility clustering (even closer to the truth compared with Model 3 in these two metrics), but it has a slimmer tail in the log return distribution as shown in Figure \ref{Fig:minute_log_return}. On the other hand, other models acting statistically like the truth include Models 4, 5, and 12. In detail, Model 4 is similar but slightly different from the truth in all these three metrics. Models 5 and 12 are close to the truth in the log return distribution and the log return autocorrelation, but deviate greatly in volatility clustering. For all the other ineligible configurations, their realism metric performances are disparate from the truth in more than two of the metrics. Some examples of this are Models 14 and 9. Model 14 is similar to the truth in log return distribution, but distinguishable in the log return autocorrelation and volatility clustering, and Model 9 is similar to the truth in log return autocorrelation, but bears a substantial difference in the log return distribution and volatility clustering. In summary, the accepted models are able to match the truth in terms of realism metrics to a great extent, while the rejected models either have a lot of difference in one of the metrics or have differences in two or more metrics.
%GIVE SOME EXAMPLES, AND ALSO MAKE CONCLUSION (THAT OUR ELIGIBLE MODELS APPEAR TO MATCH THE TRUTH IN TERMS OF REALISM METRICS)

% \subsubsection{Real Data Examples}

% \subsubsection{Applications in Decision Analytics}

\section*{Disclaimer}
This paper was prepared for informational purposes [“in part” if the work is collaborative with external partners] by the Artificial Intelligence Research group of JPMorgan Chase and Co and its affiliates (“J.P. Morgan”), and is not a product of the Research Department of J.P. Morgan.  J.P. Morgan makes no representation and warranty whatsoever and disclaims all liability, for the completeness, accuracy or reliability of the information contained herein.  This document is not intended as investment research or investment advice, or a recommendation, offer or solicitation for the purchase or sale of any security, financial instrument, financial product or service, or to be used in any way for evaluating the merits of participating in any transaction, and shall not constitute a solicitation under any jurisdiction or to any person, if such solicitation under such jurisdiction or to such person would be unlawful.

\ACKNOWLEDGMENT{We gratefully acknowledge support from the JP Morgan Chase Faculty Research Award.}

\bibliographystyle{informs2014} 
% \bibliographystyle{ormsv080}
% abbrv}

% AUTHOR: Include your bib file here
\bibliography{bibliography}

\ECSwitch
\ECHead{Supplementary Materials}
\section{Proofs of Theorems}
\proof{Proof of Theorem \ref{thm:confidence}.}
Since 
\begin{equation*}
\begin{split}
        \sup_{x\in\R}|F_n^{\theta_0}(x)-F_N^{\theta_0}(x)|&\leq \sup_{x\in\R}\left(|F_n^{\theta_0}(x)-F^{\theta_0}(x)|+|F^{\theta_0}(x)-F_N^{\theta_0}(x)|\right)\\
        &\leq \sup_{x\in\R}|F_n^{\theta_0}(x)-F^{\theta_0}(x)|+\sup_{x\in\R}|F_N^{\theta_0}(x)-F^{\theta_0}(x)|,
\end{split}
\end{equation*}
we get that 
\begin{equation*}
\begin{split}
    &\mathbb{P}\left(\sup_{x\in\R}|F_n^{\theta_0}(x)-F_N^{\theta_0}(x)|>q_{1-\alpha}/\sqrt{N}\right)\\
    \leq &\mathbb{P}\left(\sup_{x\in\R}|F_n^{\theta_0}(x)-F^{\theta_0}(x)|+\sup_{x\in\R}|F_N^{\theta_0}(x)-F^{\theta_0}(x)|>q_{1-\alpha}/\sqrt{N}\right)\\
    \leq &\mathbb{P}\left(\sup_{x\in\R}|F_n^{\theta_0}(x)-F^{\theta_0}(x)|> \lambda q_{1-\alpha}/\sqrt{N}\right)+\mathbb{P}\left(\sup_{x\in\R}|F_N^{\theta_0}(x)-F^{\theta_0}(x)|>(1-\lambda)q_{1-\alpha}/\sqrt{N}\right)
\end{split}
\end{equation*}
for any $\lambda\in(0,1)$. It is known that $\sqrt{n}\sup_{x\in\R}|F_n^{\theta_0}(x)-F^{\theta_0}(x)|\Rightarrow \sup_{t\in\R}|BB(F^{\theta_0}(t))|$ and similarly  $\sqrt{N}\sup_{x\in\R}|F_N^{\theta_0}(x)-F^{\theta_0}(x)|\Rightarrow \sup_{t\in\R}|BB(F^{\theta_0}(t))|$ as $n,N\rightarrow\infty$ where $\Rightarrow$ stands for convergence in distribution. $n=\omega(N)$ as $N\rightarrow\infty$ implies that $n/N\rightarrow\infty$ as $N\rightarrow\infty$, and thus 
\begin{equation*}
    \mathbb{P}\left(\sup_{x\in\R}|F_n^{\theta_0}(x)-F^{\theta_0}(x)|> \lambda q_{1-\alpha}/\sqrt{N}\right)=\mathbb{P}\left(\sqrt{n}\sup_{x\in\R}|F_n^{\theta_0}(x)-F^{\theta_0}(x)|> \lambda q_{1-\alpha}\sqrt{n/N}\right)\rightarrow 0
\end{equation*}
as $N\rightarrow\infty$ for any $\lambda\in (0,1)$. Hence, 
\begin{equation*}
    \begin{split}
        \limsup_{N\rightarrow\infty}\mathbb{P}\left(\sup_{x\in\R}|F_n^{\theta_0}(x)-F_N^{\theta_0}(x)|>q_{1-\alpha}/\sqrt{N}\right)&
        \leq \limsup_{N\rightarrow\infty}\mathbb{P}\left(\sup_{x\in\R}|F_N^{\theta_0}(x)-F^{\theta_0}(x)|>(1-\lambda)q_{1-\alpha}/\sqrt{N}\right)\\
        &=\mathbb{P}\left(\sup_{t\in\R}|BB(F^{\theta_0}(t))|>(1-\lambda)q_{1-\alpha}\right)\\
        &\leq \mathbb{P}\left(\sup_{t\in[0,1}|BB(t)|>(1-\lambda)q_{1-\alpha}\right).
    \end{split}
\end{equation*}
By the arbitrariness of $\lambda$ and the definition of $q_{1-\alpha}$, we finally get that 
\begin{equation*}
    \limsup_{N\rightarrow\infty}\mathbb{P}\left(\sup_{x\in\R}|F_n^{\theta_0}(x)-F_N^{\theta_0}(x)|>q_{1-\alpha}/\sqrt{N}\right)\leq \mathbb{P}\left(\sup_{t\in[0,1]}|BB(t)|\geq q_{1-\alpha}\right)=\alpha.
\end{equation*}
\hfill\Halmos
\endproof

\proof{Proof of Theorem \ref{thm:error}.}
We know that 
\begin{equation*}
    |F_n^{\theta}(x)-F_N^{\theta_0}(x)|\geq |F^{\theta}(x)-F^{\theta_0}(x)|-|F_n^{\theta}(x)-F^{\theta}(x)|-|F_N^{\theta_0}(x)-F^{\theta_0}(x)|,
\end{equation*}
and then we get that 
\begin{equation*}
    \sup_{x\in\R}|F_n^{\theta}(x)-F_N^{\theta_0}(x)|\geq \sup_{x\in\R}|F^{\theta}(x)-F^{\theta_0}(x)|-\sup_{x\in\R}|F_n^{\theta}(x)-F^{\theta}(x)|-\sup_{x\in\R}|F_N^{\theta_0}(x)-F^{\theta_0}(x)|.
\end{equation*}
Therefore,
\begin{equation*}
    \begin{split}
        &\mathbb{P}\left(\sup_{x\in\R}|F_n^{\theta}(x)-F_N^{\theta_0}(x)|\leq q_{1-\alpha}/\sqrt{N}\right)\\
        \leq & \mathbb{P}\left(\sup_{x\in\R}|F^{\theta}(x)-F^{\theta_0}(x)|-\sup_{x\in\R}|F_n^{\theta}(x)-F^{\theta}(x)|-\sup_{x\in\R}|F_N^{\theta_0}(x)-F^{\theta_0}(x)|\leq q_{1-\alpha}/\sqrt{N}\right)\\
        \leq &\mathbb{P}\left(\sup_{x\in\R}|F_n^{\theta}(x)-F^{\theta}(x)|>\varepsilon_1\right)+\mathbb{P}\left(\sup_{x\in\R}|F_N^{\theta_0}(x)-F^{\theta_0}(x)|>\varepsilon_2\right).
    \end{split}
\end{equation*}
The last inequality is obtained using the fact that $q_{1-\alpha}/\sqrt{N}<\sup_{x\in\R}|F^{\theta}(x)-F^{\theta_0}(x)|-\varepsilon_1-\varepsilon_2$ under the conditions in the theorem. By the refined Dvoretzky–Kiefer–Wolfowitz (DKW) inequality \citep{Kosorok2008}, we get that 
\begin{equation*}
    \mathbb{P}\left(\sup_{x\in\R}|F_n^{\theta}(x)-F^{\theta}(x)|>\varepsilon_1\right)\leq 2e^{-2n\varepsilon_1^2},\mathbb{P}\left(\sup_{x\in\R}|F_N^{\theta_0}(x)-F^{\theta_0}(x)|>\varepsilon_2\right)\leq 2e^{-2N\varepsilon_2^2},
\end{equation*}
which concludes the proof.
\hfill\Halmos
\endproof

\proof{Proof of Theorem \ref{thm:guarantee}.}
By applying Theorem \ref{thm:error}, we have that 
\begin{equation*}
    \begin{split}
        &\mathbb{P}\left(\exists i=1,\cdots,m \text{ s.t. } \sup_{x\in\R}|F^{\theta_i}(x)-F^{\theta_0}(x)|>\varepsilon,\theta_i\in \hat{\mathcal E}_m\right)\\
        \leq & \sum_{i=1}^m \mathbb{P}\left(\sup_{x\in\R}|F^{\theta_i}(x)-F^{\theta_0}(x)|>\varepsilon,\theta_i\in \hat{\mathcal E}_m\right)\\
        \leq & \sum_{i=1}^m \mathbb{P}\left(\sup_{x\in\R}|F_n^{\theta_i}(x)-F_N^{\theta_0}(x)|\leq q_{1-\alpha}/\sqrt{N}
        \bigg|\sup_{x\in\R}|F^{\theta_i}(x)-F^{\theta_0}(x)|>\varepsilon\right)\\
        \leq & 2m\left(e^{-2n\varepsilon_1^2}+e^{-2N\varepsilon_2^2}\right).
    \end{split}
\end{equation*}
\hfill\Halmos
\endproof

\proof{Proof of Corollary \ref{cor:guarantee1}.}
For any $\varepsilon>0$, we pick $\varepsilon_1=\varepsilon_2=\varepsilon/3$. If $\log m=o(N)$ and $n=\Omega(N)$ as $N\to\infty$, then 
\begin{equation*}
    2m\left(e^{-2n\varepsilon_1^2}+e^{-2N\varepsilon_2^2}\right)=2\left(e^{\log m-2n\varepsilon_1^2}+e^{\log m-2N\varepsilon_2^2}\right)\to 0
\end{equation*}
as $N\to\infty$. By applying Theorem \ref{thm:guarantee}, we prove the corollary.
\hfill\Halmos
\endproof

\proof{Proof of Corollary \ref{cor:guarantee2}.}
Let $\varepsilon=\sqrt{\log m/m}$ and $\varepsilon_1=\varepsilon_2=\varepsilon/3$. If $m=o(N)$ and $n=\Omega(N)$ as $N\to\infty$, then 
\begin{equation*}
    2m\left(e^{-2n\varepsilon_1^2}+e^{-2N\varepsilon_2^2}\right)=2\left(e^{\log m-2n\log m/(9m)}+e^{\log m-2N\log m/(9m)}\right)\to 0
\end{equation*}
as $N\to\infty$. By applying Theorem \ref{thm:guarantee}, we prove the corollary.
\hfill\Halmos
\endproof

\proof{Proof of Theorem \ref{thm:confidence_variant}.}
See Theorem 1 in \citet{wei2012}.
\hfill\Halmos
\endproof

\proof{Proof of Theorem \ref{thm:error_variant}.}
Follow the proof of Theorem \ref{thm:error}. We could get that 
\begin{equation*}
    \begin{split}
        &\mathbb{P}\left(\sup_{x\in\R}|F_n^{\theta}(x)-F_N^{\theta_0}(x)|\leq \sqrt{\frac{n+N}{nN}}\sqrt{-\frac{1}{2}\log(\alpha/2)}\right)\\
        \leq & \mathbb{P}\left(\sup_{x\in\R}|F^{\theta}(x)-F^{\theta_0}(x)|-\sup_{x\in\R}|F_n^{\theta}(x)-F^{\theta}(x)|-\sup_{x\in\R}|F_N^{\theta_0}(x)-F^{\theta_0}(x)|\leq \sqrt{\frac{n+N}{nN}}\sqrt{-\frac{1}{2}\log(\alpha/2)}\right)\\
        \leq &\mathbb{P}\left(\sup_{x\in\R}|F_n^{\theta}(x)-F^{\theta}(x)|>\varepsilon_1\right)+\mathbb{P}\left(\sup_{x\in\R}|F_N^{\theta_0}(x)-F^{\theta_0}(x)|>\varepsilon_2\right).
    \end{split}
\end{equation*}
where the last inequality is obtained using the fact that $\sqrt{\frac{n+N}{nN}}\sqrt{-\frac{1}{2}\log(\alpha/2)}<\sup_{x\in\R}|F^{\theta}(x)-F^{\theta_0}(x)|-\varepsilon_1-\varepsilon_2$ under the conditions in the theorem. Then we apply the refined DKW inequality to conclude the proof.
\hfill\Halmos
\endproof

\proof{Proof of Theorem \ref{thm:guarantee_variant}.}
Follow the proof of Theorem \ref{thm:guarantee} except that we apply Theorem \ref{thm:error_variant} here.
\hfill\Halmos
\endproof

\proof{Proof of Corollary \ref{cor:guarantee1_variant}.}
Follow the proof of Corollary \ref{cor:guarantee1} except that we apply Theorem \ref{thm:guarantee_variant} here.
\hfill\Halmos
\endproof

\proof{Proof of Corollary \ref{cor:guarantee2_variant}.}
Follow the proof of Corollary \ref{cor:guarantee2} except that we apply Theorem \ref{thm:guarantee_variant} here.
\hfill\Halmos
\endproof

\proof{Proof of Theorem \ref{thm:confidence_bonferroni}.}
	From Theorem \ref{thm:confidence}, we know that for any $k$,
	$$
	\limsup_{n,N\rightarrow\infty}\mathbb{P}\left(\sup_{x\in\R}|F_{n,k}^{\theta_0}(x)-F_{N,k}^{\theta_0}(x)|> q_{1-\alpha/K}/\sqrt{N}\right)\leq \alpha/K,
	$$
	and thus 
	$$
	\limsup_{n,N\rightarrow\infty}\mathbb{P}\left(\exists k,\sup_{x\in\R}|F_{n,k}^{\theta_0}(x)-F_{N,k}^{\theta_0}(x)|> q_{1-\alpha/K}/\sqrt{N}\right)\leq \alpha.
	$$
\hfill\Halmos
\endproof

\proof{Proof of Theorem \ref{thm:confidence_bonferroni_generalized}.}
	Define $\lambda=(N/n)^{1/4}/(\log K)^{1/2}$. As $K\rightarrow\infty$, $\lambda=o(1/\log K)$ and $\lambda=\omega(\sqrt{N/n})$. Without loss of generality, we assume that $0<\lambda<1$. We have that 
	$$
	\begin{aligned}
	&\limsup_{K\rightarrow\infty}\mathbb{P}\left(\exists k,\sup_{x\in\R}|F_{n,k}^{\theta_0}(x)-F_{N,k}^{\theta_0}(x)|> q_{1-\alpha/K}/\sqrt{N}\right)\\
	\leq &\limsup_{K\rightarrow\infty}\sum_{k=1}^K \mathbb{P}\left(\sup_{x\in\R}|F_{n,k}^{\theta_0}(x)-F_{N,k}^{\theta_0}(x)|> q_{1-\alpha/K}/\sqrt{N}\right)\\
	\leq&\limsup_{K\rightarrow\infty}\sum_{k=1}^K\Bigg(\mathbb{P}\left(\sup_{x\in\R}|F_{n,k}^{\theta_0}(x)-F_{k}^{\theta_0}(x)|> \lambda q_{1-\alpha/K}/\sqrt{N}\right)\\
	&+\mathbb{P}\left(\sup_{x\in\R}|F_{N,k}^{\theta_0}(x)-F_{k}^{\theta_0}(x)|> (1-\lambda) q_{1-\alpha/K}/\sqrt{N}\right)\Bigg)\\
	\leq & \limsup_{K\rightarrow\infty}\sum_{k=1}^K 2\left(e^{-2\lambda^2q_{1-\alpha/K}^2n/N}+e^{-2(1-\lambda)^2q_{1-\alpha/K}^2}\right)\\
	=&\limsup_{K\rightarrow\infty} 2K\left(e^{-2\lambda^2q_{1-\alpha/K}^2n/N}+e^{-2(1-\lambda)^2q_{1-\alpha/K}^2}\right).
	\end{aligned}
	$$
	We know that $\alpha=2\sum_{v=1}^{\infty}(-1)^{v-1}e^{-2v^2q_{1-\alpha}^2}$, and hence $\alpha/(2Ke^{-2q_{1-\alpha/K}^2})\rightarrow 1$ as $K\rightarrow\infty$. Thus, $2q_{1-\alpha/K}^2-\log K\rightarrow -\log(\alpha/2)$ as $K\rightarrow\infty$. We have that 
	$$
	Ke^{-2\lambda^2q_{1-\alpha}^2n/N}=e^{\log K-(2q_{1-\alpha/K}^2-\log K)\sqrt{n/N}/\log K-\sqrt{n/N}}\rightarrow 0
	$$
	and 
	$$
	Ke^{-2(1-\lambda)^2q_{1-\alpha/K}^2}=e^{\log K-(1-\lambda)^2(2q_{1-\alpha/K}^2-\log K)-(1-\lambda)^2\log K}\rightarrow \alpha/2.
	$$
	Therefore,
	$$
	\limsup_{K\rightarrow\infty}\mathbb{P}\left(\exists k,\sup_{x\in\R}|F_{n,k}^{\theta_0}(x)-F_{N,k}^{\theta_0}(x)|> q_{1-\alpha/K}/\sqrt{N}\right)\leq \alpha.
	$$
\hfill\Halmos
\endproof

% \proof{Proof of Theorem \ref{thm:confidence_bonferroni_variant}}
% 	We have that 
% 	\begin{eqnarray*}
% 	&&\mathbb{P}\left(\exists k,\sup_{x\in\R}|F_{n,k}^{\theta}(x)-F_{N,k}^{\theta_0}(x)|> \sqrt{\frac{n+N}{nN}}\sqrt{-\frac{1}{2}\log(\frac{\alpha}{2K})}\right)\\
% 	&\leq&\sum_{k=1}^K \mathbb{P}\left(\sup_{x\in\R}|F_{n,k}^{\theta}(x)-F_{N,k}^{\theta_0}(x)|> \sqrt{\frac{n+N}{nN}}\sqrt{-\frac{1}{2}\log(\frac{\alpha}{2K})}\right).
% 	\end{eqnarray*}
% 	By Theorem 1 in \citet{wei2012}, we get that if $n=N\geq 4$, then 
% 	$$
% 		\mathbb{P}\left(\exists k,\sup_{x\in\R}|F_{n,k}^{\theta}(x)-F_{N,k}^{\theta_0}(x)|> \sqrt{\frac{n+N}{nN}}\sqrt{-\frac{1}{2}\log(\frac{\alpha}{2K})}\right)\leq 2.16863K\frac{\alpha}{2K}\leq 1.085\alpha
% 	$$
% 	and if $n=N\geq 458$, then 
% 	$$
% 	\mathbb{P}\left(\exists k,\sup_{x\in\R}|F_{n,k}^{\theta}(x)-F_{N,k}^{\theta_0}(x)|> \sqrt{\frac{n+N}{nN}}\sqrt{-\frac{1}{2}\log(\frac{\alpha}{2K})}\right)\leq 2K\frac{\alpha}{2K}= \alpha.
% 	$$
% \endproof

\proof{Proof of Theorem \ref{thm:error_bonferroni}.}
	Suppose that $k^*=\arg\max_{1\leq k\leq K}\sup_{x\in\R}|F_k^{\theta}(x)-F_k^{\theta_0}(x)|$. Then we have that 
	$$
	\mathbb{P}\left(\sup_{x\in\R}|F_{n,k}^{\theta}(x)-F_{N,k}^{\theta_0}(x)|\leq q_{1-\alpha/K}/\sqrt{N},\forall 1\leq k\leq K\right)\leq \mathbb{P}\left(\sup_{x\in\R}|F_{n,k^*}^{\theta}(x)-F_{N,k^*}^{\theta_0}(x)|\leq q_{1-\alpha/K}/\sqrt{N}\right).
	$$
	By applying Theorem \ref{thm:error}, we get the result.
\hfill\Halmos
\endproof

\proof{Proof of Theorem \ref{thm:ssmd_correctness}.}
It is well known that by CLT, 
\begin{equation*}
    \frac{\bar{X}_k-\bar{Y}_k}{\sqrt{\left(\frac1N+\frac1n\right)var(X_{1,k})}}\Rightarrow N(0,1).
\end{equation*}
Then by Slutsky's theorem, 
\begin{equation*}
    \frac{\bar{X}_k-\bar{Y}_k}{\sqrt{\left(\frac1N+\frac1n\right)\widehat{var}_k}}\Rightarrow N(0,1),
\end{equation*}
which concludes the proof.
\hfill\Halmos
\endproof

\proof{Proof of Theorem \ref{thm:sos_1}.}
We use $\Phi$ to denote the CDF of standard normal distribution and denote $q_1=\Phi^{-1}(1-\frac{\alpha}{2})$, $q_2=\Phi^{-1}(1-\frac{\alpha}{2K})$. Then $\eta=q_1\sqrt{\frac{1}{N}+\frac{1}{n}}$ and $\eta'=q_2\sqrt{\frac{1}{N}+\frac{1}{n}}$. We have that 
$$
p_1=\Phi\left(q_1-\frac{\Delta_1}{\sqrt{\frac{1}{N}+\frac{1}{n}}}\right)-\Phi\left(-q_1-\frac{\Delta_1}{\sqrt{\frac{1}{N}+\frac{1}{n}}}\right)
$$
and
$$
p_2=\left(1-\frac{\alpha}{K}\right)^{K-1}\left(\Phi\left(q_2-\frac{\Delta_1}{\sqrt{\frac{1}{N}+\frac{1}{n}}}\right)-\Phi\left(-q_2-\frac{\Delta_1}{\sqrt{\frac{1}{N}+\frac{1}{n}}}\right)\right).
$$
For fixed $K>1$ (which implies $q_1<q_2$), as $N,n\rightarrow\infty$, we have that 
$$
\frac{p_2}{p_1}\sim \left(1-\frac{\alpha}{K}\right)^{K-1}\frac{-q_1+\frac{\Delta_1}{\sqrt{\frac{1}{N}+\frac{1}{n}}}}{-q_2+\frac{\Delta_1}{\sqrt{\frac{1}{N}+\frac{1}{n}}}}\exp\left((q_2-q_1)\frac{\Delta_1}{\sqrt{\frac{1}{N}+\frac{1}{n}}}+\frac{q_1^2-q_2^2}{2} \right),
$$
which grows exponentially in $N$ and $n$.
\hfill\Halmos
\endproof

\proof{Proof of Theorem \ref{thm:sos_2}.}
We have that
$$
p_2:=\mathbb P(\forall k,|\bar{X}_k-\bar{Y}_k|\leq \eta_k')\leq \mathbb P(|\bar{X}_k-\bar{Y}_k|\leq \eta_k')=\mathbb P\left(\left|N\left(\Delta_k,\left(\frac{1}{N}+\frac{1}{n}\right)\Sigma_{kk}\right)\right|\leq\eta_k'\right)=:p_2'.
$$

We still use the notations of $q_1$ and $q_2$. Then $\eta_k=q_1\sqrt{(\frac{1}{N}+\frac{1}{n})\Sigma_{kk}}$ and $\eta_k'=q_2\sqrt{(\frac{1}{N}+\frac{1}{n})\Sigma_{kk}}$. We have that 
$$
p_1=\Phi\left(q_1-\frac{\Delta_k}{\sqrt{(\frac{1}{N}+\frac{1}{n})\Sigma_{kk}}}\right)-\Phi\left(-q_1-\frac{\Delta_k}{\sqrt{(\frac{1}{N}+\frac{1}{n})\Sigma_{kk}}}\right)
$$
and
$$
p_2'=\Phi\left(q_2-\frac{\Delta_k}{\sqrt{(\frac{1}{N}+\frac{1}{n})\Sigma_{kk}}}\right)-\Phi\left(-q_2-\frac{\Delta_k}{\sqrt{(\frac{1}{N}+\frac{1}{n})\Sigma_{kk}}}\right).
$$

Clearly, if $\Delta_k\ne 0$, then for fixed $K>1$, as $N,n\rightarrow\infty$, both $p_1$ and $p_2'$ converge to 0 exponentially in $N$ and $n$. Compared to $p_2'$, $p_1$ decreases exponentially faster. 

Now we only analyze $p_2'$. It is known that as $K\rightarrow\infty$, $q_2=O(\sqrt{\log K})$. If $\Delta_k\ne 0$ and $N=\omega(\log K)$, $n=\omega(\log K)$ as $K\rightarrow\infty$, then 
$$
p_2'\leq \bar{\Phi}\left(-q_2+\frac{|\Delta_k|}{\sqrt{(\frac{1}{N}+\frac{1}{n})\Sigma_{kk}}}\right)\rightarrow 0.
$$
\hfill\Halmos
\endproof

\proof{Proof of Theorem \ref{thm:esmd_correctness}.}
Use $\bar X$ and $\bar Y$ to denote the sample mean vectors. It is well known that 
\begin{equation*}
    \frac{(\bar X-\bar Y)}{\sqrt{\left(\frac1N+\frac1n\right)}}\Rightarrow N(0,\Sigma_X)
\end{equation*}
and hence 
\begin{equation*}
    \frac{\sum_{k=1}^K(\bar X_k-\bar Y_k)^2}{\frac1N+\frac1n}=\frac{(\bar X-\bar Y)^T(\bar X-\bar Y)}{\frac1N+\frac1n}\Rightarrow Z^T\Sigma_X Z.
\end{equation*}
Since $\hat \Sigma\to\Sigma_X$ almost surely, we get that $\frac{\eta}{\frac1N+\frac1n}$ converges to the $(1-\alpha)$-quantile of $Z^T\Sigma_X Z$ almost surely, which concludes the proof.
\hfill\Halmos
\endproof

\section{Details on Machine Learning Models and Feature Extraction} \label{ML model details}

\noindent\textbf{Auto-encoder}:\quad Auto-encoder is a unsupervised learning method \citep{baldi2012autoencoders}. It is made up of two components, an encoder $e: \mathcal{X} \rightarrow \mathcal{H_{X}}$, and an decoder $r:\mathcal{H_{X}}\rightarrow\mathcal{X}$. Given a target dimension $K$, encoder $e$ finds the hidden feature $H\in\R^K$ of the input $X\in\mathcal{X}$, and afterward decoder $r$ reconstructs the hidden feature $H$ and outputs the reconstruction $\hat{X}\in\mathcal{X}$. The bottleneck structure allows the network to find out the hidden feature of the input by itself, and the training procedure minimizes the Euclidean distance between the input samples and the reconstructed samples. The training of the network can be summarized as follows:
\begin{equation*}
    \min_{\theta, \hat{\theta}}\sum_{X}||r(e(X; \theta); \hat{\theta}) - X||_{2}^{2},
\end{equation*}
where $\theta$, $\hat{\theta}$ are the weights of encoder $e$ and decoder $r$.\\

\noindent\textbf{GAN}:\quad GAN is another type of unsupervised learning method \citep{goodfellow2014gan}. There are two adversarial modules contained in the GAN structure, which are discriminator $D$ and generator $G$. Given the real inputs $X_1,\dots,X_N\in\mathcal{X}$ from the real distribution $p_{real}$, the GAN network aims at learning the real distribution $p_{real}$ and outputs samples similar to the real inputs. Discriminator $D$ has input $X\in\mathcal{X}$ and summarizes $X$ to a single value $D(X)$, which measures the probability that input $X$ is a sample from the real distribution. The direct output of discriminator, $D(X)$, is able to serve as a hidden feature of $X$. Otherwise, we can also extract the output from the last hidden layer. This is not as concise as the direct output $D(X)$, but preserves more information about $X$. Generator $G$ has input $z$, as a sample from the latent space $\mathcal{Z}$, and outputs $\hat{X}\in\mathcal{X}$. Similar to two players in game theory, both discriminator $D$ and generator $G$ have their own objectives. The training process is a competition between $D$ and $G$, where $D$ is trained to tell the difference between the real inputs and the simulated ones, and $G$ is trained to learn the real distribution $p_{real}$ and fool discriminator $D$. For discriminator $D$, it minimizes the cross entropy,
\begin{equation*}
    \mathcal{L}_{D} = -\mathbb{E}_{x\sim p_{real}}\log D(x)-\mathbb{E}_{z\sim \mathcal{Z}}\log(1-D(G(z))).
\end{equation*}
As a competitor against $D$, generator $G$ maximizes the objective of $D$, so the loss function of $G$ is
\begin{equation*}
    \mathcal{L}_{G} = -\mathcal{L}_{D}.
\end{equation*}
Therefore, the optimization of GAN can be describe as a minimax game:
\begin{equation*}
    \min_{G}\max_{D} \mathbb{E}_{x\sim p_{real}}\log D(x) + \mathbb{E}_{z\sim \mathcal{Z}}\log(1-D(G(z))).
\end{equation*}

\noindent\textbf{WGAN}:\quad There are many variants of GAN, and a popular one is WGAN. Specifically, with an optimal discriminator $D^{*}$, the objective of GAN quantifies the similarity between the real input distribution and the generative distribution using the Jensen–Shannon (JS) divergence. However, the JS divergence becomes less sensitive in measuring the statistical distance between distributions when their supports are on low-dimensional manifolds, which further causes the instability in training \citep{arjovski2017wgan}. Therefore, \citet{arjovski2017wgan} proposes to adopt Wasserstein-1 distance as a better metric, which is
\begin{equation*}
    W(p_{real}, p_{gen}) = \inf_{\gamma\in\Pi(p_{real}, p_{gen})}\mathbb{E}_{(x,y)\sim\gamma}[||x-y||]
\end{equation*}
where $\Pi(p_{real},p_{gen})$ is the set of all possible joint distributions with marginal distributions $p_{real}$ and $p_{gen}$. However, due to the intractability of Wasserstein distance calculation, an alternative is to leverage the Kantorovich-Rubinstein duality, where the problem becomes
\begin{equation*}
    W(p_{real}, p_{gen}) = \sup_{||f||_{L}\leq1}\mathbb{E}_{x\sim p_{real}}[f(x)] - \mathbb{E}_{x\sim p_{gen}}[f(x)].
\end{equation*}
In practice, one can always extend the $1$-Lipschitz class functions to $L$-Lipschitz and the objective becomes $L\cdot W(p_{real}, p_{gen})$. Meanwhile, the $L$-Lipschitz continuity of neural network can be obtained by weight clipping \citep{arjovski2017wgan}. Therefore, the optimization of WGAN can be described as follows:
\begin{equation*}
    \min_{G}\max_{||D||_{L}\leq L} \mathbb{E}_{x\sim p_{real}}[D(x)] - \mathbb{E}_{z\sim \mathcal{Z}}[D(G(z))].
\end{equation*}

\noindent Since WGAN has a similar architecture to GAN, we can also use the direct output from the critic function or output from the last hidden layer as the hidden feature of the input.

\section{Experimental Details}\label{sec:details}

% \subsection{Section \ref{sec:basic_example}}

% The $17$ configurations in section \ref{sec:basic_example} are shown in Table \ref{Tab:configs_detail}:

\begin{table}[h]
\centering
%\begin{tabular}{ |c|c|c|c|c|c| } 
\begin{tabular}{|P{2.2cm}|P{2.2cm}|P{2.2cm}|P{2.2cm}|P{2.2cm}|P{2.2cm}|}
 \hline
 Config. & $m$ & $n$ & $\overline{r}$ & $\kappa$ & $\lambda_a$\\ 
 \hline
 1 & $100$ & $1000$ & $10^{5}$ & $1.67\times10^{-12}$ & $10^{-13}$\\
 2 & $105$ & $1050$ & $10^{5}$ & $1.67\times10^{-12}$ & $10^{-14}$\\
 3 & $90$ & $900$ & $10^{5}$ & $1.67\times10^{-12}$ & $10^{-13}$\\
 4 & $70$ & $700$ & $10^{5}$ & $1.67\times10^{-12}$ & $10^{-13}$\\
 5 & $95$ & $950$ & $1.1\times10^{5}$ & $1.5\times10^{-12}$ & $1.1\times10^{-13}$\\
 6 & $500$ & $1000$ & $10^{5}$ & $1.67\times10^{-12}$ & $10^{-13}$\\
 7 & $70$ & $700$ & $10^{5}$ & $8\times10^{-1}$ & $10^{-12}$\\
 8 & $100$ & $1000$ & $10^{5}$ & $5\times10^{-2}$ & $10^{-12}$\\
 9 & $200$ & $2000$ & $10^{5}$ & $1.67\times10^{-12}$ & $10^{-13}$\\
 10 & $10$ & $100$ & $10^{5}$ & $1.67\times10^{-12}$ & $10^{-13}$\\
 11 & $50$ & $500$ & $10^{5}$ & $1.67\times10^{-12}$ & $10^{-13}$\\
 12 & $105$ & $1050$ & $9\times10^{4}$ & $1.8\times10^{-12}$ & $8\times10^{-14}$\\
 13 & $100$ & $3000$ & $10^{5}$ & $1.67\times10^{-12}$ & $10^{-13}$\\ 
 14 & $10$ & $10$ & $10^{5}$ & $1.67\times10^{-12}$ & $10^{-12}$\\
 15 & $200$ & $1500$ & $10^{5}$ & $1.67\times10^{-12}$ & $10^{-12}$\\ 
 16 & $10$ & $10$ & $10^{5}$ & $1.67\times10^{-12}$ & $10^{-13}$\\ 
 17 & $50$ & $500$ & $10^{5}$ & $1.67\times10^{-12}$ & $10^{-11}$\\ 
 \hline
\end{tabular}
 \caption{Agent configurations}
 \label{Tab:configs_detail}
\end{table}

\begin{table}[h]
\centering
%\begin{tabular}{ |c|c|c|c|c|c| } 
\begin{tabular}{ |c|c| } 
 \hline
 Layer &  Network Architecture \\ 
 \hline
 1 - 3 & \makecell{Convolutional 1D (kernel size = $4$, stride = $2$)\\ Leaky ReLU (slope = 0.2), Dropout (probability = 0.2)}\\ 
 \hline
 4 - 6 & \makecell{Convolutional 1D (kernel size = $4$, stride = $2$)\\ Leaky ReLU (slope = 0.2), Dropout (probability = 0.2)}\\ 
 \hline
 7 - 9 & \makecell{Convolutional 1D (kernel size = $4$, stride = $2$)\\ Leaky ReLU (slope = 0.2), Dropout (probability = 0.2)}\\ 
 \hline
 10 - 12 & \makecell{Convolutional 1D (kernel size = $4$, stride = $2$)\\ Leaky ReLU (slope = 0.2), Dropout (probability = 0.2)}\\ 
 \hline
 13 - 15 & \makecell{Convolutional 1D (kernel size = $4$, stride = $2$)\\ Leaky ReLU (slope = 0.2), Dropout (probability = 0.2)}\\ 
 \hline
 16 - 18 & \makecell{Transposed Convolutional 1D (kernel size = $5$, stride = $2$)\\ Leaky ReLU (slope = 0.2), Dropout (probability = 0.2)}\\ 
 \hline
 19 - 21 & \makecell{Transposed Convolutional 1D (kernel size = $5$, stride = $2$)\\ Leaky ReLU (slope = 0.2), Dropout (probability = 0.2)}\\ 
 \hline
 22 - 24 & \makecell{Transposed Convolutional 1D (kernel size = $5$, stride = $2$)\\ Leaky ReLU (slope = 0.2), Dropout (probability = 0.2)}\\ 
 \hline
 25 - 27 & \makecell{Transposed Convolutional 1D (kernel size = $5$, stride = $2$)\\ Leaky ReLU (slope = 0.2), Dropout (probability = 0.2)}\\ 
 \hline
 28 - 30 & \makecell{Transposed Convolutional 1D (kernel size = $5$, stride = $2$)\\ Leaky ReLU (slope = 0.2), Dropout (probability = 0.2)}\\ 
 \hline
 31 - 32 & \makecell{Convolutional 1D (kernel size = $3$, stride = $2$), sigmoid}\\ 
 \hline
\end{tabular}
 \caption{Autoencoder architecture}
 \label{Tab:autoencoder_architecture}
\end{table}

\begin{table}[h]
\centering
%\begin{tabular}{ |c|c|c|c|c|c| } 
\begin{tabular}{ |c|c| } 
 \hline
 Layer &  Network Architecture \\ 
 \hline
 1 - 3 &  \makecell{Transposed Convolutional 1D (kernel size = $4$, stride = $2$)\\ Batch normalization, Leaky ReLU (slope = 0.2)}\\ 
 \hline
 4 - 6 &  \makecell{Transposed Convolutional 1D (kernel size = $4$, stride = $2$)\\ Batch normalization, Leaky ReLU (slope = 0.2)}\\ 
 \hline
 7 - 9 &  \makecell{Transposed Convolutional 1D (kernel size = $4$, stride = $2$)\\ Batch normalization, Leaky ReLU (slope = 0.2)}\\ 
 \hline
 10 - 12 &  \makecell{Transposed Convolutional 1D (kernel size = $4$, stride = $3$)\\ Batch normalization, Leaky ReLU (slope = 0.2))}\\ 
 \hline
 13 - 15 &  \makecell{Transposed Convolutional 1D (kernel size = $4$, stride = $3$)\\ Batch normalization, Leaky ReLU (slope = 0.2)}\\ 
 \hline
 16 - 18 & \makecell{Convolutional 1D (kernel size = $4$, stride = $2$)\\ Leaky ReLU (slope = 0.2), Dropout (probability = 0.2)}\\ 
 \hline
 19 - 21 & \makecell{Convolutional 1D (kernel size = $4$, stride = $2$)\\ Leaky ReLU (slope = 0.2), Dropout (probability = 0.2)}\\ 
 \hline
 22 - 24 & \makecell{Convolutional 1D (kernel size = $4$, stride = $2$)\\ Leaky ReLU (slope = 0.2), Dropout (probability = 0.2)}\\ 
 \hline
 25 - 27 & \makecell{Convolutional 1D (kernel size = $4$, stride = $2$)\\ Leaky ReLU (slope = 0.2), Dropout (probability = 0.2)}\\ 
 \hline
 28 - 30 & \makecell{Convolutional 1D (kernel size = $4$, stride = $2$)\\ Leaky ReLU (slope = 0.2), Dropout (probability = 0.2)}\\ 
 \hline
 31 - 32 & \makecell{Convolutional 1D (kernel size = $100$, stride = $1$), Dense (output dimension is $1$), Sigmoid}\\ 
 \hline
 
\end{tabular}
 \caption{GAN architecture}
 \label{Tab:GAN_architecture}
\end{table}

\begin{table}[h]
\centering
%\begin{tabular}{ |c|c|c|c|c|c| } 
\begin{tabular}{ |c|c| } 
 \hline
 Layer &  Network Architecture \\ 
 \hline
 1 - 3 &  \makecell{Transposed Convolutional 1D (kernel size = $4$, stride = $2$)\\ Batch normalization, Leaky ReLU (slope = 0.2)}\\ 
 \hline
 4 - 6 &  \makecell{Transposed Convolutional 1D (kernel size = $4$, stride = $2$)\\ Batch normalization, Leaky ReLU (slope = 0.2)}\\ 
 \hline
 7 - 9 &  \makecell{Transposed Convolutional 1D (kernel size = $4$, stride = $2$)\\ Batch normalization, Leaky ReLU (slope = 0.2)}\\ 
 \hline
 10 - 12 &  \makecell{Transposed Convolutional 1D (kernel size = $4$, stride = $3$)\\ Batch normalization, Leaky ReLU (slope = 0.2))}\\ 
 \hline
 13 - 15 &  \makecell{Transposed Convolutional 1D (kernel size = $4$, stride = $3$)\\ Batch normalization, Leaky ReLU (slope = 0.2)}\\ 
 \hline
 16 - 18 & \makecell{Convolutional 1D (kernel size = $4$, stride = $2$)\\ Leaky ReLU (slope = 0.2), Dropout (probability = 0.2)}\\ 
 \hline
 19 - 21 & \makecell{Convolutional 1D (kernel size = $4$, stride = $2$)\\ Leaky ReLU (slope = 0.2), Dropout (probability = 0.2)}\\ 
 \hline
 22 - 24 & \makecell{Convolutional 1D (kernel size = $4$, stride = $2$)\\ Leaky ReLU (slope = 0.2), Dropout (probability = 0.2)}\\ 
 \hline
 25 - 27 & \makecell{Convolutional 1D (kernel size = $4$, stride = $2$)\\ Leaky ReLU (slope = 0.2), Dropout (probability = 0.2)}\\ 
 \hline
 28 - 30 & \makecell{Convolutional 1D (kernel size = $4$, stride = $2$)\\ Leaky ReLU (slope = 0.2), Dropout (probability = 0.2)}\\ 
 \hline
 31 - 32 & \makecell{Convolutional 1D (kernel size = $100$, stride = $1$), Dense (output dimension is $1$), Linear}\\ 
 \hline
 
\end{tabular}
 \caption{WGAN architecture}
 \label{Tab:WGAN_architecture}
\end{table}

\end{document}